\pdfoutput=1
\documentclass[11pt,twoside,a4paper,cmspaper,final,collab]{cms-tdr}
\def\svnVersion{13:14138M}\def\cmsCernNo{2010-019}\def\cmsCernDate{2010/07/14}\def\cmsMessage{Submitted to the European Physical Journal C}

\begin{document}\cmsNoteHeader{TRK-10-001}
%
%
%

%
%
\hyphenation{env-iron-men-tal}
\hyphenation{had-ron-i-za-tion}
\hyphenation{cal-or-i-me-ter}
\hyphenation{de-vices}
%
\RCS$Revision: 14044 $
\RCS$HeadURL: svn+ssh://alverson@svn.cern.ch/reps/tdr2/papers/TRK-10-001/trunk/TRK-10-001.tex $
\RCS$Id: TRK-10-001.tex 14044 2010-07-12 03:20:23Z stenson $
%
%
%

\providecommand {\etal}{\mbox{et al.}\xspace} 
\providecommand {\ie}{\mbox{i.e.}\xspace}     
\providecommand {\eg}{\mbox{e.g.}\xspace}     
\providecommand {\etc}{\mbox{etc.}\xspace}     
\providecommand {\vs}{\mbox{\sl vs.}\xspace}      
\providecommand {\mdash}{\ensuremath{\mathrm{-}}} 

\providecommand {\Lone}{Level-1\xspace} 
\providecommand {\Ltwo}{Level-2\xspace}
\providecommand {\Lthree}{Level-3\xspace}

\providecommand{\ACERMC} {\textsc{AcerMC}\xspace}
\providecommand{\ALPGEN} {{\textsc{alpgen}}\xspace}
\providecommand{\CHARYBDIS} {{\textsc{charybdis}}\xspace}
\providecommand{\CMKIN} {\textsc{cmkin}\xspace}
\providecommand{\CMSIM} {{\textsc{cmsim}}\xspace}
\providecommand{\CMSSW} {{\textsc{cmssw}}\xspace}
\providecommand{\COBRA} {{\textsc{cobra}}\xspace}
\providecommand{\COCOA} {{\textsc{cocoa}}\xspace}
\providecommand{\COMPHEP} {\textsc{CompHEP}\xspace}
\providecommand{\EVTGEN} {{\textsc{evtgen}}\xspace}
\providecommand{\FAMOS} {{\textsc{famos}}\xspace}
\providecommand{\GARCON} {\textsc{garcon}\xspace}
\providecommand{\GARFIELD} {{\textsc{garfield}}\xspace}
\providecommand{\GEANE} {{\textsc{geane}}\xspace}
\providecommand{\GEANTfour} {{\textsc{geant4}}\xspace}
\providecommand{\GEANTthree} {{\textsc{geant3}}\xspace}
\providecommand{\GEANT} {{\textsc{geant}}\xspace}
\providecommand{\HDECAY} {\textsc{hdecay}\xspace}
\providecommand{\HERWIG} {{\textsc{herwig}}\xspace}
\providecommand{\HIGLU} {{\textsc{higlu}}\xspace}
\providecommand{\HIJING} {{\textsc{hijing}}\xspace}
\providecommand{\IGUANA} {\textsc{iguana}\xspace}
\providecommand{\ISAJET} {{\textsc{isajet}}\xspace}
\providecommand{\ISAPYTHIA} {{\textsc{isapythia}}\xspace}
\providecommand{\ISASUGRA} {{\textsc{isasugra}}\xspace}
\providecommand{\ISASUSY} {{\textsc{isasusy}}\xspace}
\providecommand{\ISAWIG} {{\textsc{isawig}}\xspace}
\providecommand{\MADGRAPH} {\textsc{MadGraph}\xspace}
\providecommand{\MCATNLO} {\textsc{mc@nlo}\xspace}
\providecommand{\MCFM} {\textsc{mcfm}\xspace}
\providecommand{\MILLEPEDE} {{\textsc{millepede}}\xspace}
\providecommand{\ORCA} {{\textsc{orca}}\xspace}
\providecommand{\OSCAR} {{\textsc{oscar}}\xspace}
\providecommand{\PHOTOS} {\textsc{photos}\xspace}
\providecommand{\PROSPINO} {\textsc{prospino}\xspace}
\providecommand{\PYTHIA} {{\textsc{pythia}}\xspace}
\providecommand{\SHERPA} {{\textsc{sherpa}}\xspace}
\providecommand{\TAUOLA} {\textsc{tauola}\xspace}
\providecommand{\TOPREX} {\textsc{TopReX}\xspace}
\providecommand{\XDAQ} {{\textsc{xdaq}}\xspace}

\providecommand {\DZERO}{D\O\xspace}     


\providecommand{\de}{\ensuremath{^\circ}}
\providecommand{\ten}[1]{\ensuremath{\times \text{10}^\text{#1}}}
\providecommand{\unit}[1]{\ensuremath{\text{\,#1}}\xspace}
\providecommand{\mum}{\ensuremath{\,\mu\text{m}}\xspace}
\providecommand{\micron}{\ensuremath{\,\mu\text{m}}\xspace}
\providecommand{\cm}{\ensuremath{\,\text{cm}}\xspace}
\providecommand{\mm}{\ensuremath{\,\text{mm}}\xspace}
\providecommand{\mus}{\ensuremath{\,\mu\text{s}}\xspace}
\providecommand{\keV}{\ensuremath{\,\text{ke\hspace{-.08em}V}}\xspace}
\providecommand{\MeV}{\ensuremath{\,\text{Me\hspace{-.08em}V}}\xspace}
\providecommand{\GeV}{\ensuremath{\,\text{Ge\hspace{-.08em}V}}\xspace}
\providecommand{\TeV}{\ensuremath{\,\text{Te\hspace{-.08em}V}}\xspace}
\providecommand{\PeV}{\ensuremath{\,\text{Pe\hspace{-.08em}V}}\xspace}
\providecommand{\keVc}{\ensuremath{{\,\text{ke\hspace{-.08em}V\hspace{-0.16em}/\hspace{-0.08em}}c}}\xspace}
\providecommand{\MeVc}{\ensuremath{{\,\text{Me\hspace{-.08em}V\hspace{-0.16em}/\hspace{-0.08em}}c}}\xspace}
\providecommand{\GeVc}{\ensuremath{{\,\text{Ge\hspace{-.08em}V\hspace{-0.16em}/\hspace{-0.08em}}c}}\xspace}
\providecommand{\TeVc}{\ensuremath{{\,\text{Te\hspace{-.08em}V\hspace{-0.16em}/\hspace{-0.08em}}c}}\xspace}
\providecommand{\keVcc}{\ensuremath{{\,\text{ke\hspace{-.08em}V\hspace{-0.16em}/\hspace{-0.08em}}c^\text{2}}}\xspace}
\providecommand{\MeVcc}{\ensuremath{{\,\text{Me\hspace{-.08em}V\hspace{-0.16em}/\hspace{-0.08em}}c^\text{2}}}\xspace}
\providecommand{\GeVcc}{\ensuremath{{\,\text{Ge\hspace{-.08em}V\hspace{-0.16em}/\hspace{-0.08em}}c^\text{2}}}\xspace}
\providecommand{\TeVcc}{\ensuremath{{\,\text{Te\hspace{-.08em}V\hspace{-0.16em}/\hspace{-0.08em}}c^\text{2}}}\xspace}

\providecommand{\pbinv} {\mbox{\ensuremath{\,\text{pb}^\text{$-$1}}}\xspace}
\providecommand{\fbinv} {\mbox{\ensuremath{\,\text{fb}^\text{$-$1}}}\xspace}
\providecommand{\nbinv} {\mbox{\ensuremath{\,\text{nb}^\text{$-$1}}}\xspace}
\providecommand{\percms}{\ensuremath{\,\text{cm}^\text{$-$2}\,\text{s}^\text{$-$1}}\xspace}
\providecommand{\lumi}{\ensuremath{\mathcal{L}}\xspace}
\providecommand{\Lumi}{\ensuremath{\mathcal{L}}\xspace}
%
\providecommand{\LvLow}  {\ensuremath{\mathcal{L}=\text{10}^\text{32}\,\text{cm}^\text{$-$2}\,\text{s}^\text{$-$1}}\xspace}
\providecommand{\LLow}   {\ensuremath{\mathcal{L}=\text{10}^\text{33}\,\text{cm}^\text{$-$2}\,\text{s}^\text{$-$1}}\xspace}
\providecommand{\lowlumi}{\ensuremath{\mathcal{L}=\text{2}\times \text{10}^\text{33}\,\text{cm}^\text{$-$2}\,\text{s}^\text{$-$1}}\xspace}
\providecommand{\LMed}   {\ensuremath{\mathcal{L}=\text{2}\times \text{10}^\text{33}\,\text{cm}^\text{$-$2}\,\text{s}^\text{$-$1}}\xspace}
\providecommand{\LHigh}  {\ensuremath{\mathcal{L}=\text{10}^\text{34}\,\text{cm}^\text{$-$2}\,\text{s}^\text{$-$1}}\xspace}
\providecommand{\hilumi} {\ensuremath{\mathcal{L}=\text{10}^\text{34}\,\text{cm}^\text{$-$2}\,\text{s}^\text{$-$1}}\xspace}


\providecommand{\zp}{\ensuremath{\mathrm{Z}^\prime}\xspace}


\providecommand{\kt}{\ensuremath{k_{\mathrm{T}}}\xspace}
\providecommand{\BC}{\ensuremath{{\mathrm{B}_{\mathrm{c}}}}\xspace}
\providecommand{\bbarc}{\ensuremath{{\overline{\mathrm{b}}\mathrm{c}}}\xspace}
\providecommand{\bbbar}{\ensuremath{{\mathrm{b}\overline{\mathrm{b}}}}\xspace}
\providecommand{\ccbar}{\ensuremath{{\mathrm{c}\overline{\mathrm{c}}}}\xspace}
\providecommand{\JPsi}{\ensuremath{{\mathrm{J}}\hspace{-.08em}/\hspace{-.14em}\psi}\xspace}
\providecommand{\bspsiphi}{\ensuremath{\mathrm{B}_\mathrm{s} \to \JPsi\, \phi}\xspace}
\providecommand{\AFB}{\ensuremath{A_\text{FB}}\xspace}
\providecommand{\EE}{\ensuremath{\mathrm{e}^+\mathrm{e}^-}\xspace}
\providecommand{\MM}{\ensuremath{\mu^+\mu^-}\xspace}
\providecommand{\TT}{\ensuremath{\tau^+\tau^-}\xspace}
\providecommand{\wangle}{\ensuremath{\sin^{2}\theta_{\text{eff}}^\text{lept}(M^2_\mathrm{Z})}\xspace}
\providecommand{\ttbar}{\ensuremath{{\mathrm{t}\overline{\mathrm{t}}}}\xspace}
\providecommand{\stat}{\ensuremath{\,\text{(stat.)}}\xspace}
\providecommand{\syst}{\ensuremath{\,\text{(syst.)}}\xspace}

\providecommand{\HGG}{\ensuremath{\mathrm{H}\to\gamma\gamma}}
\providecommand{\gev}{\GeV}
\providecommand{\GAMJET}{\ensuremath{\gamma + \text{jet}}}
\providecommand{\PPTOJETS}{\ensuremath{\mathrm{pp}\to\text{jets}}}
\providecommand{\PPTOGG}{\ensuremath{\mathrm{pp}\to\gamma\gamma}}
\providecommand{\PPTOGAMJET}{\ensuremath{\mathrm{pp}\to\gamma + \mathrm{jet}}}
\providecommand{\MH}{\ensuremath{M_{\mathrm{H}}}}
\providecommand{\RNINE}{\ensuremath{R_\mathrm{9}}}
\providecommand{\DR}{\ensuremath{\Delta R}}


\providecommand{\PT}{\ensuremath{p_{\mathrm{T}}}\xspace}
\providecommand{\pt}{\ensuremath{p_{\mathrm{T}}}\xspace}
\providecommand{\ET}{\ensuremath{E_{\mathrm{T}}}\xspace}
\providecommand{\HT}{\ensuremath{H_{\mathrm{T}}}\xspace}
\providecommand{\et}{\ensuremath{E_{\mathrm{T}}}\xspace}
\providecommand{\Em}{\ensuremath{E\hspace{-0.6em}/}\xspace}
\providecommand{\Pm}{\ensuremath{p\hspace{-0.5em}/}\xspace}
\providecommand{\PTm}{\ensuremath{{p}_\mathrm{T}\hspace{-1.02em}/}\xspace}
\providecommand{\PTslash}{\ensuremath{{p}_\mathrm{T}\hspace{-1.02em}/}\xspace}
\providecommand{\ETm}{\ensuremath{E_{\mathrm{T}}^{\text{miss}}}\xspace}
\providecommand{\ETslash}{\ensuremath{E_{\mathrm{T}}\hspace{-1.1em}/}\xspace}
\providecommand{\MET}{\ensuremath{E_{\mathrm{T}}^{\text{miss}}}\xspace}
\providecommand{\ETmiss}{\ensuremath{E_{\mathrm{T}}^{\text{miss}}}\xspace}
\providecommand{\VEtmiss}{\ensuremath{{\vec E}_{\mathrm{T}}^{\text{miss}}}\xspace}

\providecommand{\dd}[2]{\ensuremath{\frac{\mathrm{d} #1}{\mathrm{d} #2}}}

%

\providecommand{\ga}{\ensuremath{\gtrsim}}
\providecommand{\la}{\ensuremath{\lesssim}}
\providecommand{\swsq}{\ensuremath{\sin^2\theta_\mathrm{W}}\xspace}
\providecommand{\cwsq}{\ensuremath{\cos^2\theta_\mathrm{W}}\xspace}
\providecommand{\tanb}{\ensuremath{\tan\beta}\xspace}
\providecommand{\tanbsq}{\ensuremath{\tan^{2}\beta}\xspace}
\providecommand{\sidb}{\ensuremath{\sin 2\beta}\xspace}
\providecommand{\alpS}{\ensuremath{\alpha_S}\xspace}
\providecommand{\alpt}{\ensuremath{\tilde{\alpha}}\xspace}

\providecommand{\QL}{\ensuremath{\mathrm{Q}_\mathrm{L}}\xspace}
\providecommand{\sQ}{\ensuremath{\tilde{\mathrm{Q}}}\xspace}
\providecommand{\sQL}{\ensuremath{\tilde{\mathrm{Q}}_\mathrm{L}}\xspace}
\providecommand{\ULC}{\ensuremath{\mathrm{U}_\mathrm{L}^\mathrm{C}}\xspace}
\providecommand{\sUC}{\ensuremath{\tilde{\mathrm{U}}^\mathrm{C}}\xspace}
\providecommand{\sULC}{\ensuremath{\tilde{\mathrm{U}}_\mathrm{L}^\mathrm{C}}\xspace}
\providecommand{\DLC}{\ensuremath{\mathrm{D}_\mathrm{L}^\mathrm{C}}\xspace}
\providecommand{\sDC}{\ensuremath{\tilde{\mathrm{D}}^\mathrm{C}}\xspace}
\providecommand{\sDLC}{\ensuremath{\tilde{\mathrm{D}}_\mathrm{L}^\mathrm{C}}\xspace}
\providecommand{\LL}{\ensuremath{\mathrm{L}_\mathrm{L}}\xspace}
\providecommand{\sL}{\ensuremath{\tilde{\mathrm{L}}}\xspace}
\providecommand{\sLL}{\ensuremath{\tilde{\mathrm{L}}_\mathrm{L}}\xspace}
\providecommand{\ELC}{\ensuremath{\mathrm{E}_\mathrm{L}^\mathrm{C}}\xspace}
\providecommand{\sEC}{\ensuremath{\tilde{\mathrm{E}}^\mathrm{C}}\xspace}
\providecommand{\sELC}{\ensuremath{\tilde{\mathrm{E}}_\mathrm{L}^\mathrm{C}}\xspace}
\providecommand{\sEL}{\ensuremath{\tilde{\mathrm{E}}_\mathrm{L}}\xspace}
\providecommand{\sER}{\ensuremath{\tilde{\mathrm{E}}_\mathrm{R}}\xspace}
\providecommand{\sFer}{\ensuremath{\tilde{\mathrm{f}}}\xspace}
\providecommand{\sQua}{\ensuremath{\tilde{\mathrm{q}}}\xspace}
\providecommand{\sUp}{\ensuremath{\tilde{\mathrm{u}}}\xspace}
\providecommand{\suL}{\ensuremath{\tilde{\mathrm{u}}_\mathrm{L}}\xspace}
\providecommand{\suR}{\ensuremath{\tilde{\mathrm{u}}_\mathrm{R}}\xspace}
\providecommand{\sDw}{\ensuremath{\tilde{\mathrm{d}}}\xspace}
\providecommand{\sdL}{\ensuremath{\tilde{\mathrm{d}}_\mathrm{L}}\xspace}
\providecommand{\sdR}{\ensuremath{\tilde{\mathrm{d}}_\mathrm{R}}\xspace}
\providecommand{\sTop}{\ensuremath{\tilde{\mathrm{t}}}\xspace}
\providecommand{\stL}{\ensuremath{\tilde{\mathrm{t}}_\mathrm{L}}\xspace}
\providecommand{\stR}{\ensuremath{\tilde{\mathrm{t}}_\mathrm{R}}\xspace}
\providecommand{\stone}{\ensuremath{\tilde{\mathrm{t}}_1}\xspace}
\providecommand{\sttwo}{\ensuremath{\tilde{\mathrm{t}}_2}\xspace}
\providecommand{\sBot}{\ensuremath{\tilde{\mathrm{b}}}\xspace}
\providecommand{\sbL}{\ensuremath{\tilde{\mathrm{b}}_\mathrm{L}}\xspace}
\providecommand{\sbR}{\ensuremath{\tilde{\mathrm{b}}_\mathrm{R}}\xspace}
\providecommand{\sbone}{\ensuremath{\tilde{\mathrm{b}}_1}\xspace}
\providecommand{\sbtwo}{\ensuremath{\tilde{\mathrm{b}}_2}\xspace}
\providecommand{\sLep}{\ensuremath{\tilde{\mathrm{l}}}\xspace}
\providecommand{\sLepC}{\ensuremath{\tilde{\mathrm{l}}^\mathrm{C}}\xspace}
\providecommand{\sEl}{\ensuremath{\tilde{\mathrm{e}}}\xspace}
\providecommand{\sElC}{\ensuremath{\tilde{\mathrm{e}}^\mathrm{C}}\xspace}
\providecommand{\seL}{\ensuremath{\tilde{\mathrm{e}}_\mathrm{L}}\xspace}
\providecommand{\seR}{\ensuremath{\tilde{\mathrm{e}}_\mathrm{R}}\xspace}
\providecommand{\snL}{\ensuremath{\tilde{\nu}_L}\xspace}
\providecommand{\sMu}{\ensuremath{\tilde{\mu}}\xspace}
\providecommand{\sNu}{\ensuremath{\tilde{\nu}}\xspace}
\providecommand{\sTau}{\ensuremath{\tilde{\tau}}\xspace}
\providecommand{\Glu}{\ensuremath{\mathrm{g}}\xspace}
\providecommand{\sGlu}{\ensuremath{\tilde{\mathrm{g}}}\xspace}
\providecommand{\Wpm}{\ensuremath{\mathrm{W}^{\pm}}\xspace}
\providecommand{\sWpm}{\ensuremath{\tilde{\mathrm{W}}^{\pm}}\xspace}
\providecommand{\Wz}{\ensuremath{\mathrm{W}^{0}}\xspace}
\providecommand{\sWz}{\ensuremath{\tilde{\mathrm{W}}^{0}}\xspace}
\providecommand{\sWino}{\ensuremath{\tilde{\mathrm{W}}}\xspace}
\providecommand{\Bz}{\ensuremath{\mathrm{B}^{0}}\xspace}
\providecommand{\sBz}{\ensuremath{\tilde{\mathrm{B}}^{0}}\xspace}
\providecommand{\sBino}{\ensuremath{\tilde{\mathrm{B}}}\xspace}
\providecommand{\Zz}{\ensuremath{\mathrm{Z}^{0}}\xspace}
\providecommand{\sZino}{\ensuremath{\tilde{\mathrm{Z}}^{0}}\xspace}
\providecommand{\sGam}{\ensuremath{\tilde{\gamma}}\xspace}
\providecommand{\chiz}{\ensuremath{\tilde{\chi}^{0}}\xspace}
\providecommand{\chip}{\ensuremath{\tilde{\chi}^{+}}\xspace}
\providecommand{\chim}{\ensuremath{\tilde{\chi}^{-}}\xspace}
\providecommand{\chipm}{\ensuremath{\tilde{\chi}^{\pm}}\xspace}
\providecommand{\Hone}{\ensuremath{\mathrm{H}_\mathrm{d}}\xspace}
\providecommand{\sHone}{\ensuremath{\tilde{\mathrm{H}}_\mathrm{d}}\xspace}
\providecommand{\Htwo}{\ensuremath{\mathrm{H}_\mathrm{u}}\xspace}
\providecommand{\sHtwo}{\ensuremath{\tilde{\mathrm{H}}_\mathrm{u}}\xspace}
\providecommand{\sHig}{\ensuremath{\tilde{\mathrm{H}}}\xspace}
\providecommand{\sHa}{\ensuremath{\tilde{\mathrm{H}}_\mathrm{a}}\xspace}
\providecommand{\sHb}{\ensuremath{\tilde{\mathrm{H}}_\mathrm{b}}\xspace}
\providecommand{\sHpm}{\ensuremath{\tilde{\mathrm{H}}^{\pm}}\xspace}
\providecommand{\hz}{\ensuremath{\mathrm{h}^{0}}\xspace}
\providecommand{\Hz}{\ensuremath{\mathrm{H}^{0}}\xspace}
\providecommand{\Az}{\ensuremath{\mathrm{A}^{0}}\xspace}
\providecommand{\Hpm}{\ensuremath{\mathrm{H}^{\pm}}\xspace}
\providecommand{\sGra}{\ensuremath{\tilde{\mathrm{G}}}\xspace}
\providecommand{\mtil}{\ensuremath{\tilde{m}}\xspace}
\providecommand{\rpv}{\ensuremath{\rlap{\kern.2em/}R}\xspace}
\providecommand{\LLE}{\ensuremath{LL\bar{E}}\xspace}
\providecommand{\LQD}{\ensuremath{LQ\bar{D}}\xspace}
\providecommand{\UDD}{\ensuremath{\overline{UDD}}\xspace}
\providecommand{\Lam}{\ensuremath{\lambda}\xspace}
\providecommand{\Lamp}{\ensuremath{\lambda'}\xspace}
\providecommand{\Lampp}{\ensuremath{\lambda''}\xspace}
\providecommand{\spinbd}[2]{\ensuremath{\bar{#1}_{\dot{#2}}}\xspace}

\providecommand{\MD}{\ensuremath{{M_\mathrm{D}}}\xspace}
\providecommand{\Mpl}{\ensuremath{{M_\mathrm{Pl}}}\xspace}
\providecommand{\Rinv} {\ensuremath{{R}^{-1}}\xspace} 
\cmsNoteHeader{TRK-10-001} 
\title{CMS Tracking Performance Results from Early LHC Operation}

\address[cern]{CERN}
\address[neu]{Northeastern University}
\author[neu]{George Alverson}\author[neu]{Lucas Taylor}\author[cern]{A. Cern Person}

\date{\today}

\abstract{
The first LHC pp collisions at centre-of-mass energies of 0.9 and 2.36\TeV were recorded
by the CMS detector in December 2009.  The trajectories of charged particles produced in
the collisions were reconstructed using the all-silicon Tracker and their momenta were
measured in the 3.8\,T axial magnetic field.  Results from the Tracker commissioning
are presented including studies of timing, efficiency, signal-to-noise,
resolution, and ionization energy.  Reconstructed tracks are used to benchmark the performance in terms 
of track and vertex resolutions, reconstruction of decays, estimation of ionization energy loss, as well as identification of photon conversions, nuclear interactions, and heavy-flavour decays.}
\hypersetup{%
pdfauthor={CMS Collaboration},%
pdftitle={CMS Tracking Performance Results from early LHC Running},%
pdfsubject={CMS},%
pdfkeywords={CMS, physics, software, computing}}

\maketitle 

\section{Introduction}
\label{sec:introduction}
The Compact Muon Solenoid (CMS)~\cite{CMS} is a general purpose 
detector at the Large Hadron
Collider (LHC) of CERN\@. It has been designed primarily to perform
new physics studies at the highest energies achievable with the
LHC\@. The main components of CMS are a muon detection system,
electromagnetic and hadronic calorimeters, and an inner tracking
system (Tracker). The Tracker provides robust, efficient, and precise
reconstruction of the charged particle trajectories inside a 3.8~T
axial magnetic field. The nominal momentum resolution is typically 0.7
(5.0)\% at 1 (1000)\,\GeVc in the central region 
and the impact parameter resolution for high-momentum tracks is
typically 10\micron.

The reconstructed tracks of charged particles are among the most fundamental objects in
the reconstruction of pp~collisions.  Tracks are used in the reconstruction of electrons,
muons, hadrons, taus, and jets as well as in the determination of the primary interaction
vertices.  In addition, tracks may be used to identify b jets, in particular through
evidence of a displaced vertex associated with a given jet.

This paper describes the performance of the Tracker, which was evaluated with
collision data from early LHC operations at centre-of-mass
energies of 0.9 and 2.36~TeV\@.  The next section contains a brief
description of the Tracker.  Section~\ref{sec:data_samples}
illustrates the LHC data and conditions that underlie the analysis.
Results obtained from the commissioning of the Pixel and Silicon Strip
detectors are described in Section~\ref{sec:commissioning}.
Section~\ref{sec:reconstruction} describes the track reconstruction
and Section~\ref{sec:tracking} presents tracking results
demonstrating the overall performance of the Tracker.  In particular,
reconstructed tracks are used for
track and vertex resolution measurements, the reconstruction of hadron decays,
the estimation of ionization energy loss, the identification of photon
conversions and nuclear interactions, and b tagging.  Finally,
conclusions are presented in Section~\ref{sec:conclusion}.

\section{Tracker Description}
\label{sec:tracker_description}
The CMS experiment uses a right-handed coordinate system, with the
origin at the nominal interaction point, the $x$ axis pointing to the
centre of the LHC ring, the $y$ axis pointing up (perpendicular to the LHC
plane) and the $z$ axis along the anticlockwise-beam direction.  The
azimuthal angle $\phi$ is measured in the $xy$ plane, with
$\phi= 0$ along the positive $x$ axis and $\phi = \pi/2$ along the
positive $y$ axis.

\begin{figure}[hbtp]
\begin{center}
    \mbox{
{\scalebox{0.95}{
	  \includegraphics[width=\linewidth]{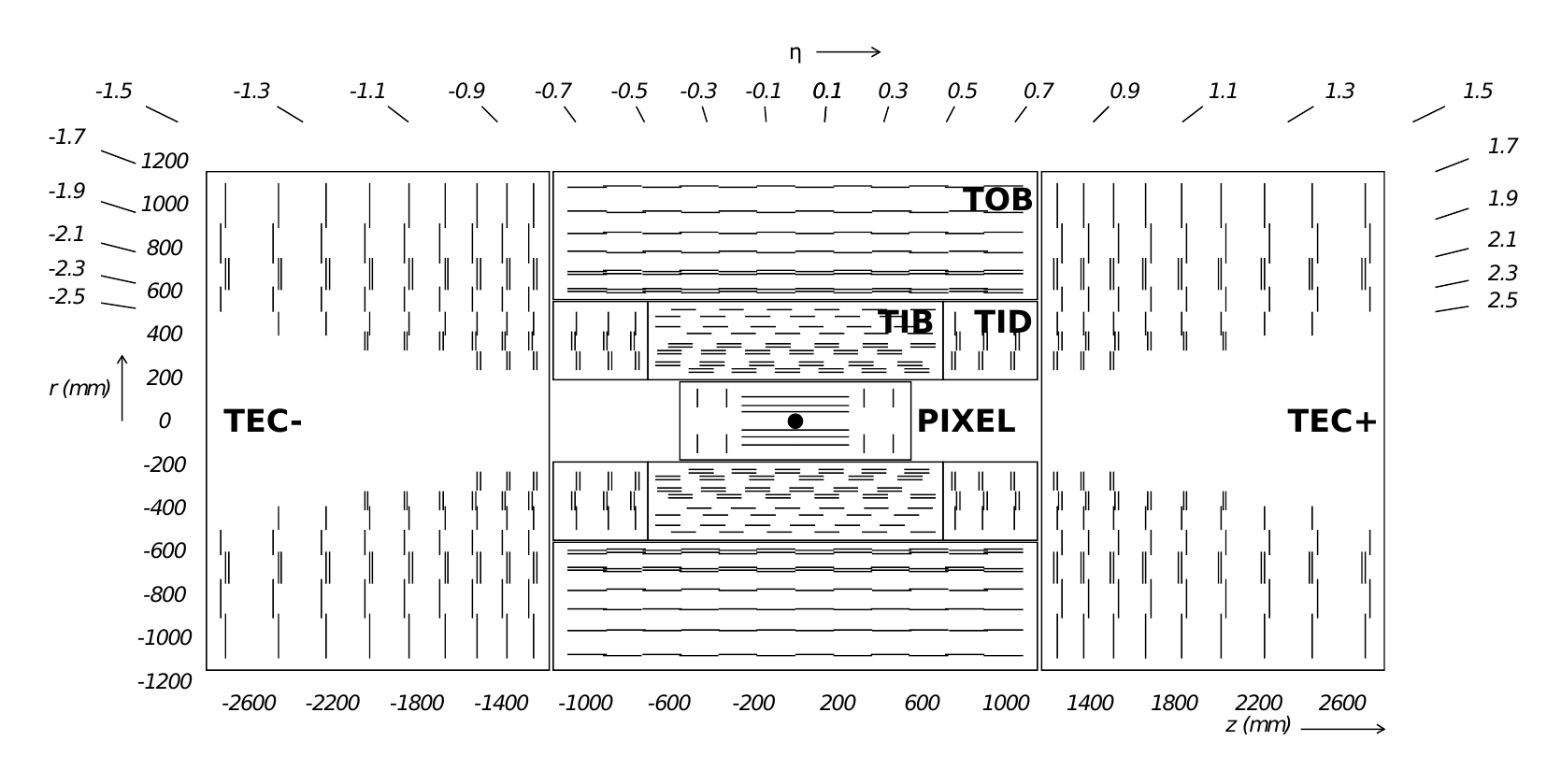}
	  \label{fig:general_layout} 
      }}
    }
\caption{
$r$-$z$ slice of the CMS Tracker.}
\label{fig:detector} 
\end{center}
\end{figure}

The CMS Tracker \cite{CMS}, shown in Fig.~\ref{fig:detector}, consists
of two main detectors: a silicon pixel detector, covering the region
from 4 to 15\,cm in radius, and 49\,cm on either side of the
collision point along the LHC beam axis, and a silicon strip detector,
covering the region from 25 to 110\,cm in radius, and within 280\,cm
on either side of the collision point along the LHC beam axis.

The CMS silicon pixel detector has 66 million active elements
instrumenting a surface area of about 1\,m$^2$.  It is designed to
provide the determination of three high precision three-dimensional
points on track trajectories.  The detector consists of three
concentric cylindrical barrel layers and four fan-blade disks which
close the barrel ends.  The barrel layers have an active length of
53\,cm and are located at average radii of 4.3, 7.3, and 10.2\,cm.
The endcap disks instrument the regions between radii 4.8 and
14.4\,cm at mean longitudinal distances of 35.5 and 48.5\,cm from the
interaction point.  The system provides efficient three-hit coverage
in the region of pseudorapidity $|\eta|<2.2$ and efficient two-hit
coverage in the region $|\eta|<2.5$.  The active elements are n-in-n
100\,$\mu$m$\times$150\,$\mu$m pixels~\cite{CMS} which are oriented with the
smaller pitch in the azimuthal direction in the barrel and the radial
direction in the disks.  The 3.8\,T magnetic field in CMS causes
significant azimuthal Lorentz drift of the collected electrons in the
pixel barrel which enhances the azimuthal charge sharing and therefore
improves the resolution in that direction.  The blades of the endcap
disks are rotated by 20 degrees about their radial axes with respect
to the disk planes to produce azimuthal charge sharing and radial
Lorentz drift which enhances the radial charge sharing.  The charge
sharing improves the endcap resolution in both planes.


The CMS silicon strip detector has 9.3 million active elements
instrumenting a surface area of 198\,m$^2$.  The detector consists of
three large subsystems. The Tracker Inner Barrel and Disks (TIB/TID)
extend in radius to 55\,cm and are composed of four barrel layers,
supplemented by three disks at each end.  The TIB/TID delivers up to
four $r$-$\phi$ measurements on a trajectory using 320\,$\mu$m thick
silicon microstrip sensors, which have their strips oriented parallel
to the beam axis in the barrel and oriented radially in the disks. The
strip pitch is 80\,$\mu$m in the inner pair of TIB layers and
120\,$\mu$m in the outer pair of TIB layers. In the TID, the mean
pitch varies between 100\,$\mu$m and 141\,$\mu$m. The TIB/TID is
enclosed within the Tracker Outer Barrel (TOB), which has an outer
radius of 116\,cm.  The TOB consists of six barrel layers of
500\,$\mu$m thick microstrip sensors with strip pitches of 183\,$\mu$m
in the first four layers and 122\,$\mu$m in the last pair of
layers. The TOB extends to $\pm$118\,cm in $z$. Beyond this $z$ range,
the Tracker EndCaps (TEC) instrument the region 124 $< |z| <$ 280\,cm
and 22.0 $< r <$ 113.5\,cm. Each TEC is composed of nine disks that
are instrumented with up to seven rings of radial-strip silicon
detectors.  The sensor thicknesses are {\sl thin} (320\,$\mu$m) in the
inner four rings and {\sl thick} (500\,$\mu$m) in the outer three
rings; the average radial strip pitch varies from 97\,$\mu$m to
184\,$\mu$m.  The inner two layers of the TIB and TOB, the inner two
rings of the TID and TEC, and the fifth ring of the TEC include a
second microstrip detector module that is mounted back-to-back at a
stereo angle of 100\,mrad and that enables a measurement of the
orthogonal coordinate.  Assuming fully efficient planes and not
counting hits in stereo modules, there are from 8 to 14 high precision
measurements of track impact points for $|\eta|<2.4$.

\section{Data Samples}
\label{sec:data_samples}
The results presented in this paper were obtained from data samples collected by the CMS
experiment during LHC operation in December 2009 at proton-proton centre-of-mass energies
of 0.9 and 2.36~TeV\@.  The CMS axial magnetic field was maintained at the nominal
value of 3.8\,T and the silicon pixel and silicon strip detectors were biased at their
nominal voltages.  Due to the relatively low LHC luminosity, the CMS readout was triggered
by the coincidence of signals from the beam scintillator counter (BSC) minimum bias
trigger and the beam pick-up timing detector which detects the passage of the beam
bunches~\cite{dNdeta}.  The BSC minimum bias trigger requires that the arrival times of the
signals from the forward and backward arms of the BSC $(3.23<|\eta|< 4.65)$ be consistent
with the passage of particles emerging from a pp collision in the middle of CMS.  In contrast, the
BSC beam-gas trigger, used to veto non-collision events, 
requires that the arrival times be consistent with the passage of
particles traversing the detector from one end to the other in time with particles from
either beam.
The total number of selected minimum bias events is approximately 305\,000.

Prior to the LHC pp collisions, the CMS experiment was commissioned using 
events containing cosmic muons during Cosmic Run At Four Tesla (CRAFT)~\cite{craft_paper}.  
The detector and magnetic field conditions during CRAFT were quite similar to the conditions 
during pp collisions.  Thus, the results obtained from CRAFT provided good initial 
operating points for the pixel detector~\cite{craft_pixel_paper}, the strip 
detector~\cite{craft_strip_paper}, the tracker alignment~\cite{craft_alignment_paper}, and 
the magnetic field~\cite{craft_bfield_paper}.  The data used in the referenced CRAFT papers
were obtained in the fall of 2008, more than one year before the pp collisions.  
In most cases, more recent CRAFT data were used to improve on these results.


\section{Tracker Commissioning}
\label{sec:commissioning}

The following two subsections describe the operating characteristics and performance of the silicon pixel and silicon strip detectors, respectively.

\subsection{Silicon Pixel Detector}
\label{sec:pixels}
\subsubsection{Operating Conditions}
In order to make maximal use of experience gained from the operation of the pixel detector with cosmic rays during summer/autumn 2009, the operating conditions were not changed for the December 2009 data taking period.  The coolant temperature was kept constant at 7$^\circ$C.  The bias potential applied to the 285\micron thick p-spray barrel sensors \cite{Allkofer:2007ek} was a uniform 150~V.  The bias potential applied to the 270\micron thick p-stop endcap sensors \cite{Arndt:2003ck} was a uniform 300~V.  Small fractions of the barrel (1.0\%) and endcap (3.1\%) detectors were inactive resulting in a net operational fraction of 98.4\% for the entire detector.

The calibration procedures described in Ref.~\cite{craft_pixel_paper} were used to determine the ADC gains and pedestals for all channels.  Iterative tuning reduced the mean (spread) of the readout threshold distributions for the pixel Readout Chips (ROCs) from the values measured during the 2008 cosmic ray commissioning \cite{craft_pixel_paper} to 2733\,$e$ (196\,$e$) in the barrel detector and 2483\,$e$ (163\,$e$) in the endcap detectors, where $e$ is the magnitude of the electron charge.  These measured threshold values apply only to the calibration procedure.  Because the bandwidth of the preamplifiers is limited by power considerations, small signals can take more than a bunch crossing time (25 ns) to fire the zero-crossing discriminator that triggers the storage of the signal.  This causes some smaller signals to be associated with the wrong bunch crossing and to be ignored by the readout system.  The net result is that the effective or ``in-time'' thresholds are larger than the set values. 

The effective thresholds are estimated by comparing the distribution of measured cluster $x$-sizes (azimuthal direction in the barrel detector and radial direction in the endcap detectors) with those predicted by the detailed pixel simulation, \textsc{pixelav}~\cite{Chiochia:2004qh, Swartz:2005vp}. The cluster sizes are sensitive to the effective thresholds.  To avoid highly ionizing particles, the tracks used in this analysis were required to have momenta larger than 4\GeVc.  This selection ensures that even protons and deuterons produce signals that are within a few percent of the ionization minimum.   By varying the simulated thresholds until the measured and simulated distributions agree, the average effective thresholds are found to be approximately 3500\,$e$ in the barrel detector and 3000\,$e$ in the endcap detectors.  

A study of the pixel hit reconstruction efficiency using a technique similar to the strip detector technique described in Section~\ref{sec:strip_effs} suggests that the efficiency is larger than 99\% for the live regions of the detector and is consistent with earlier work \cite{dNdeta}.

\subsubsection{Pixel Timing Scan}

\begin{figure}[hbtp]
\begin{center}
    \mbox{
{\scalebox{0.50}{
	  \includegraphics[width=\linewidth]{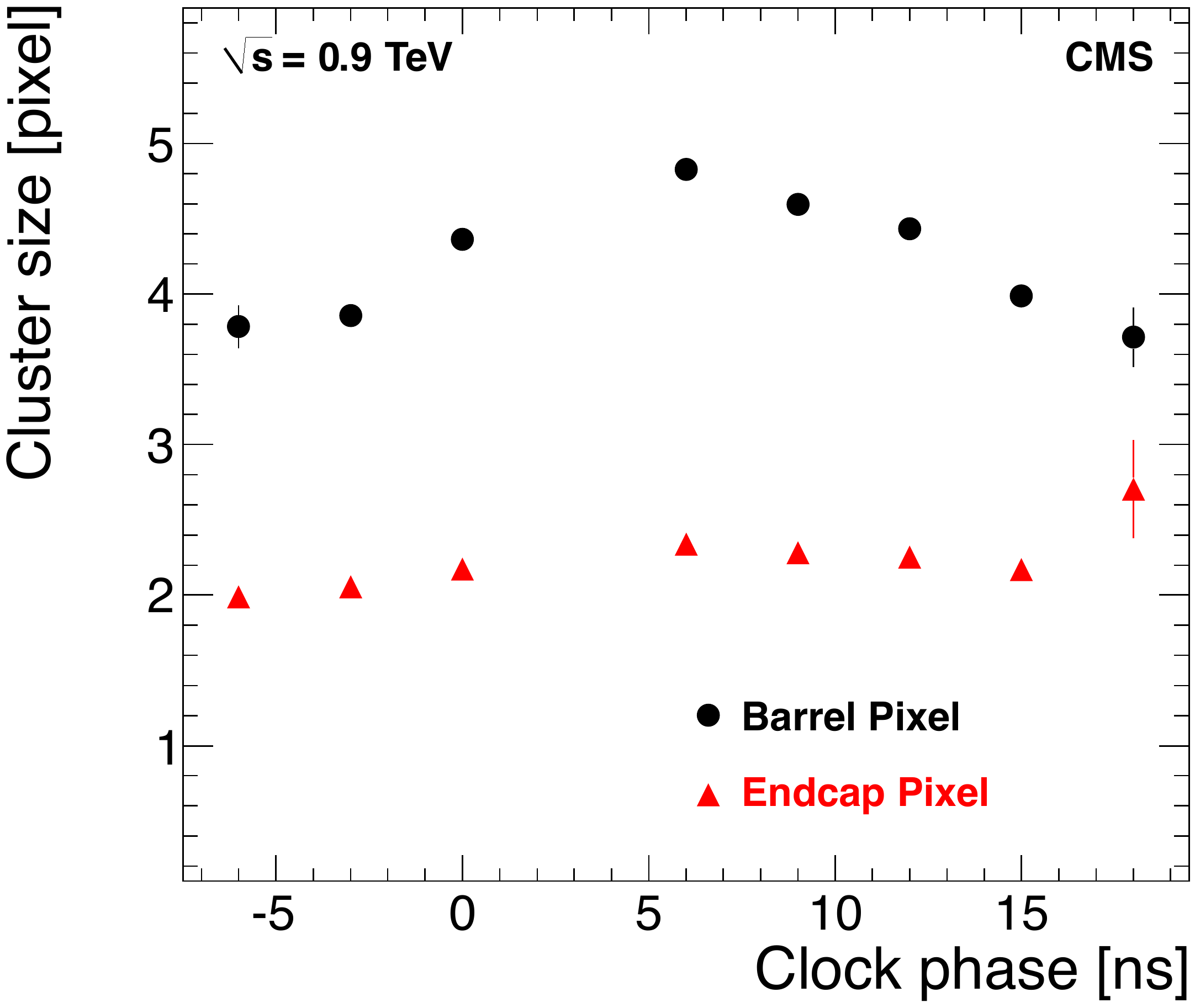}
      }}
    }
\caption{
  The average cluster size distributions for the barrel and endcap pixel detectors in minimum bias events are plotted versus clock phase.}
   \label{fig:pixel_timing_size} 
\end{center}
\end{figure}

The pixel detector readout system uses the 40~MHz LHC clock as input.  Signals from the CMS trigger system must arrive at the correct time within the 25~ns clock cycle to associate the correct bunch crossing time stamp with any signal above the readout threshold.  An optimally phased clock signal will maximize the number of pixels observed in clusters.  The overall trigger timing was adjusted by varying the clock phase until the average barrel and endcap cluster sizes as measured in minimum bias triggers were maximized.  These quantities are plotted versus clock phase in Fig.~\ref{fig:pixel_timing_size}.  The clock phase setting of 6~ns was found to optimize the smoothly varying detector averages.  A finer module-by-module adjustment of the clock phase will be performed when higher trigger rates become available.

\subsubsection{Operating Characteristics with Minimum Bias Triggers}
\label{sssec:pixel_operting_minbias}

\begin{figure}[hbtp] 
\begin{center}
    \mbox{
{\scalebox{0.90}{
         \includegraphics[width=\linewidth]{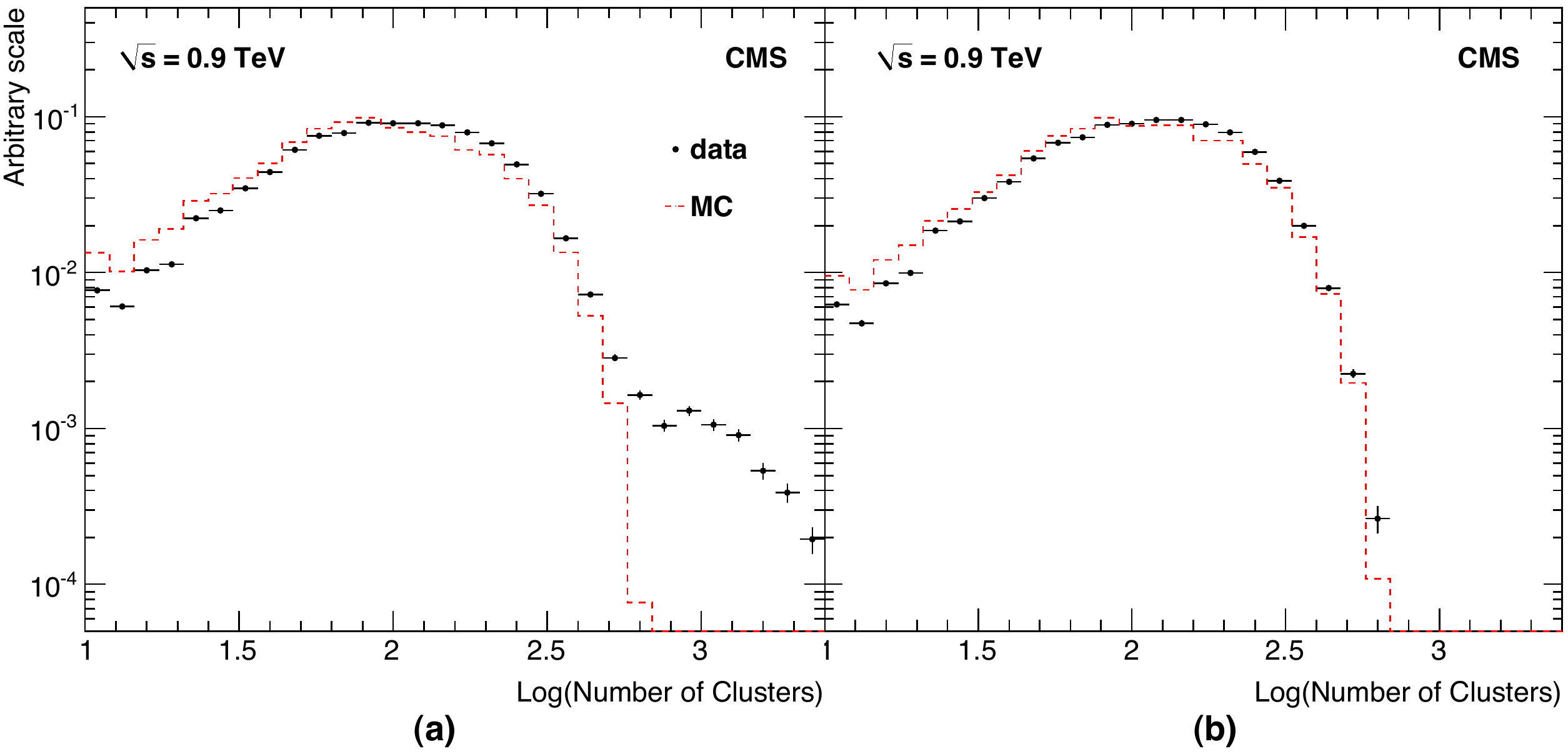}
     }}
  }
\caption{The cluster multiplicity of (a) all minimum bias triggered events and (b) those that do not trigger the beam-gas veto in the 0.9\TeV data sample.  The histograms show the similar distribution for a sample of simulated data.}
\label{fig:clN_900GeV}
\end{center}
\end{figure}

The distributions of the number of clusters observed in 0.9\TeV events selected by the minimum bias trigger are shown in Fig~\ref{fig:clN_900GeV}.  The observed data, shown as solid dots, are compared with fully simulated data, shown as histograms, that were generated with a recent tuning of the \PYTHIA event generator~\cite{Moraes:2007rq}.  The left plot shows the distribution for all events, whereas the right plot shows the distribution after removing events that also satisfy the beam-gas trigger.  There is an excess of large multiplicity events that are removed by the beam-gas trigger requirement.  The source of these events could be beam-gas interactions or beam scraping in the beam transport system near the interaction point.  After removal of the beam background events, the measured distributions are approximately consistent with preliminary expectations.  The measured average cluster multiplicities per layer (barrel detector) and per disk (endcap detector) are listed in Table~\ref{tab:pixel_clN}.  They are compared with the expectation from the simulation and are found to be in rough agreement.  It should be noted that the event generator is based on an event model that has not yet been tuned in detail and is not expected to provide accurate predictions.

During the extremely low luminosity run in December 2009 (the instantaneous luminosity was typically in the range 10$^{26}$--10$^{27}$~cm$^{-2}$s$^{-1}$), the beam background events occurred at a rate that was roughly comparable to the rate of minimum bias triggers.  Because they are characterized by particle trajectories that are nearly parallel to one of the beams, most background events ($\sim$90\%) do not fire the minimum bias trigger but do have clusters in the endcap detectors and elongated clusters in the first two layers of the barrel detector.  At the beam energies of the December 2009 run, the pixel detector occupancies associated with the background events were typically five times larger than those associated with minimum bias events.  The beam-gas trigger veto effectively removes background events, as do cluster shape, track quality, and vertex requirements.  

\begin{table}[htbp]
\caption{\label{tab:pixel_clN} The average cluster multiplicity per layer/disk in 0.9\TeV minimum bias triggers.  The simulation errors are entirely statistical and do not represent the uncertainties in the event modelling.  The asymmetry seen in the forward and backward endcaps is caused by an offset in the luminous region along the beam axis.}
\begin{center}
\begin{minipage}{0.45\textwidth}
\begin{tabular}{ccc}
\hline
\multicolumn{3}{c}{Barrel Pixel: clusters/layer}\\
Layer & Measured & Simulation \\
\hline
1  & 35.2$\pm$0.9 &  31.6$\pm$1.2 \\
2  & 30.6$\pm$0.8 &  27.8$\pm$1.1 \\
3  & 27.4$\pm$0.8 &  24.8$\pm$1.0 \\
\hline
\end{tabular}
\end{minipage}
\hfil
\begin{minipage}{0.45\textwidth}
\begin{center}
\begin{tabular}{ccc}
\hline
\multicolumn{3}{c}{Endcap Pixel: clusters/disk}\\
Disk & Measured & Simulation \\
\hline
$-$2 &  8.0$\pm$0.1 &  7.3$\pm$0.2 \\
$-$1 &  7.8$\pm$0.1 &  7.2$\pm$0.2 \\
1 &  8.1$\pm$0.1 &  7.7$\pm$0.2 \\
2 &  8.6$\pm$0.1 &  8.1$\pm$0.2 \\
\hline
\end{tabular}
\end{center}
\newpage
\end{minipage}
\end{center}
\end{table}

\begin{figure}[hbtp]
\begin{center}
    \mbox{
      \subfigure[]
{\scalebox{0.40}{
	  \includegraphics[width=\linewidth]{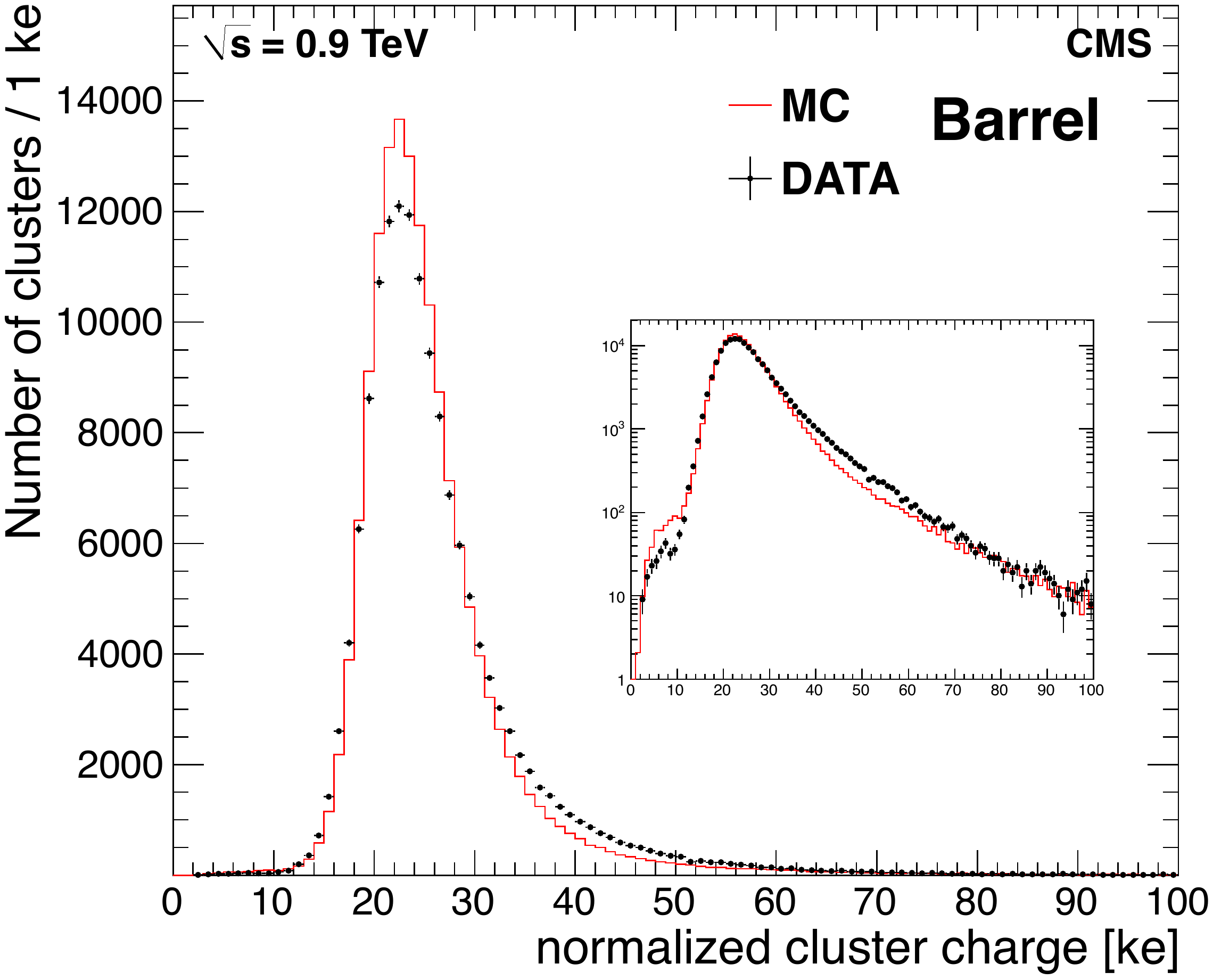}
	  \label{fig:cc_norm_bpix} 
     }}
    }
    \mbox{
      \subfigure[]
{\scalebox{0.40}{
	  \includegraphics[width=\linewidth]{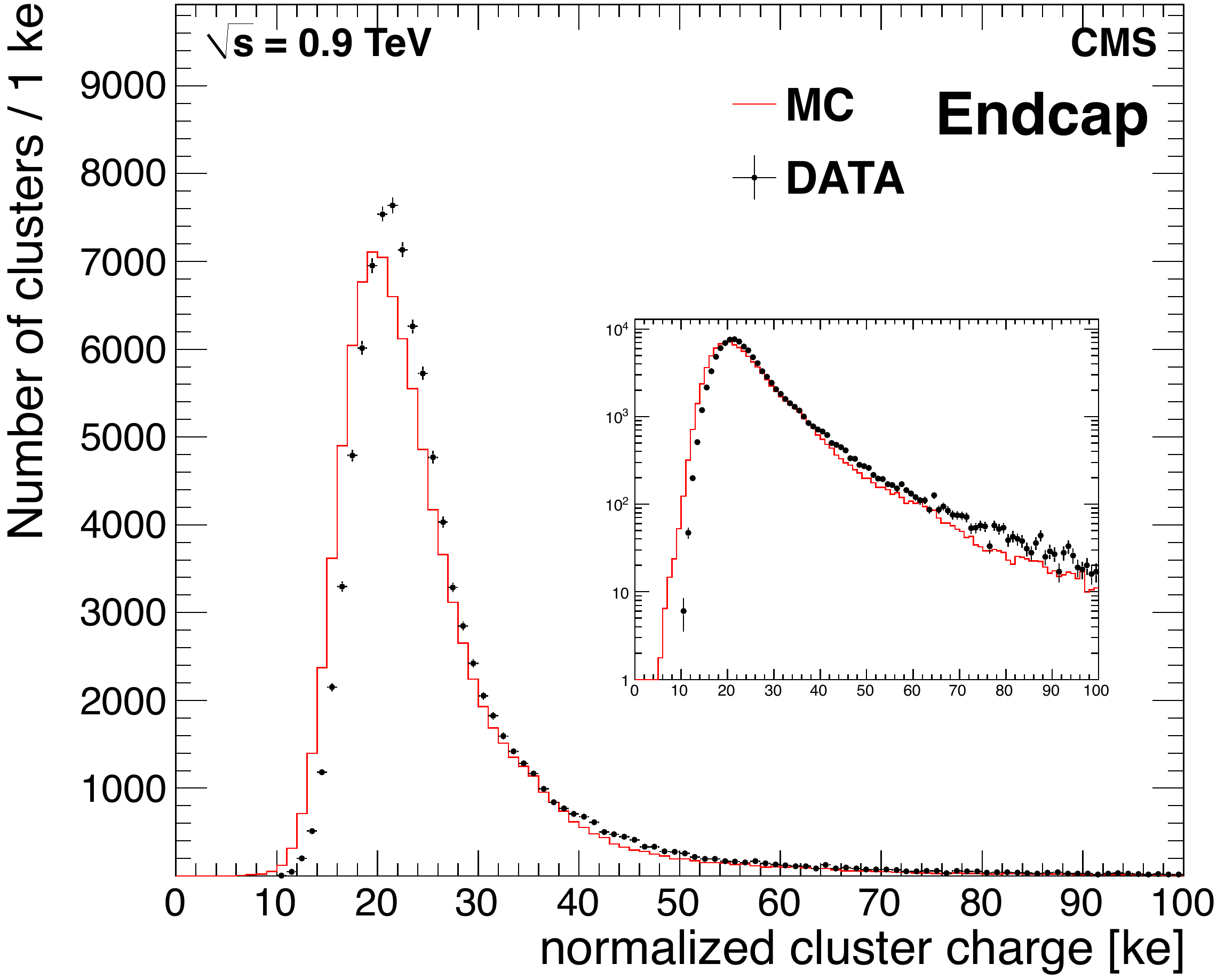}
	  \label{fig:cc_norm_fpix} 
      }}
    }
\caption{
The normalized cluster charge measured in the (a) barrel and (b) endcap pixel detectors for the sample of 0.9\TeV minimum bias events.  The insets show the same distributions on semi-log scales.}
	  \label{fig:cc_norm} 
\end{center}
\end{figure}

The cluster charge distributions measured in the barrel and endcap detectors with the 0.9\TeV sample are shown as solid dots in Fig.~\ref{fig:cc_norm}.  Each entry is scaled by the ratio of the pixel sensor thickness to the track path length in the sensor.  The solid histograms represent the expectations from the \PYTHIA-based, full detector simulation.  The measured and simulated barrel distributions have similar peaks but the measured distribution is somewhat broader than the simulated one.  This may be due to residual pixel-to-pixel gain variation resulting from the use of a single gain for all 80 channels in each ROC column or residual module-to-module clock phase variation.  The corresponding distributions for the endcap detectors have similar widths but indicate a 5\% charge-scale mismatch.

\subsubsection{Lorentz Angle Calibration}

The use of n-in-n pixel technology and the large magnetic field in CMS imply that pixel hit 
reconstruction involves large Lorentz drift corrections (the typical bias corrections are 53\micron in the barrel and 10\micron in the endcap).
The estimation of track impact coordinates from pixel clusters is performed with two different algorithms.  The simpler, faster ``Generic Algorithm'' \cite{CMS_NOTE_2002_049} uses the Lorentz width $W_L$ to estimate the projected cluster size and bias correction.  The Lorentz width is the product of the effective thickness of the sensor $T_\mathrm{eff}$ and the tangent of the average Lorentz angle $\theta_L$: $W_L = T_\mathrm{eff}\tan{\theta_L}$.  Due to the focusing of the electric field at the n+ implants, the charge sharing near the n+ side of the sensors is reduced.  This is modelled by the effective thickness which is 5--10\% smaller than the physical thickness of the sensor substrate.  The detailed \textsc{pixelav} simulation is used to extract the Lorentz width by applying the Generic Algorithm to a sample of simulated clusters and by adjusting $W_L$ to minimize the bias and maximize the resolution.  The slower, more sophisticated ``Template Algorithm'' \cite{CMS_NOTE_2007_033} fits pre-computed cluster shapes to the measured clusters.  The Lorentz-drift effects are encoded in the cluster shapes and the same \textsc{pixelav} simulation is used to compute them.  Therefore, the actual Lorentz calibration procedure is to tune the detailed simulation to agree with data and then to generate a Lorentz width for the Generic Algorithm and cluster shapes for the Template Algorithm.

\begin{figure}[hbtp]
\begin{center}
    \mbox{
      \subfigure[]
{\scalebox{0.45}{
	  \includegraphics[width=\linewidth]{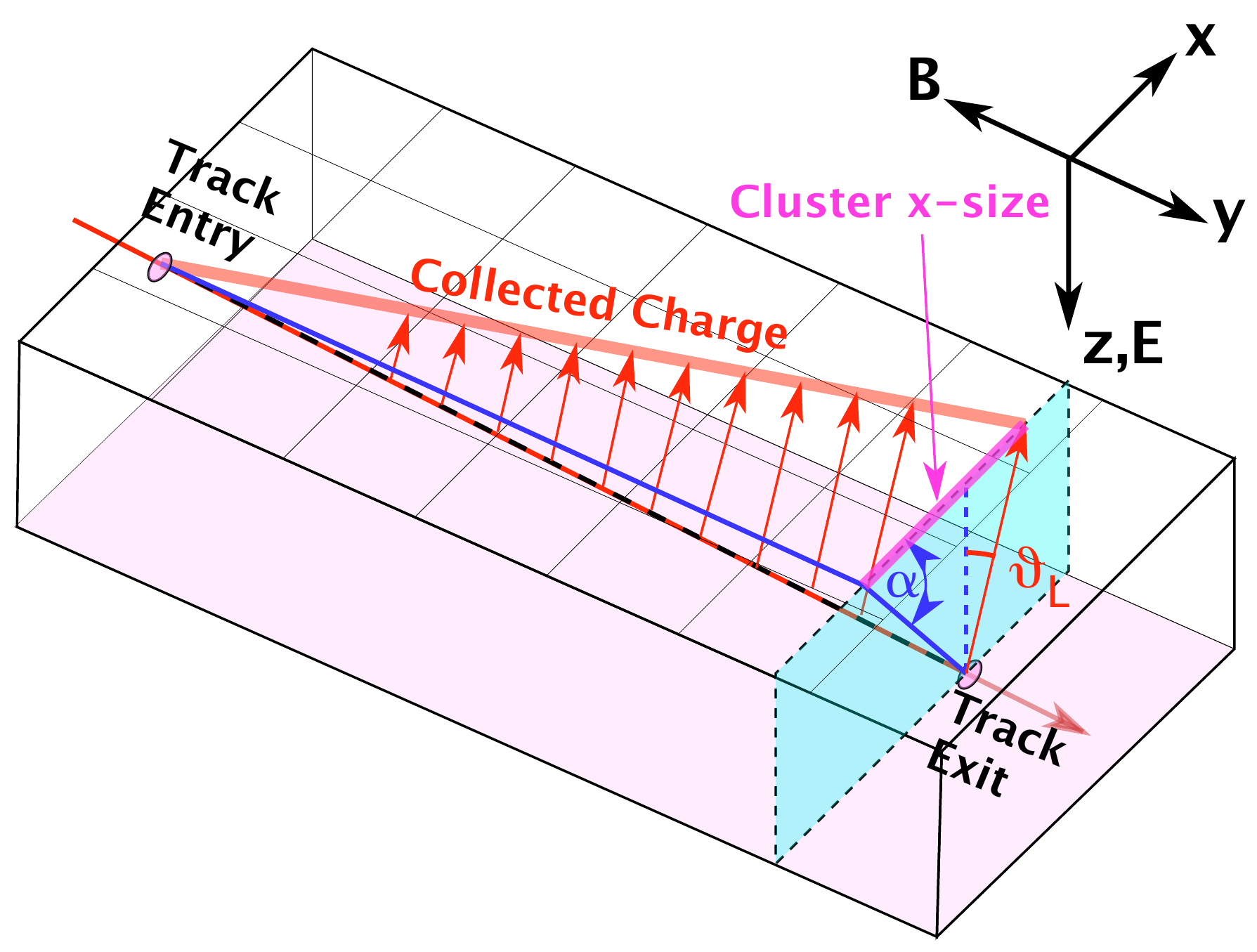}
	   \label{fig:pixel_coords}
      }}
    }
    \mbox{
      \subfigure[]
{\scalebox{0.45}{
	  \includegraphics[width=\linewidth]{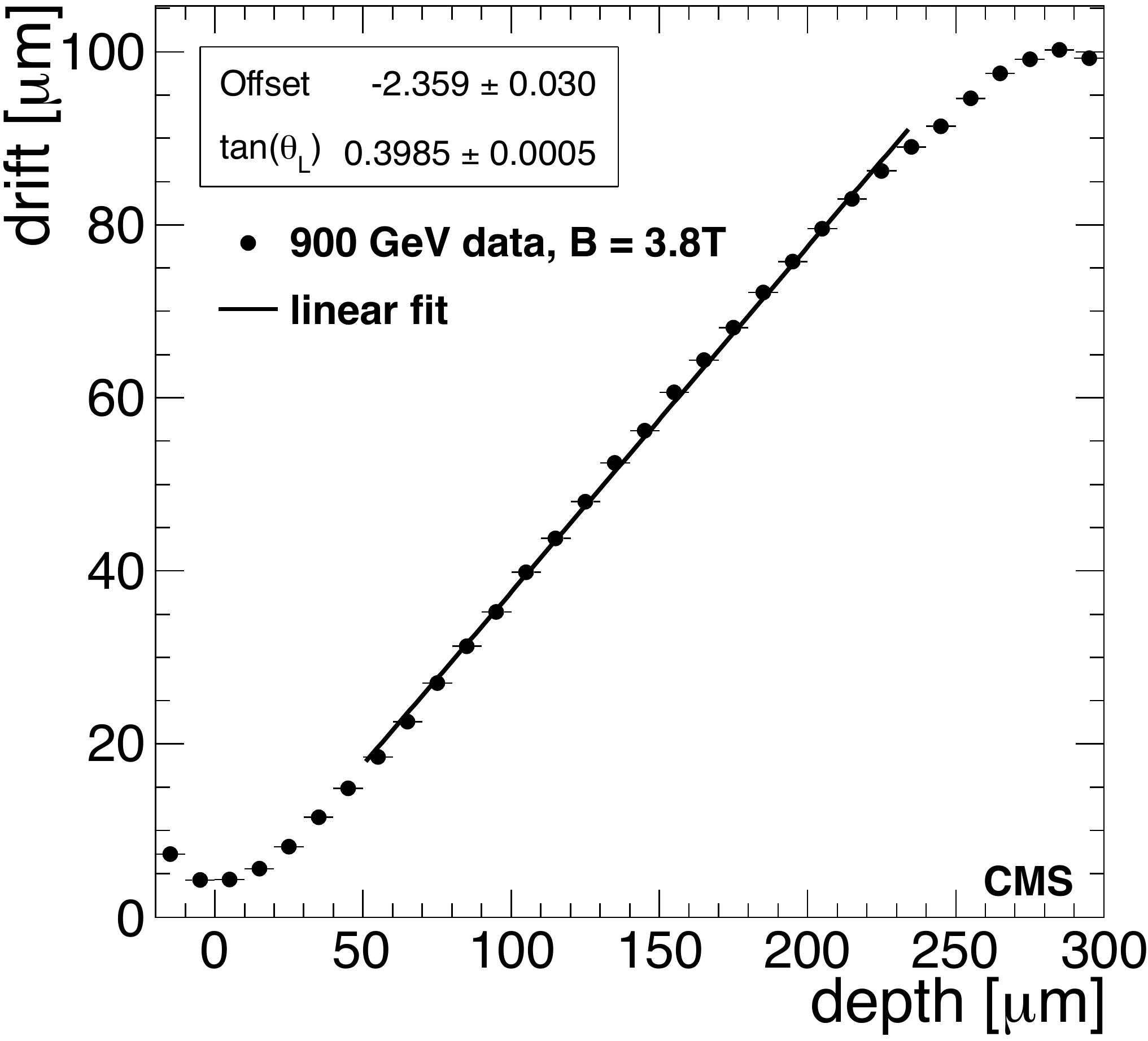}
	  \label{fig:pixel_la_grazing_angle} 
      }}
    }
\caption{
  (a) The pixel local coordinate system and track angle definitions.  The local $z$ axis coincides with the sensor electric field $\vec E$.  The local $x$ axis is chosen to be parallel to $\vec E\times\vec B$ where $\vec B$ is the axial magnetic field.  The local $y$ axis is defined to make a right-handed coordinate system.  The angle $\alpha$ is the angle between the $x$ axis and the track projection on the local $xz$ plane.  (b) The transverse cluster displacement of highly inclined barrel clusters as a function of depth for a sample of 0.9\TeV minimum bias events at a magnetic field of 3.8~T.  The tangent of the Lorentz angle is given by the slope of a linear fit which is shown as the solid line.}
\end{center}
\end{figure}

Two different techniques have been used to perform the calibration.  The 2008 cosmic ray data were calibrated by measuring the cluster $x$-sizes as functions of $\cot{\alpha}$ (see Fig.~\ref{fig:pixel_coords} for definitions) and by determining the locations of the cluster-size minimum $\cot{\alpha}_\mathrm{min}$\cite{craft_pixel_paper}.  In the pixel barrel, $-\cot{\alpha}_\mathrm{min}$ is equal to $\tan{\theta_L} = r_H\bar\mu B$, where $r_H$ is the electron Hall factor, $\bar\mu$ is the average electron mobility, and $B$ is the magnetic field.  The 2008 cosmic ray measurements suggested that the value of the electron Hall factor used in \textsc{pixelav} should be increased to 1.05 from the 1.02 value determined in test beam measurements \cite{Dorokhov:2003if}.  In 2009, the temperature of the detector was lowered and the bias voltage of the pixel barrel was increased, which changed the average Lorentz angles in both barrel and endcap detectors.  New cosmic ray based determinations are reported in Table~\ref{tab:pixel_la} and are compared with the tuned simulation. 

\begin{table}[htbp]
\caption{\label{tab:pixel_la} The tangent of the Lorentz angle  $\tan{\theta_L}$ as determined by 2009 calibrations. }
\begin{center}
\begin{tabular}{ccccc}
\hline
\multicolumn{5}{c}{2009 Lorentz Angle Measurements}\\ 
Sample & Detector & Technique & Measured $\tan{\theta_L}$ & Simulation \\
\hline
Cosmic Ray & Barrel & Cluster Size & 0.409$\pm$0.002(stat) & 0.407$\pm$0.002(stat) \\
Cosmic Ray & Endcap & Cluster Size & 0.081$\pm$0.005(stat) & 0.080$\pm$0.004(stat) \\
Minimum Bias & Barrel & Grazing Angle  & 0.3985$\pm$0.0005(stat) &  0.4006$\pm$0.0005(stat) \\
Minimum Bias & Barrel & Cluster Size  &  0.409$\pm$0.002(stat)  &   0.411$\pm$0.005(stat)\\
\hline
\end{tabular}
\end{center}
\end{table}

The barrel calibration was repeated with collision data in December 2009 using a new ``grazing angle'' technique \cite{CMS_NOTE_2008_012}.  This technique makes use of the two-dimensional pixel segmentation to simultaneously measure the average transverse displacement of the charge carriers as a function of distance along clusters produced by a sample of highly inclined tracks.  Since longitudinal position in the cluster is completely correlated with depth in the junction, this technique determines the average transverse carrier displacement as a function of depth as shown graphically in Fig.~\ref{fig:pixel_la_grazing_angle}.  The average Lorentz angle, extracted from the linear fit shown in the figure, is compared with the detailed simulation in Table~\ref{tab:pixel_la}.  The extremely large population of highly curved, low transverse momentum tracks observed in minimum bias triggers spans the $\cot{\alpha}$ region needed to determine the minimum projected cluster size in the pixel barrel.  This enables the use of the cluster size technique as a cross check which is also reported in Table~\ref{tab:pixel_la}.  Note that the two techniques are affected by different systematic effects and that a better than 1\% consistency is observed between the real and simulated measurements in all cases.  A variation of fitting procedures suggests that the total systematic uncertainty on the Lorentz angle calibration is less than 2\%.

\subsubsection{Resolution Study}

The intrinsic position resolution in a limited range of the angular acceptance was measured using tracks from minimum bias triggers that traverse overlapping sensors in the barrel layers.  A similar analysis was performed in a very different angular region with 2008 cosmic ray data \cite{craft_pixel_paper} using the measurement technique given in Ref.~\cite{:2009mq}. Tracks passing through two overlapping modules in the same layer are used to compare the hit position with the expected position from the track trajectory. 
Because it is insensitive to alignment uncertainties, the difference of the local track impact points on a fitted trajectory is known about ten times more precisely than are the individual predicted hit positions. A double difference is formed by taking the difference between the measured hit position difference in the two modules and the predicted trajectory position difference. The width of this double difference distribution is insensitive to translational misalignment of the overlapping modules.

To limit the effect of multiple scattering, a
minimum track momentum of 2.5\GeVc is required.  Clusters with measured charge
below 10\,000\,$e$ or containing pixels on the sensor edges are 
excluded.  The double difference widths are fitted with a
Gaussian and the uncertainty from the trajectory prediction is
subtracted quadratically to recover the hit resolution on the position
difference.  With the assumption of equal resolution for each of the modules in the overlap, the final fit values for the resolution for a single module are $12.8\pm0.9$\micron along $x$ and $32.4\pm1.4$\micron along $y$.
The \textsc{pixelav} simulation is used to generate a sample of clusters that has the same distribution of impact angles as the measured sample. Since the simulation does not include the double-size pixels that span the gaps between the sixteen readout chips which tile each module, a subsample of the overlap data sample is used to determine single-size-pixel resolutions of $12.7\pm2.3$\micron along $x$ and $28.2\pm1.9$\micron along $y$.  These numbers can be directly compared with those extracted from Gaussian fits to the simulated residual distributions.  The simulated resolutions are $14.1\pm0.5$\micron and $24.1\pm0.5$\micron along $x$ and $y$, respectively, and agree reasonably well with the measured resolutions.  Because overlaps occur only at the edges of the track $\alpha$-angle acceptance where the $x$ sizes of the clusters deviate from the optimal size of two, the measured and simulated $x$ resolutions are somewhat worse than the typical $x$ resolution (less than 10\micron) expected for most collision-related clusters.  The measured and simulated $y$ resolutions are expected to be typical of the detector performance.

\subsection{Silicon Strip Detector}
\label{sec:strips}
\subsubsection{Operating Conditions}
All of the modules in the strip tracker were biased at 300~V in the
early collision running. This is the same setting that was used in the
CRAFT studies and is well above the full depletion voltage for the
sensors. Similarly, the coolant temperature was set at
4--6\,$^{\circ}$C, the same as in the CRAFT study. This meant that the
$p^+$ on $n$ sensors~\cite{Agram:2004} were approximately at room
temperature.

As described in the Technical Design Report for the CMS
Tracker~\cite{Tracker_TDR,Tracker_TDR_add}, there are two main modes
of operation for the strip tracker analogue pipeline integrated circuits
(APV25~\cite{French:2001}): peak and deconvolution. In deconvolution
mode, the output charge for each strip represents a weighted sum of
three consecutive pipeline cells~\cite{Gadomski:1992}.  Although
deconvolution mode was designed to avoid signal pile-up in high
(design) luminosity operations, it will be necessary to run in this
mode whenever the expected separation between proton bunches will be
less than a few hundred nanoseconds. The luminosity in the early
collision running was very low and the bunches well separated; most of
the strip data were collected in peak mode, which is based on the
signal in a single pipeline cell.  All of the data, whether in peak or
deconvolution mode, were zero suppressed, meaning that only strips
which were part of clusters were read out for each event.

Many of the starting parameters for the strip tracker during the early
collision running had been established in the preceding CRAFT
period. For example, the timing of the tracker subsystems (in peak
mode) with respect to CMS triggers was set during the cosmic-ray muon
studies. Similarly, the alignment parameters for the strip detector
modules were derived from the same studies. 

As part of the alignment process, offsets had been determined for
cluster positions in sensors due to the Lorentz drift of holes and
electrons under the influence of the solenoid field.
For barrel layers, the Lorentz angle correction for cluster positions
during track reconstruction is about 10\,$\mu$m, which is
significantly larger than the 3--4\,$\mu$m alignment precision
achieved in the cosmic ray studies~\cite{craft_alignment_paper}.


\subsubsection{Strip Timing Scan}

As the strip tracker was operated in peak mode at the start of the
early collision running, the trigger timing established in the
preceding CRAFT period could be used.  In CRAFT the sampling time of
the APV25's was set within each subsystem by means of a dedicated
synchronization signal, adjusted according to the measured readout
fibre lengths. The synchronization of the subsystems was obtained
using the signal from cosmic-ray muon tracks.  Details on how the scan was
done can be found in Ref.~\cite{CMS_NOTE_2008_007}.  Toward the end of
the data collection period the APV25 mode was changed from peak to
deconvolution and since timing is more critical in the latter, a
fine-delay scan was made following the mode change.  For expediency
only one layer (TOB L3) was used in the study.

\begin{figure}[hbtp] 
\begin{center}
    \mbox{
{\scalebox{0.40}{
         \includegraphics[width=\linewidth]{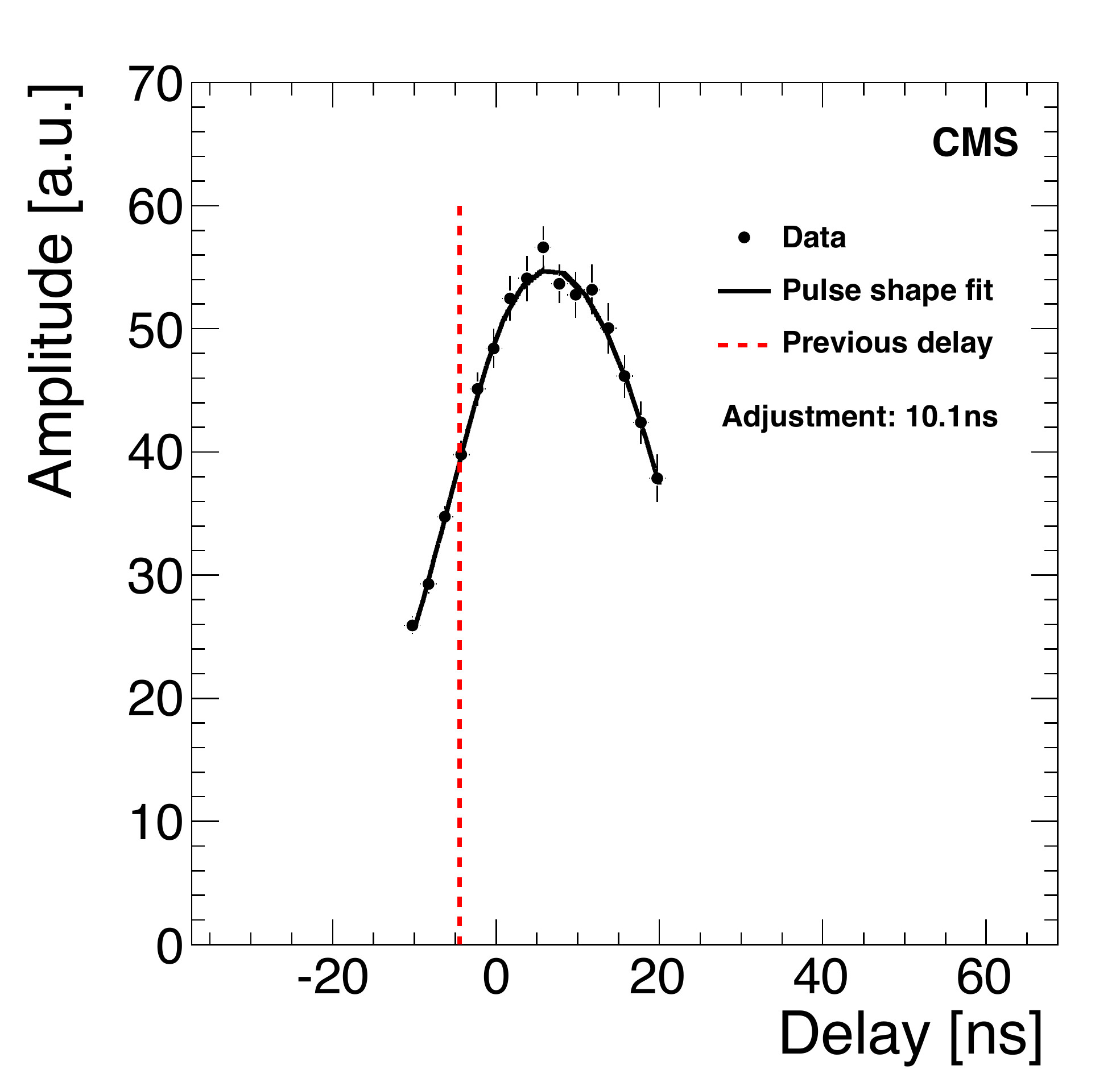}
     }}
  }
\caption{Normal-incidence-scaled charge (arbitrary units) of the 
highest pulse height strip in a cluster as a function of the readout 
delay with respect to the CMS trigger, in deconvolution mode.
The dashed vertical line corresponds to the setting prior to
the timing scan.}
\label{fig:strip_timing}
\end{center}
\end{figure}

Figure~\ref{fig:strip_timing} shows the result of the fine-delay
timing scan. The timing adjustment for the clock and trigger signals
is set on the front-end hybrids and the smallest step size is
1.04\,ns. From the figure it can be seen that the timing prior to the
scan had been off by about 10\,ns from ideal.  This level of mistiming
resulted in an estimated 2.5\% decrease in Signal-to-Noise (S/N) in
the strip modules during the peak mode running, where the delay timing
is less critical. The amplitude that is measured in the timing scan
represents the signal of the highest pulse height strip in a cluster
scaled by the ratio of the sensor thickness to the path length of the
track in the sensor.

Following the scan, the timing offsets for all of the strip tracker
subsystems were updated and some data were collected in deconvolution mode.
No data samples were collected in peak mode with the new delays.

\subsubsection{Signal-to-Noise Measurements}

Signal-to-Noise measurements were made in both peak and deconvolution
modes. In peak mode, the S/N ratio was determined at both
centre-of-mass energies, 0.9 and 2.36\,TeV, whereas deconvolution mode
is restricted to 2.36\,TeV. The ratio is evaluated on the basis of
charge clusters associated with reconstructed tracks, where the
individual strip noise values are taken from calibration runs. For
track angles that are not normal to the surface of modules the signal
values are scaled by the cosine of the angle relative to the local
normal.  This is done to give the same expectation value per cluster
for modules of the same type.  Cluster noise, which takes into account
the noise of each strip within a cluster, is used as the denominator
in the S/N ratio. When all strips within a cluster have the same
noise, cluster noise is equivalent to the noise of a single
strip. Further details on the determination of the S/N ratio can be
found in Ref.~\cite{Eklund:1999sn}.
 
\begin{figure}[hbtp] 
\begin{center}
    \mbox{
     \subfigure[]
{\scalebox{0.40}{
         \includegraphics[width=\linewidth]{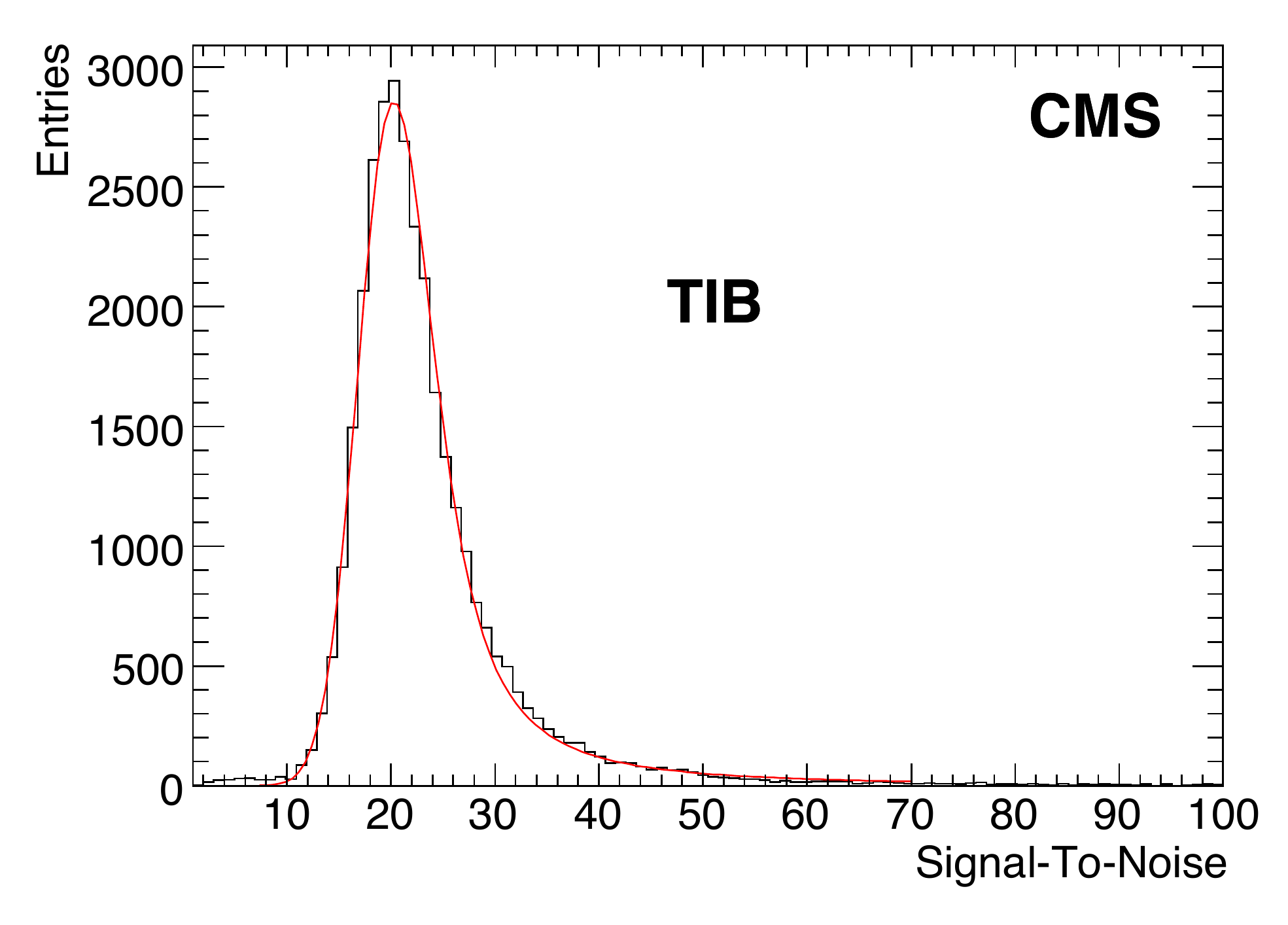}
         \label{fig:TIBsn}
     }}
  }
    \mbox{
     \subfigure[]
{\scalebox{0.40}{
         \includegraphics[width=\linewidth]{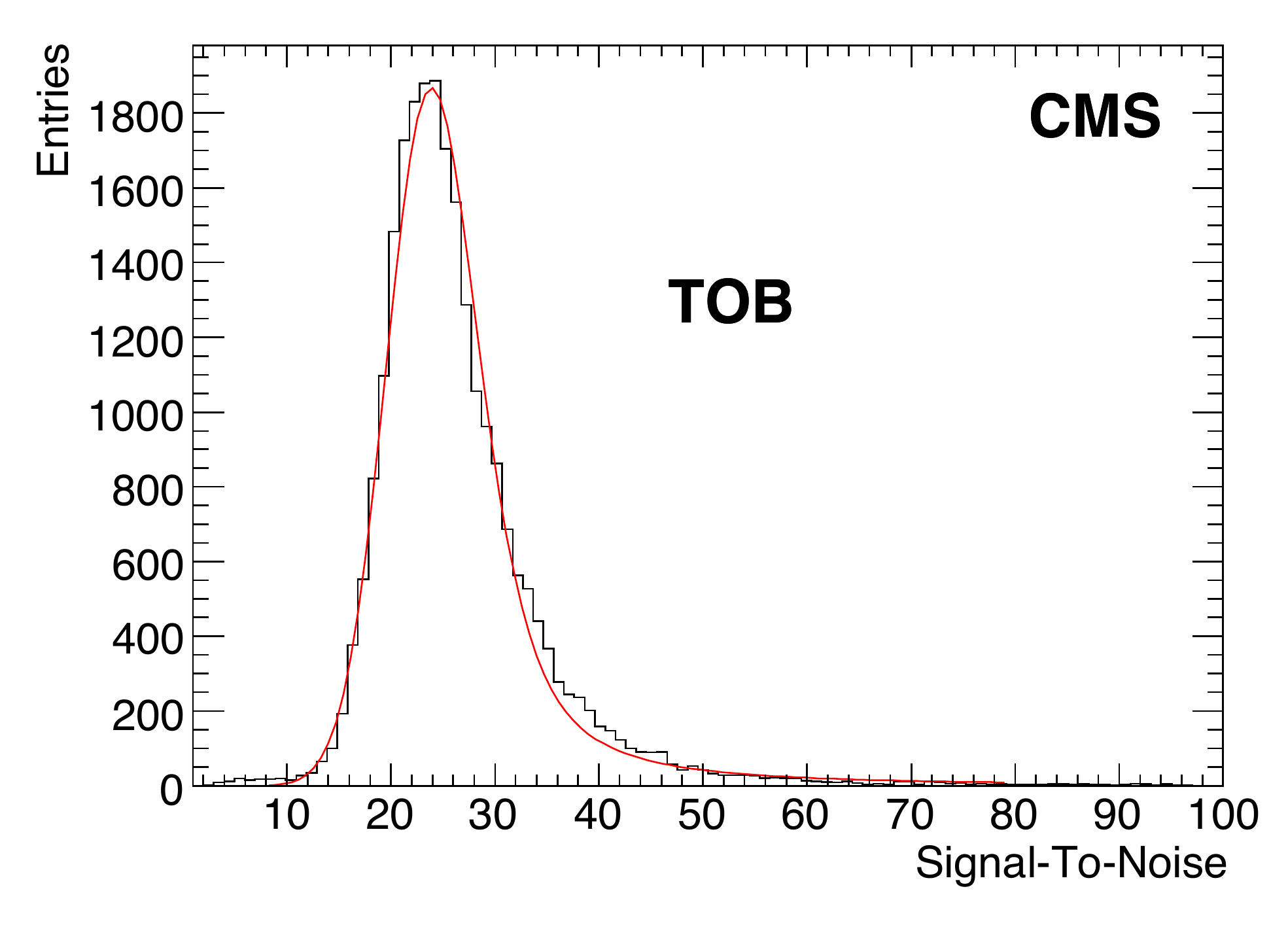}
         \label{fig:TOBsn}
     }}
  }
\caption{
Signal-to-Noise distributions in deconvolution mode for (a) (thin
sensor) TIB and (b) (thick sensor) TOB modules. The curves are 
results of the fits to a Landau distribution convoluted with a Gaussian 
distribution.}
\end{center}
\end{figure}

Figures~\ref{fig:TIBsn} and \ref{fig:TOBsn} show the S/N distributions
for the TIB and TOB modules, respectively, in deconvolution
mode. Included with each distribution is the result of the fit to a
Landau distribution convolved with a Gaussian distribution. The most
probable value of the fitted curves is taken to be the S/N value and
results for all of the strip tracker subdetectors are summarized in
Table~\ref{tab:ston} for all three running conditions.  Peak values
shown in the table have not been corrected for the 2.5\% loss due to
non-optimal timing. They are comparable with results obtained in the
CRAFT studies and in earlier cosmic ray studies.  The difference in
peak and deconvolution mode S/N values stems largely from the higher
noise in deconvolution.  After calibration there is some variation in
signal values (measured in electrons) for the two modes, but this has
been shown to be within 10\%. The S/N ratio should not depend on the
centre-of-mass energy and this is confirmed by the table entries.

Although it is not possible to directly compare channel noise
distributions in the early collision data with results from
calibration runs given the zero suppression, the frequency and
distribution of clusters in empty LHC buckets provide an indirect
cross-check of the calibration results and assumptions about the
Gaussian and uncorrelated nature of the noise.  For example, with bad
modules excluded from the readout the mean number of clusters in empty
buckets, out of some 9 million channels, was 4.2. This is consistent
with the clustering rules, which require a certain number of standard
deviations (five for the total charge in a cluster), and Gaussian
probabilities. By way of contrast, there were $\sim$1200 clusters per
minimum bias trigger in the 0.9\,TeV data.


\begin{table}[htbp]
\caption{\label{tab:ston}Summary of strip tracker Signal-to-Noise measurements. The peak mode ratios have not been corrected for the estimated 2.5\% decrease in signal from the trigger mistiming, as described in the text.} 
\begin{center}
\begin{tabular}{lccccc}\hline
Conditions & TIB & TID & TOB & TEC thin & TEC thick \\ \hline 0.9\TeV,
peak mode & 27.4 & 26.7 & 34.1 & 28.8 & 35.7 \\ 2.36\TeV, peak mode &
27.4 & 26.8 & 34.1 & 28.8 & 35.7 \\ 2.36\TeV, deco mode & 20.3 & 19.2
& 23.9 & 20.3 & 26.1 \\ \hline
\end{tabular}
\end{center}
\end{table}

\subsubsection{Strip Layer Efficiencies}
\label{sec:strip_effs}

Efficiencies for strip tracker layers were determined using events
that were collected in peak mode.  Reconstructed tracks in these
events were required to have a minimum of 8 hits in order to be used
in the efficiency measurements. To avoid inactive regions and allow for
alignment imprecision, trajectories passing near the edges of sensors
were excluded. The presence of a hit anywhere within the non-excluded
region of a traversed module was counted as a positive response;
efficiency is determined by the ratio of positive responses to the
total number of traversing tracks. Layers under study were not removed
from the track reconstruction and could in fact count toward the
minimum hit requirement. The total integrated hit efficiency during
the early collision period was measured to be 97.8\%, which is
essentially explained by the number of bad modules in the strip
tracker. That is, about 2.2\% of the modules have been excluded from
the readout because of problems with high voltage short circuits,
control ring failures, or other issues.  With known problem modules
excluded, the overall hit efficiency is 99.8\%, consistent with the
$\sim$0.2\% bad channel rate from the construction process.  Detailed
simulations, used to determine track reconstruction efficiency, take
into account inactive regions in addition to the measured
efficiencies. The efficiency measurements for the collision data
include an estimated 0.04\% systematic error due to the use of the
layers under study in the reconstruction process and the wide search
windows within modules.


\subsubsection{Energy Loss Measurement}
\label{sec:strips_dedx}

Although the primary function of the strip tracker is to provide hit
position information for track reconstruction and precise momentum
determination, the wide linear range of the strip channel output also
provides a measure of energy loss. That is, the charge collected in a
hit cluster is directly proportional to energy lost by a particle,
largely through ionization, while traversing the silicon. For
reconstructed tracks the angle $\theta$ between the track direction
and the axis normal to module sensor is well defined for each hit on
the track.  The instantaneous energy loss per unit path length
($dE/dx$) in the silicon is then approximated by the quantity $\Delta
E/(\Delta L \cdot \sec\theta)$, where $\Delta E$ is the
cluster charge expressed in units of MeV and $\Delta L$ is the
normal-angle thickness of the active volume of the silicon sensor.
All of the TIB and TID modules and the modules on rings 1--4 of the
TEC have silicon sensors that are 320\,$\mu$m thick, whereas the TOB and
TEC ring 5--7 modules have 500\,$\mu$m thick sensors.  Some 30\,$\mu$m
of the nominal thicknesses for both thin and thick types is inactive
material, i.e., does not contribute to the charge collection.

In zero-suppressed readout, which was used exclusively in the early
collision period, there are 8 ADC bits for the charge on each channel
within a cluster. Channel gains are set such that a single ADC count
corresponds to about one-quarter of the average noise and full scale
corresponds to approximately three times the average loss expected
from normally incident minimum ionizing particles. The highest two ADC
values have a special significance: 254 implies a value between 254
and 1024 counts, and 255 indicates that the actual value was in excess
of 1024 counts. The $dE/dx$ algorithm includes the saturated values
but without any special treatment.

The main point in determining energy loss per unit path length is
that, for a given medium, $dE/dx$ depends largely on the velocity
($\beta$) of the traversing particle. By combining $dE/dx$ information
with the measured momentum $p$ of a track, one can determine the mass
of the traversing particle. On the scale of charged particle momenta
in CMS collisions, there is only a limited range near the low end
where the difference in $\beta$ values is significant enough to
distinguish among long-lived hadrons. The momentum range where pions
would have relatively large energy loss is such that tracks tend to
curl up in the 3.8~T solenoid field and thus fail to be reconstructed.
 
\begin{figure}[hbtp] 
\begin{center}
    \mbox{
     \subfigure[]
{\scalebox{0.40}{
         \includegraphics[width=\linewidth]{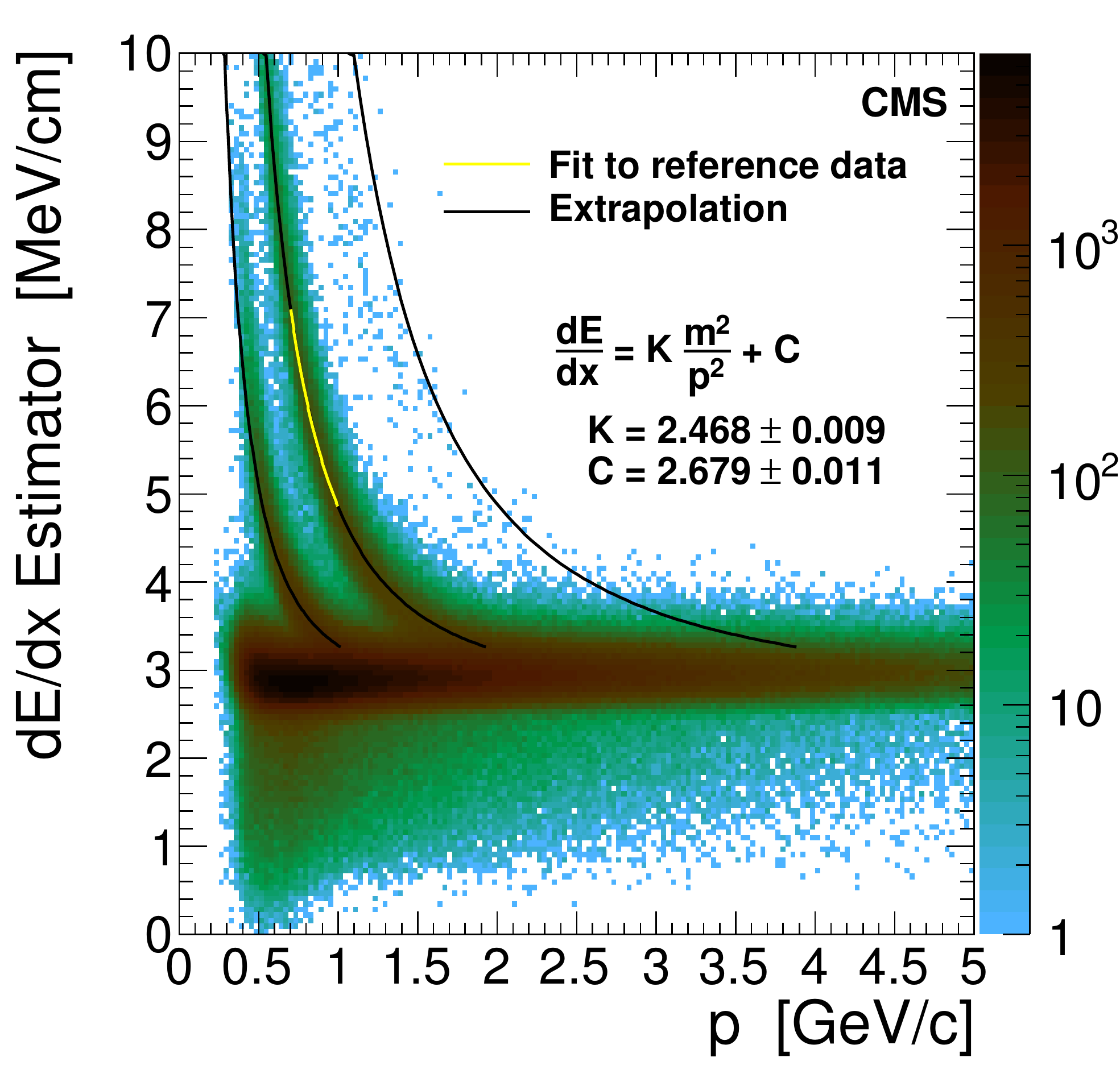}
         \label{fig:dedx_p}
     }}
  }
    \mbox{
     \subfigure[]
{\scalebox{0.40}{
         \includegraphics[width=\linewidth]{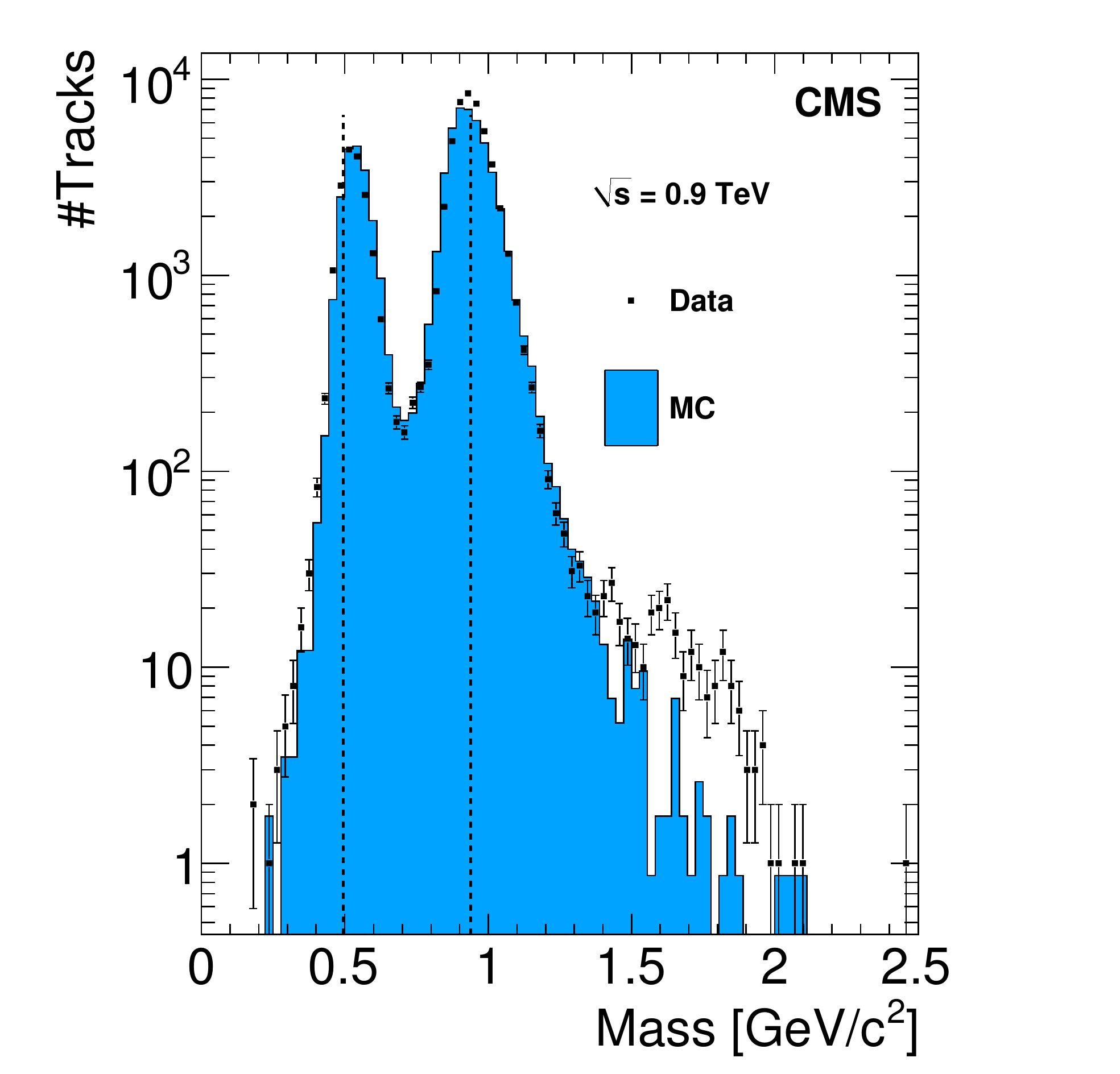}
         \label{fig:dedx_m}
     }}
  }
\caption{
Energy loss versus the momentum of tracks (a) and frequency of tracks 
as a function of track mass as determined from the 
measured energy loss and momentum (b). The lightly shaded line
in the (a) indicates the fit in the reference range of the proton band 
while the darker lines correspond to extrapolations for kaons, protons,
and deuterons based on the fit parameters.}
\end{center}
\end{figure}

The strip hits on reconstructed tracks represent independent measures
of $dE/dx$, ignoring the negligible loss of energy in traversing the
tracker.  Although pixel hits are included in the track
reconstruction, they are not used in the $dE/dx$ calculation due to
their more limited linear range.  Several methods have been used to
determine an estimate for the most probable $dE/dx$ value based
on the measurements in the strip tracker modules traversed by a
track. Figure~\ref{fig:dedx_p}, for example, shows the relationship
between the Harmonic-2 $dE/dx$ estimator~\cite{CMS_NOTE_2008_005} and
momentum for 0.9\,TeV data taken in peak mode. In the figure, clear
bands can be seen for kaons and protons and to a much lesser extent
for deuterons. 

An estimate of the mass of each candidate can be obtained using the
particle momentum and the measurement of the ionization energy loss
provided by the $dE/dx$ estimators.  To this end the following
relation between $dE/dx$, $p$, and $m$ is assumed for the momenta below
the minimum-ionizing region:
\begin{equation}
\frac{dE}{dx}= K\frac{m^2}{p^2}+C  \;.
\label{eq:bethebloch}
\end{equation}
The proton line in Fig.~\ref{fig:dedx_p} is used to extract the
parameters $K$ and $C$ in Eq.~\ref{eq:bethebloch}.  The
0.7--1.0\,GeV$/c$ range in the proton band is used for the reference
data fit, while extrapolations based on the same $K$ and $C$ values
yield a good agreement for protons with momenta above and below the
reference range and for kaons.

The mass spectrum that results from inverting Eq.~\ref{eq:bethebloch} for
all tracks with $dE/dx > 4.15$\,MeV/cm and $p<2$\,GeV$/c$ is shown in
Fig.~\ref{fig:dedx_m}. From the frequency plot one can observe clear
kaon and proton peaks as well as good agreement for the peaks from a
Monte Carlo simulation. There is also evidence for a deuteron peak in
data, although saturation of the ADC scale is particularly pronounced
for deuterons given their reduced $\beta$ values and relatively higher
$|\eta|$ values. That the deuteron peak is poorly modelled by the
simulation is partly understood as the underlying generator, \PYTHIA,
does not produce deuterons by design, although they can be produced in
the subsequent \GEANT~\cite{Geant} hadron showers.

\section{Track Reconstruction}
\label{sec:reconstruction}
The track reconstruction algorithms rely on a good estimate of the proton-proton
interaction region, referred to as the beamspot.  The beamspot is used as a precise
estimate of the primary interaction point (in the transverse direction) prior to primary
vertex reconstruction and as the sole primary interaction point if no primary vertex is found.
When the beamspot centre is displaced from the expected position there is a correlation between
the transverse impact parameter ($d_{xy}$) and the angle of the track at the point of
closest approach ($\phi_0$).  The beamspot fitter~\cite{CMS_NOTE_2007_021} uses an
iterative $\chi^2$ fitter to exploit this correlation between $d_{xy}$ and $\phi_0$,
looping over a sample of reconstructed tracks (using the old beamspot) to determine the new beamspot parameters.
After the beamspot is measured, the standard track reconstruction is performed.
During the 2009 data-taking, a beamspot was fitted during each LHC fill; fill-to-fill variations were 
at the level of $\sim$0.5\mm in $x$ and $y$, and $\sim$2\cm in $z$.

Starting from the location of the beamspot, an initial round of track and vertex reconstruction is
performed using only pixel hits.  The pixel vertices found at this stage are used in the
standard tracking.  The standard track reconstruction at CMS is performed by the
combinatorial track finder (CTF)~\cite{CMS_NOTE_2006_041}.  Tracks are seeded from either
triplets of hits in the tracker or pairs of hits with an additional constraint from the
beamspot or a pixel vertex, yielding an initial estimate of the trajectory, including its
uncertainty.  The seed is then propagated outward in a search for compatible hits.  As
hits are found, they are added to the trajectory and the track parameters and
uncertainties are updated.  This search continues until either the boundary of the tracker is
reached or no more compatible hits can be found.  An additional search for hits is
performed starting from the outermost hits and propagating inward.  In the final step, the
collection of hits is fit to obtain the best estimate of the track parameters.

The current implementation of the CTF performs six iterations.  Between each
iteration, hits that can be unambiguously assigned to tracks in the previous iteration are
removed from the collection of tracker hits to create a smaller collection that can be
used in the subsequent iteration.  At the end of each iteration, the reconstructed tracks
are filtered to remove tracks that are likely fakes and to provide a means of quantifying
the quality of the remaining tracks.  The filtering uses information on the number of
hits, the normalized $\chi^2$ of the track, and the compatibility of the track originating
from a pixel vertex.  Tracks that pass the tightest selection are labelled
\textit{highPurity}.  The first two iterations use pixel triplets and pixel pairs as seeds
to find prompt tracks with $\pt>0.9\GeVc$.  The next iteration uses pixel triplet seeds
to reconstruct low-momentum prompt tracks.  The following iteration uses combinations of 
pixel and strip layers as seeds, and is primarily intended to find displaced tracks.  
The final two iterations use seeds of strip pairs to reconstruct tracks lacking pixel hits.


\section{Tracking Performance}
\label{sec:tracking}

The results presented here come from the sample described in
Section~\ref{sec:data_samples}, using data taken at both centre-of-mass energies
(0.9 and 2.36\TeV), unless stated otherwise.  To reduce the background from beam-gas events, discussed
in Section~\ref{sssec:pixel_operting_minbias},
and to select useful events for tracking studies, two additional criteria are imposed for
most of the results in this section.  First, more than 20\% of the reconstructed tracks in
an event must be flagged as \textit{highPurity} if there are at least 10 tracks in the event.
Second, a primary vertex must be reconstructed in the region of pp interactions (see Section~\ref{sec:pvtx}).


The alignment parameters for the Tracker were computed from approximately two million cosmic ray
muon tracks collected during CRAFT running in November 2009 as described in Section~\ref{sec:data_samples}. 
The nominal values of the alignment parameter errors
have been used in the track reconstruction. Since the applied procedure was similar to the
one discussed in Ref.~\cite{craft_alignment_paper}, the resulting precision is also very
similar.
In particular, the width of the distribution of the mean of the residuals (taken as a
measure of the local alignment precision) in the pixel barrel local $x$ and $y$ coordinates is
3\micron and 4\micron, respectively.

The simulated events are minimum-bias events produced with the \PYTHIA
6.4~\cite{Pythia} event generator, tune D6T~\cite{D6T}, at centre-of-mass energies of
0.9 and 2.36\TeV (10 million events each) and processed with a simulation of the CMS
detector response based on \GEANTfour.  The misalignment, miscalibration, and
dead-channel map corresponding to the detector status and calibration accuracy at the time
of the first LHC collisions have been included in the simulation.  The longitudinal
distribution of the primary collision vertices has been adjusted to match the data.

\subsection{Basic Tracking Distributions}
\label{sec:basic}
\begin{figure}[hbtp]
\begin{center}
    \mbox{
      \subfigure[]
{\scalebox{0.313}{
	  \includegraphics[angle=90,width=\linewidth]{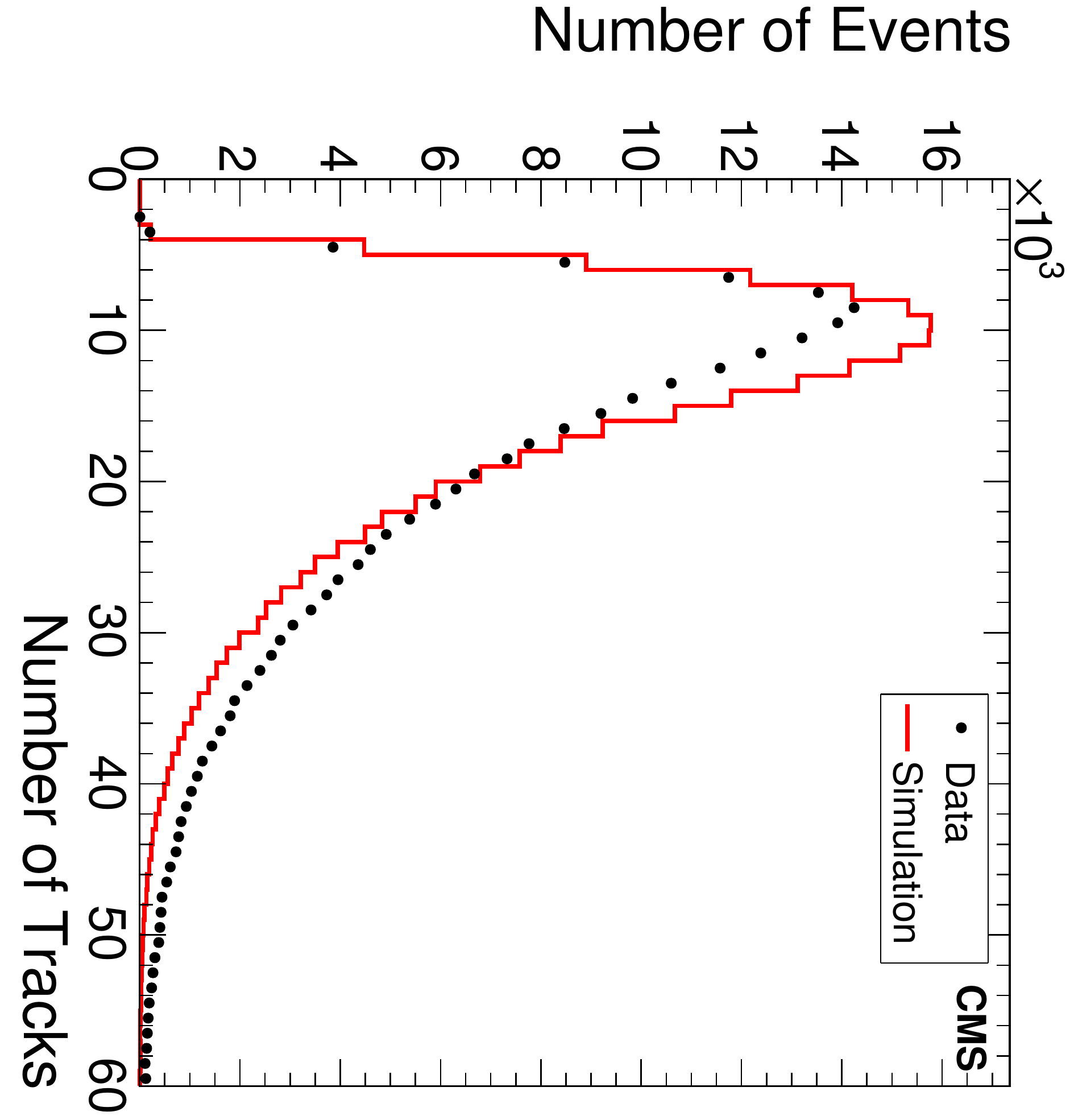}
   \label{fig:trk_n} 
      }}
    }
    \mbox{
      \subfigure[]
{\scalebox{0.313}{
	  \includegraphics[angle=90,width=\linewidth]{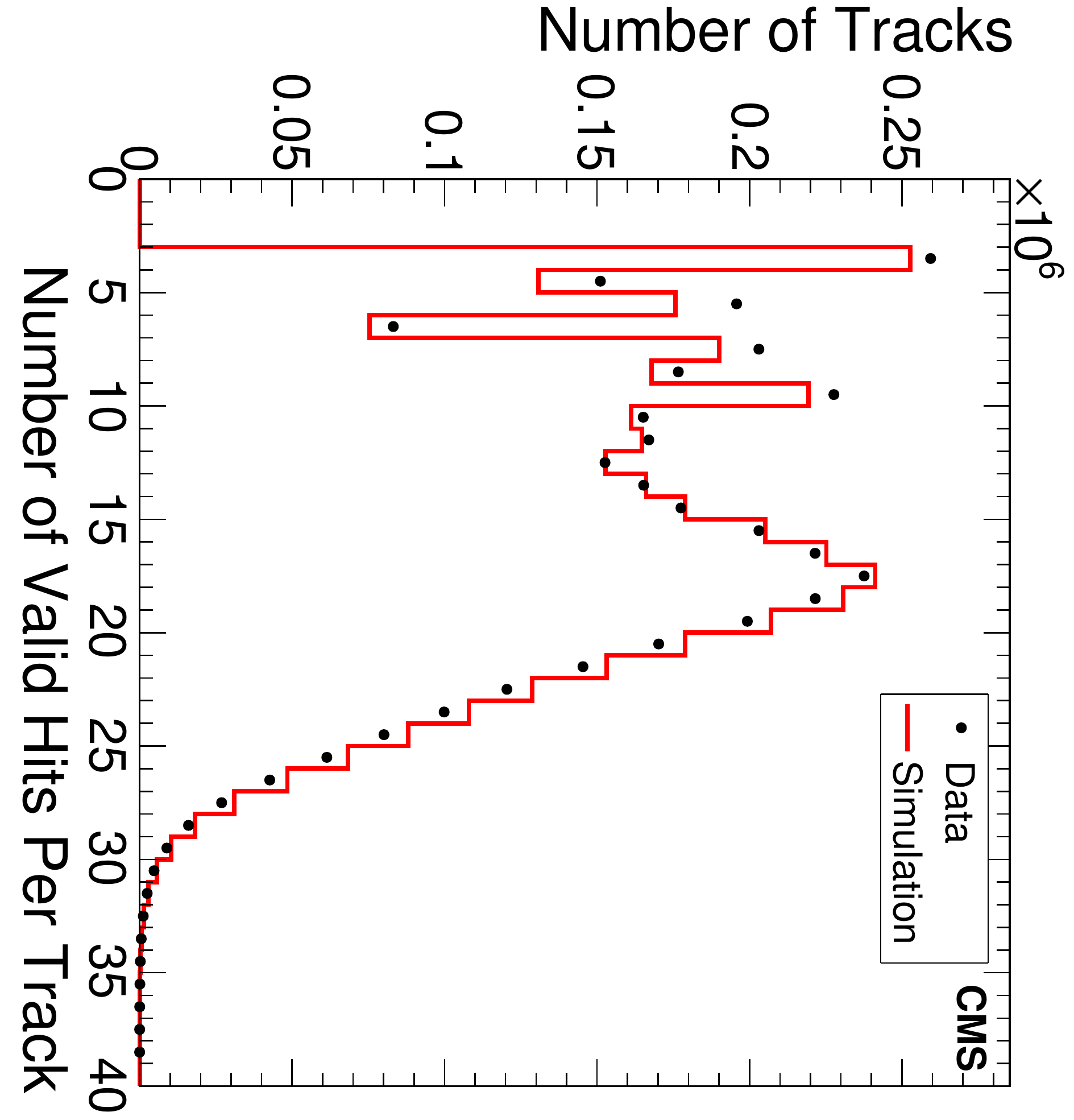}
   \label{fig:trk_nHit} 
      }}
    }
    \mbox{
      \subfigure[]
{\scalebox{0.313}{
	  \includegraphics[angle=90,width=\linewidth]{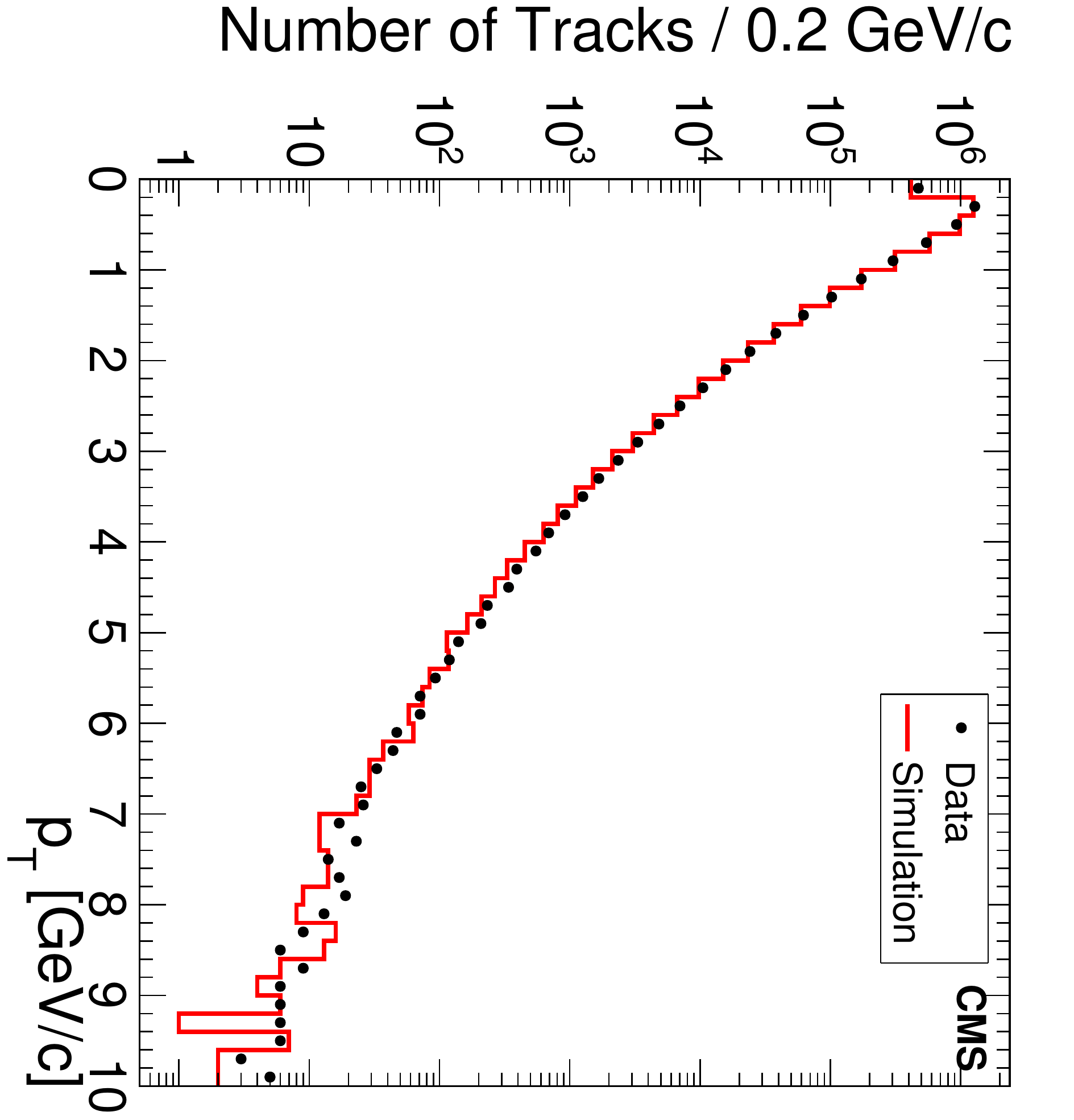}
   \label{fig:trk_pt} 
      }}
    }
    \mbox{
      \subfigure[]
{\scalebox{0.313}{
	  \includegraphics[angle=90,width=\linewidth]{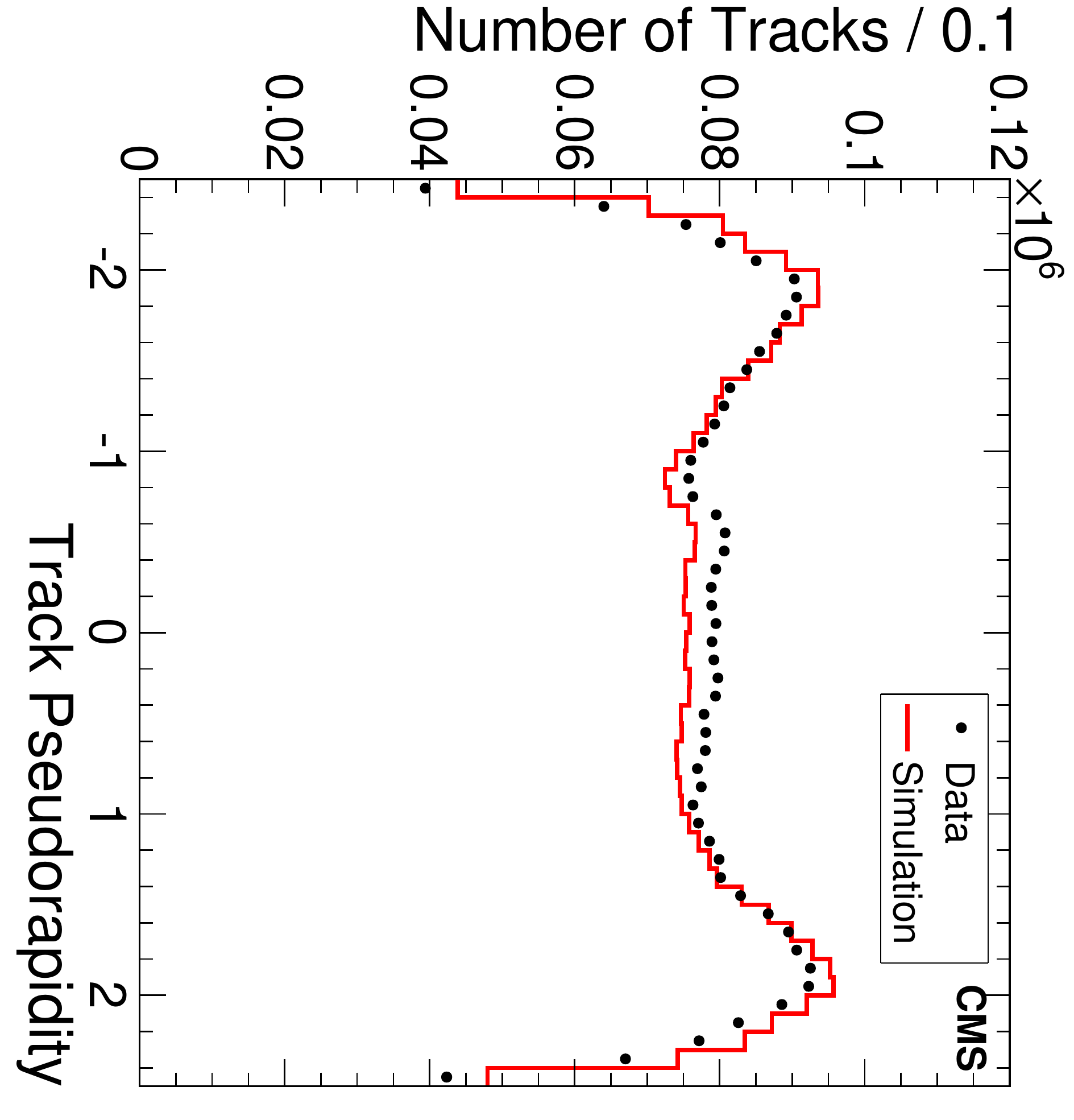}
   \label{fig:trk_eta} 
      }}
    }
    \mbox{
      \subfigure[]
{\scalebox{0.313}{
	  \includegraphics[angle=90,width=\linewidth]{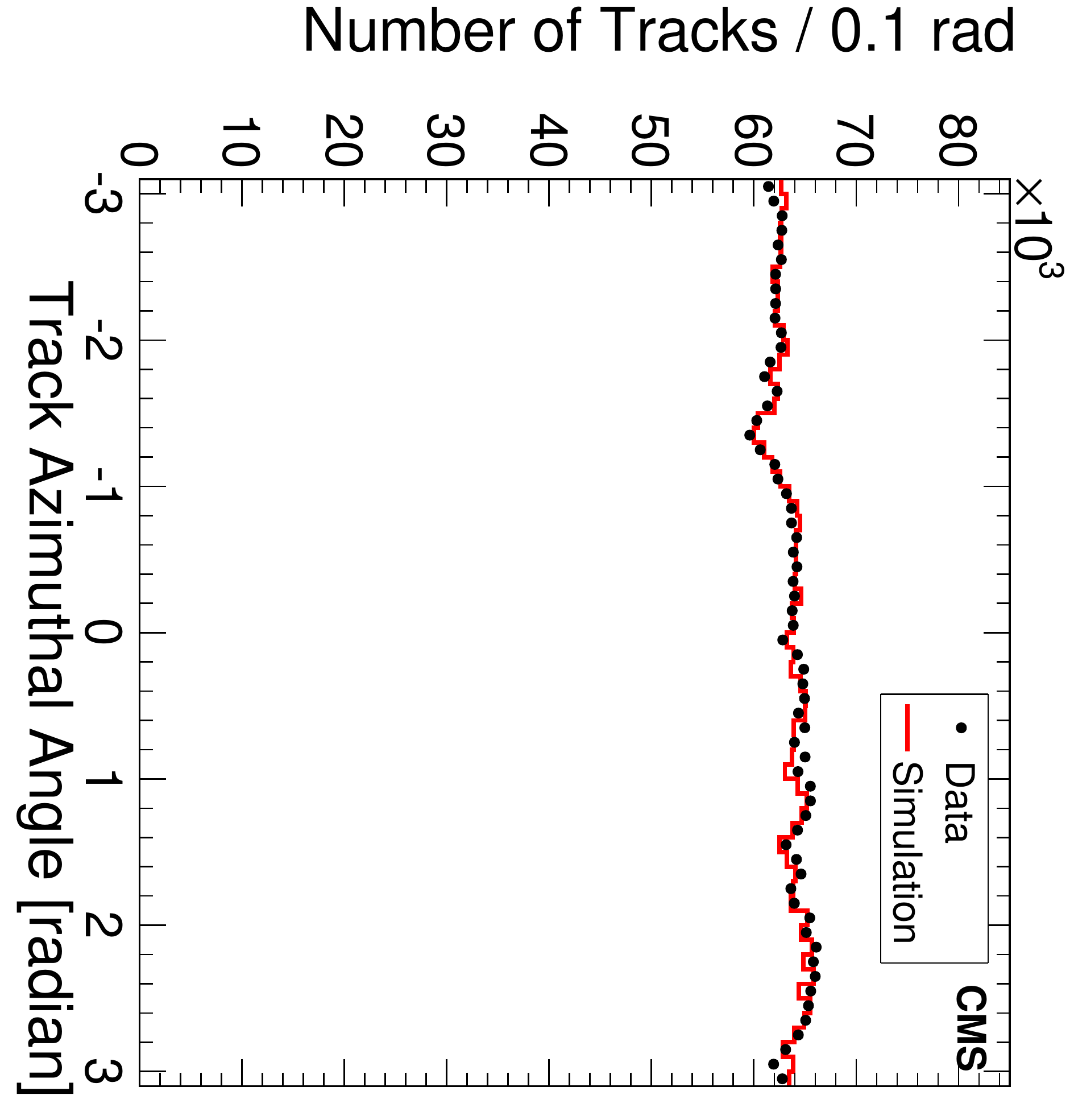}
   \label{fig:trk_phi} 
      }}
    }
    \mbox{
      \subfigure[]
{\scalebox{0.313}{
	  \includegraphics[angle=90,width=\linewidth]{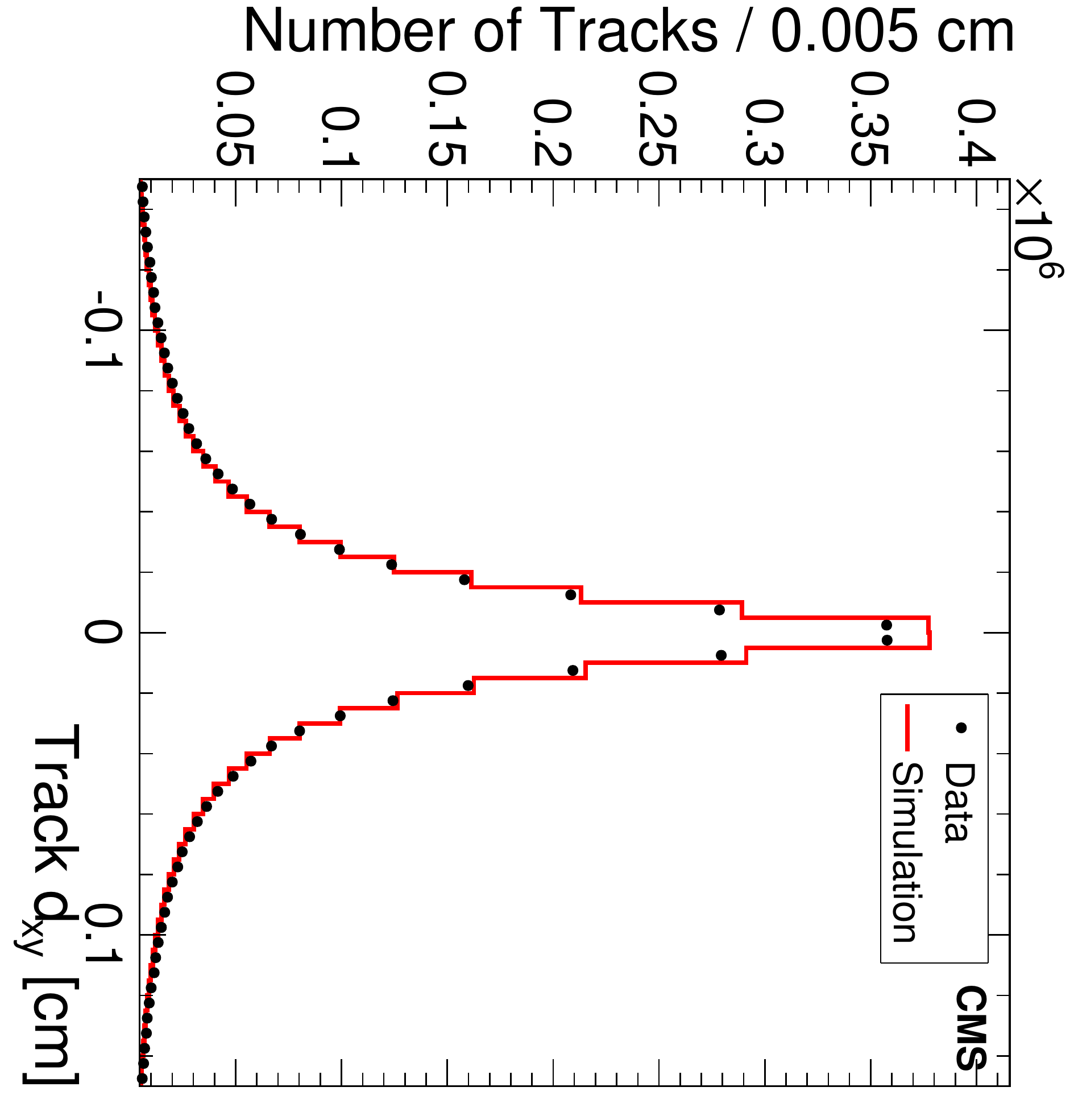}
   \label{fig:trk_dxyCorr} 
      }}
    }
    \mbox{
      \subfigure[]
{\scalebox{0.313}{
	  \includegraphics[angle=90,width=\linewidth]{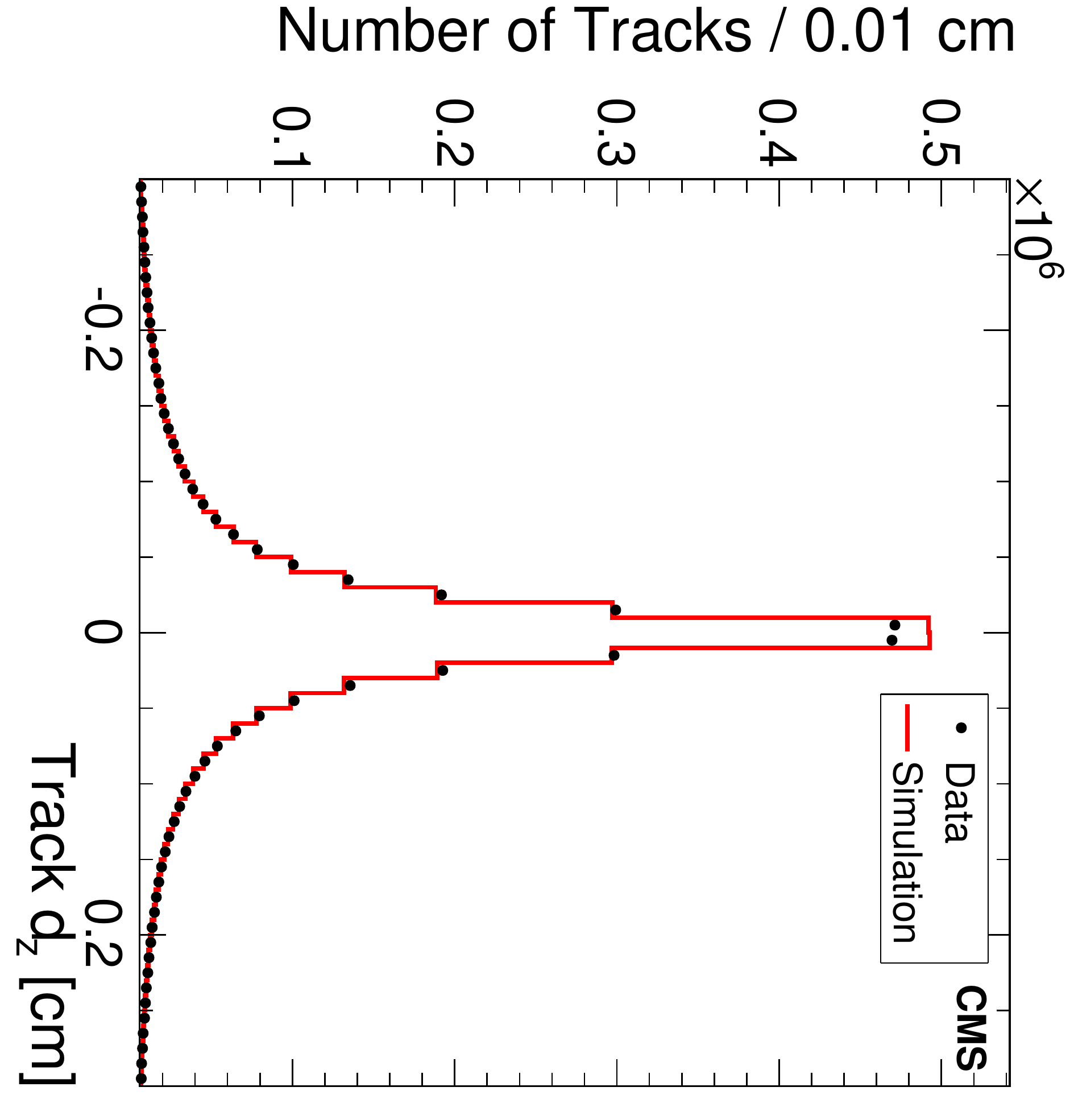}
   \label{fig:trk_dzCorr} 
      }}
    }
    \mbox{
      \subfigure[]
{\scalebox{0.313}{
	  \includegraphics[angle=90,width=\linewidth]{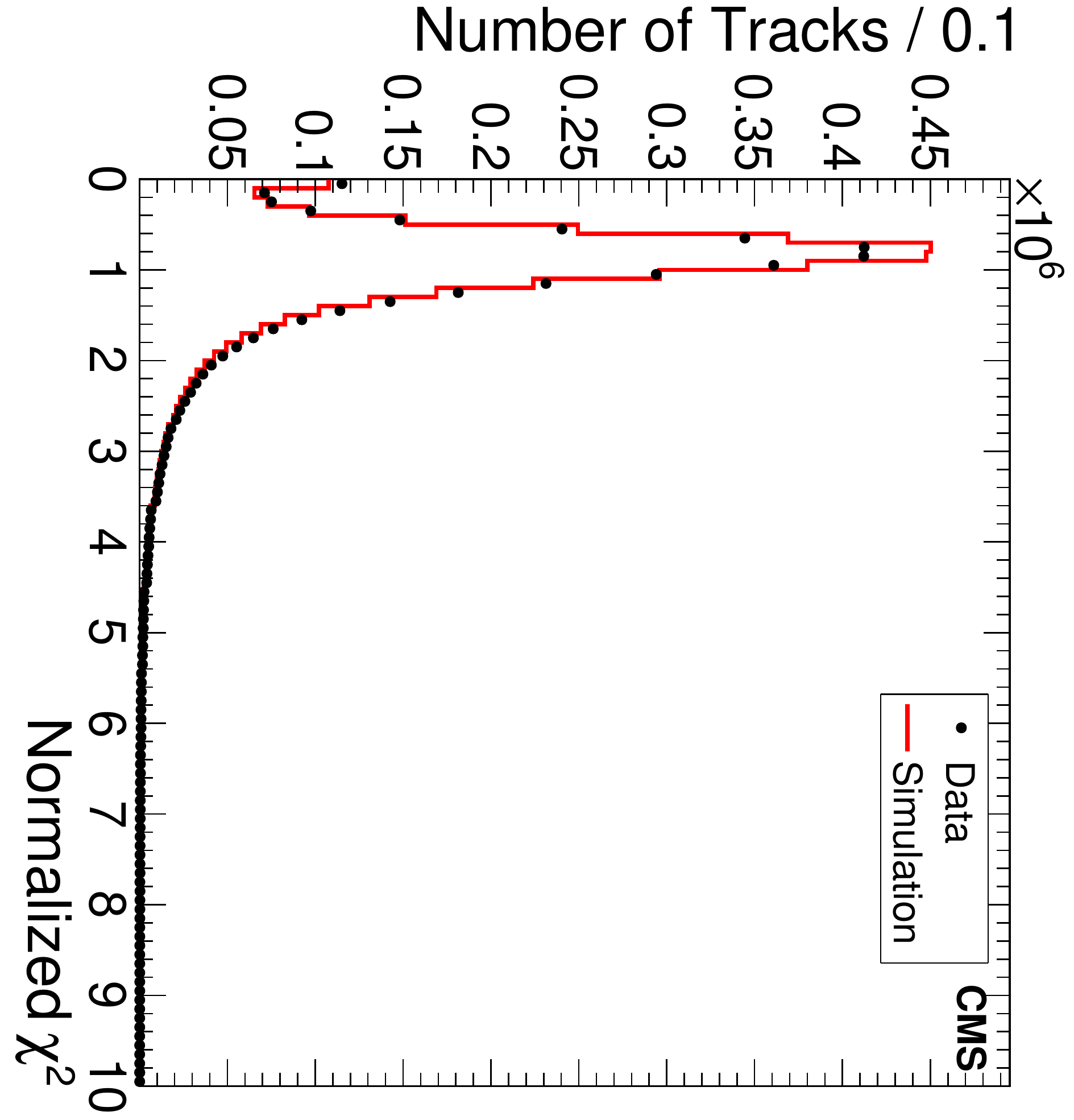}
   \label{fig:trk_chi2ndof} 
      }}
    }
\caption{Comparison of the data (points) and simulation (histogram) distributions of tracking parameters: (a)
number of tracks per event, (b) number of hits used per track, transverse (c)
momentum \pt, (d) track pseudorapidity $\eta$, (e) azimuthal angle $\phi$, (f) transverse impact
parameter $d_{xy}$ with respect to the primary vertex, (g) longitudinal impact parameter $d_z$
with respect to the primary vertex, and (h) normalized $\chi^2$.  The simulated distributions
are normalized by area to the data distributions.}
    \label{fig:basic_dist}
\end{center}
\end{figure}

The \textit{highPurity} tracks are selected, with additional requirements of 
$|d_z| < 10\,\sigma_z$ (where $d_z$ is the longitudinal impact parameter with respect to the primary 
vertex and $\sigma_z$ is the combined track and primary vertex uncertainty in $z$) 
and $\sigma_{\pt}/\pt < 10\%$, to compare
the data and simulation.  Figure~\ref{fig:basic_dist} shows the results of this comparison 
for several important track parameters.   
The distribution of the number of tracks per event, shown in Fig.~\ref{fig:trk_n}, has been normalized
to the number of events.  The data clearly have more tracks per event than 
are present in the simulated data.  This is believed to be due to an as-yet 
unoptimized tune of the \PYTHIA generator.
To be able to compare shapes, the other
distributions have been normalized to
the number of reconstructed tracks in the data.  There is general agreement
between the data and simulation distribution shapes for all other tracking variables.
In particular, the features in the $\phi$ distribution, due to inactive modules, 
are well modelled by the simulation.


\subsection{Primary Vertex Resolution}
\label{sec:pvtx}

The reconstruction of the primary interaction vertex in the event starts from the track
collection.  The tracks are clustered based on the $z$ coordinate of the track at the 
point of closest approach to the beamline.  The clusters are fit
with an adaptive vertex fit~\cite{CMS_NOTE_2007_008}, 
where tracks in the vertex are assigned a weight between 0
and 1 based on their proximity to the common vertex.

\begin{figure}[hbtp]
\begin{center}
    \mbox{
      \subfigure[]
{\scalebox{0.313}{
	  \includegraphics[width=\linewidth]{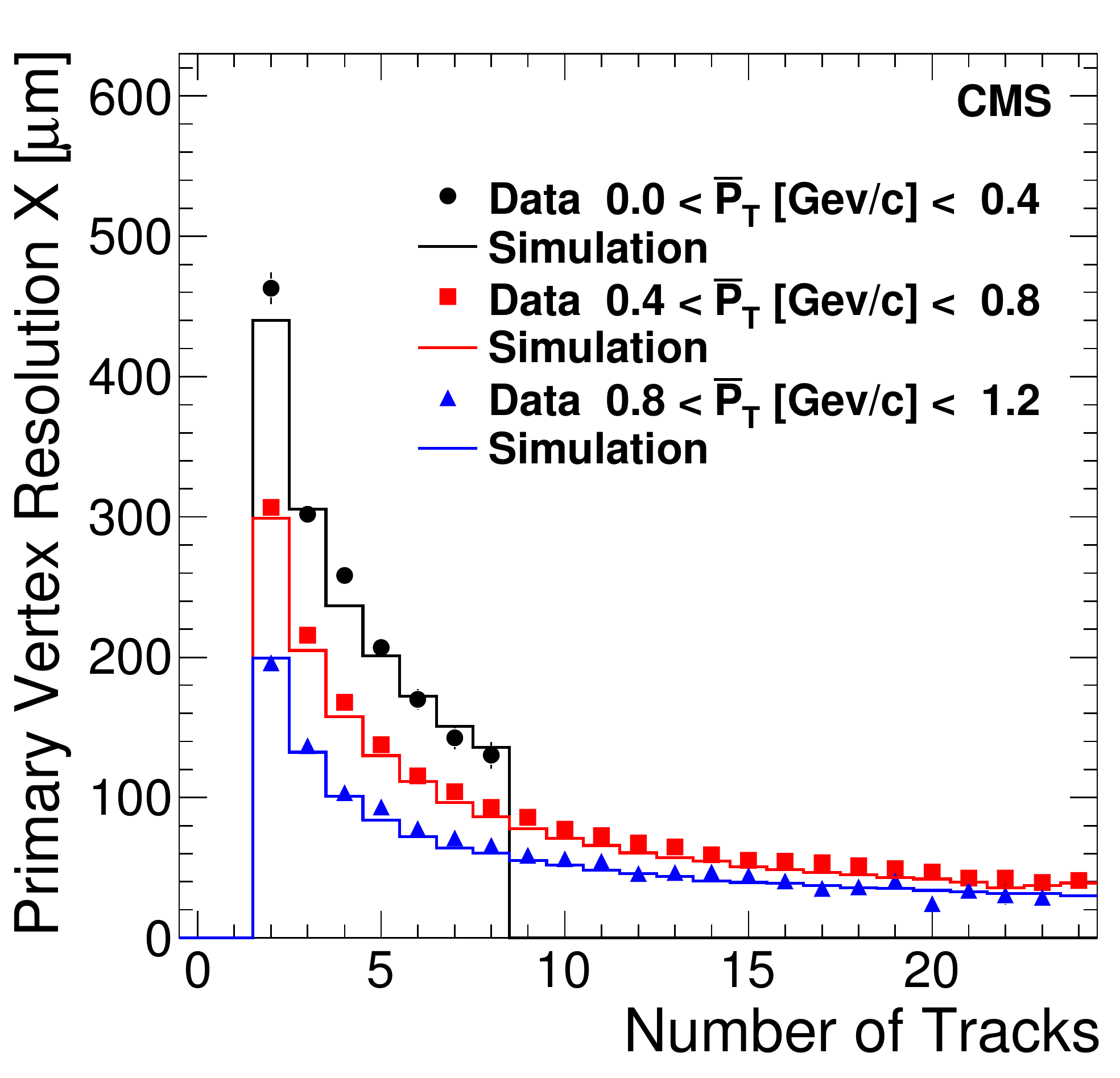}
   \label{fig:pvtx_respt_x} 
      }}
    }
    \mbox{
      \subfigure[]
{\scalebox{0.313}{
	  \includegraphics[width=\linewidth]{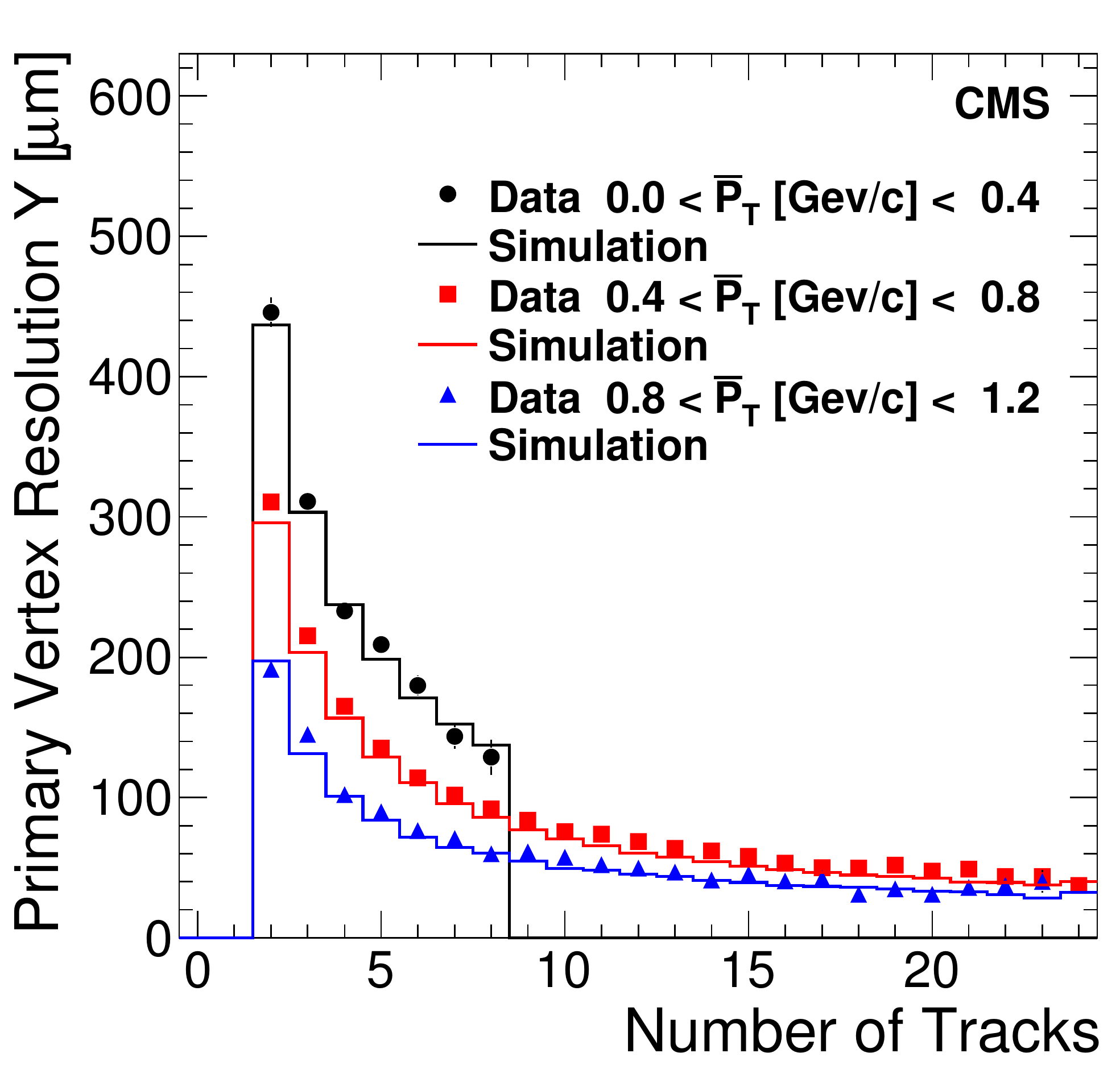}
   \label{fig:pvtx_respt_y} 
      }}
    }
    \mbox{
      \subfigure[]
{\scalebox{0.313}{
	  \includegraphics[width=\linewidth]{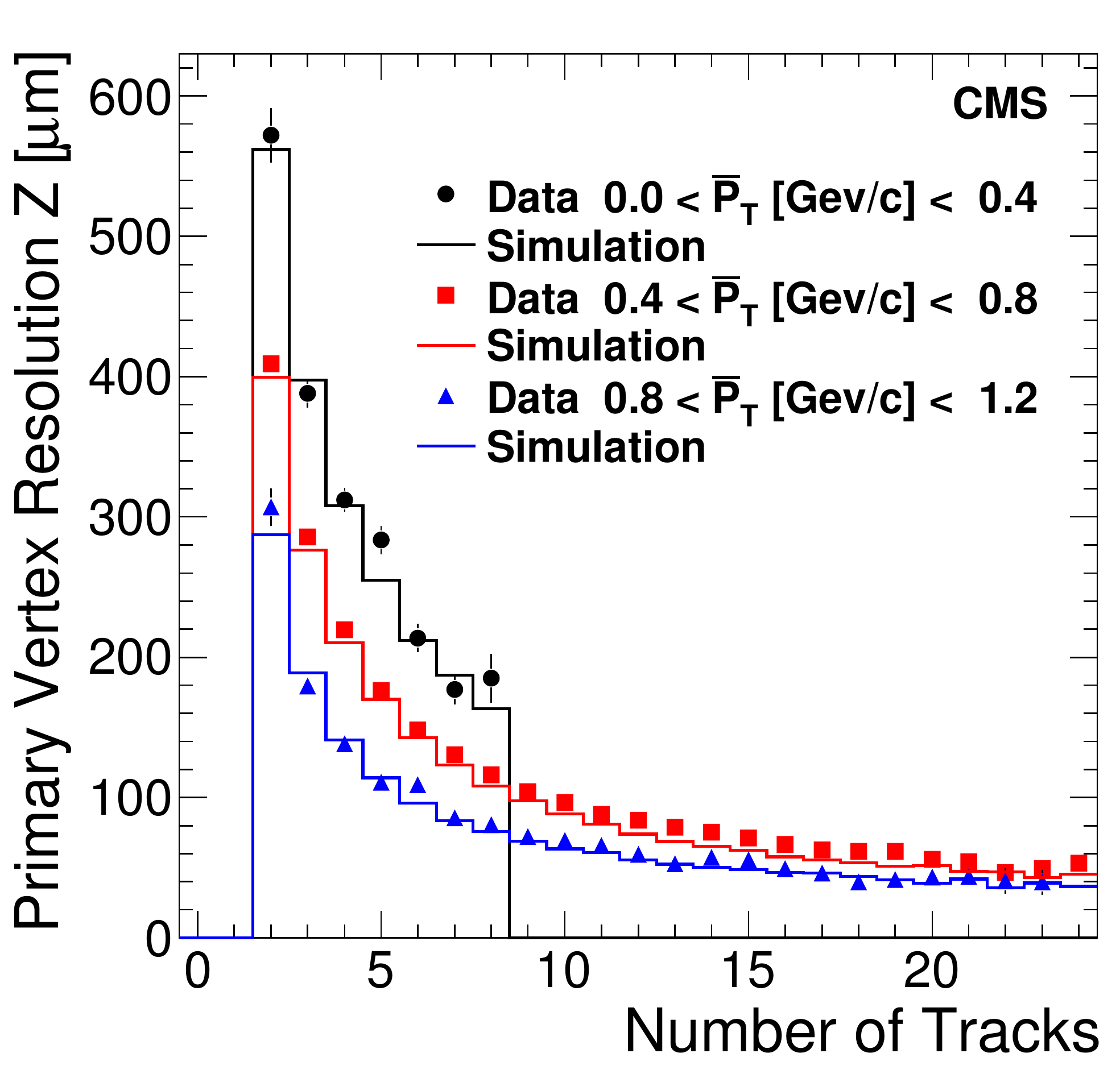}
   \label{fig:pvtx_respt_z} 
      }}
    }
\caption{Primary vertex resolution distributions in (a) $x$, (b) $y$, and (c) $z$ versus 
number of tracks.  The three sets of results in each plot show different average \pt ranges and 
within each \pt range, data and simulation are compared.}
\label{fig:pvtx_respt} 
\end{center}
\end{figure}

The primary vertex resolution strongly depends on the number of tracks used in fitting the vertex
and on their \pt.  To measure the resolution, 
the tracks in an event with only one vertex are randomly split into two different sets
and used to independently fit the primary vertex.  The distribution of the difference in
the fitted vertex positions can then be used to extract the resolution by fitting a 
Gaussian to it and dividing $\sigma$ by $\sqrt{2}$.
To examine the effect of the \pt\ of the tracks in the vertex, we study the resolution
versus the number of tracks in the vertex for different average \pt\ of tracks in the
vertex.  Figure~\ref{fig:pvtx_respt} shows the $x$, $y$, and $z$ resolutions for 
different average \pt ranges.  While the resolution differs considerably depending on \pt 
and multiplicity, the simulation accurately reproduces the data results.

\subsection{Reconstruction of Particle Decays}
\label{sec:V0s}

\subsubsection{V$^0$ Reconstruction}
\label{sec:V0reco}

V$^0$ particles are long-lived ($c\tau > 1$\cm) neutral particles reconstructed
by their decay to two charged particles\footnote{Charge conjugate states are implied throughout the paper.}: 
K$_\mathrm{S}^0 \to \pi^+\pi^-$ and $\Lambda^0 \to p\pi^-$.
Reconstruction of V$^0$ decays requires finding oppositely charged tracks that are 
detached from the primary vertex and form a good secondary vertex with an appropriate invariant mass.
For the $\Lambda^0$, the lowest momentum track is assumed to be the pion.  As no further particle
identification is required, a V$^0$ candidate can appear in both K$_\mathrm{S}^0$ and $\Lambda^0$ samples.
To be considered as a V$^0$ decay track, a track must have at least 6 hits, a normalized
$\chi^2$ less than 5, and a transverse impact parameter with respect to the beamspot
greater than $0.5\sigma_{IP}$, where $\sigma_{IP}$ is the calculated uncertainty (including
beamspot and track uncertainties).  The reconstructed V$^0$ decay vertex must have a normalized 
$\chi^2$ less than 7
and a transverse separation from the beamspot greater than 15$\sigma_T$, where $\sigma_T$
is the calculated uncertainty (including beamspot and vertex uncertainties).
In addition, the V$^0$ candidate is discarded if either of the daughter tracks has hits that are more 
than 4$\sigma_{3D}$ from the V$^0$ vertex, towards the primary vertex, where $\sigma_{3D}$ is the 
uncertainty in the vertex position.

\begin{figure}[hbtp]
\begin{center}
    \mbox{
      \subfigure[]
{\scalebox{0.42}{
	  \includegraphics[width=\linewidth]{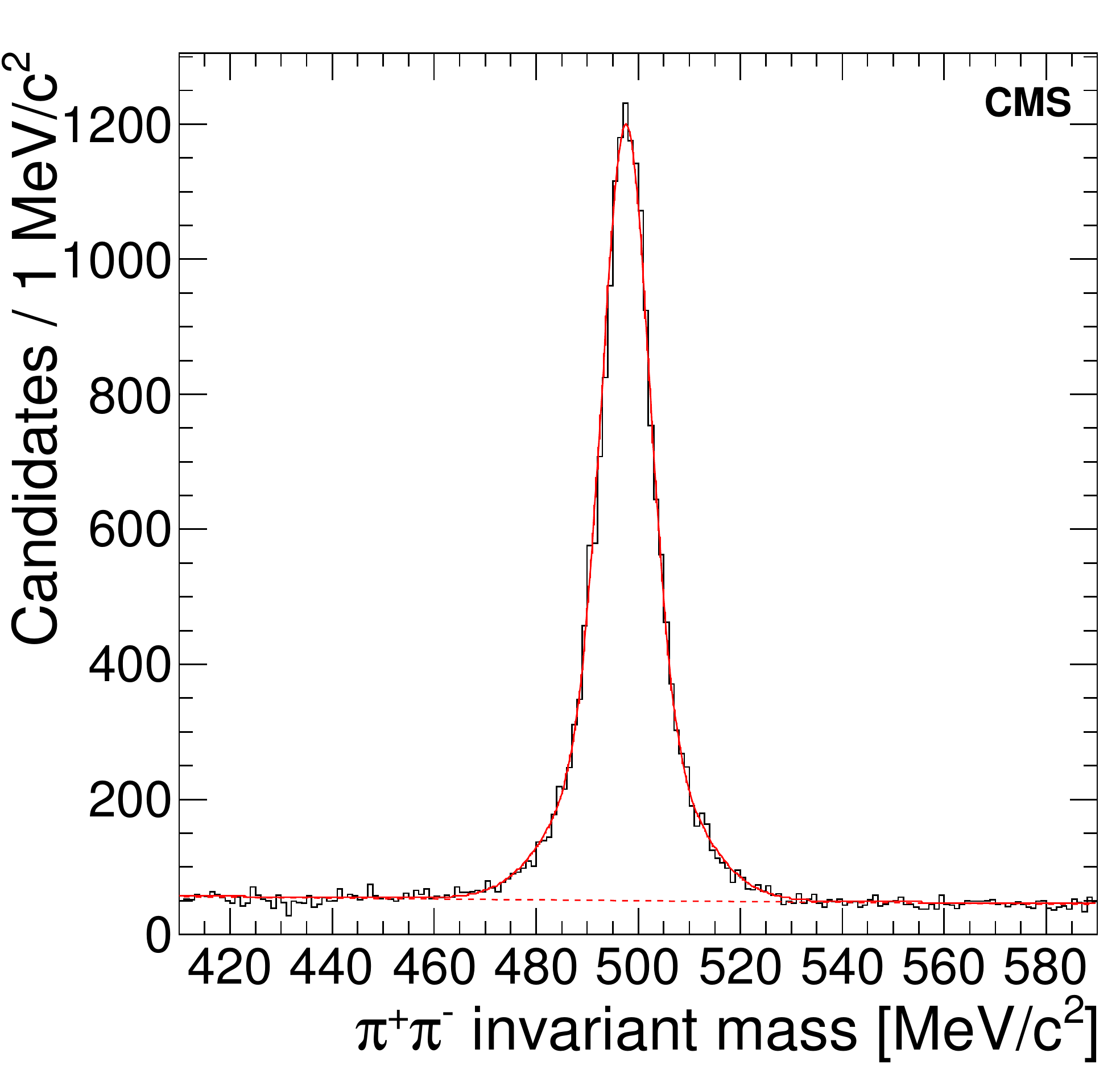}
   \label{fig:kshort_mass} 
      }}
    }
    \mbox{
      \subfigure[]
{\scalebox{0.42}{
	  \includegraphics[width=\linewidth]{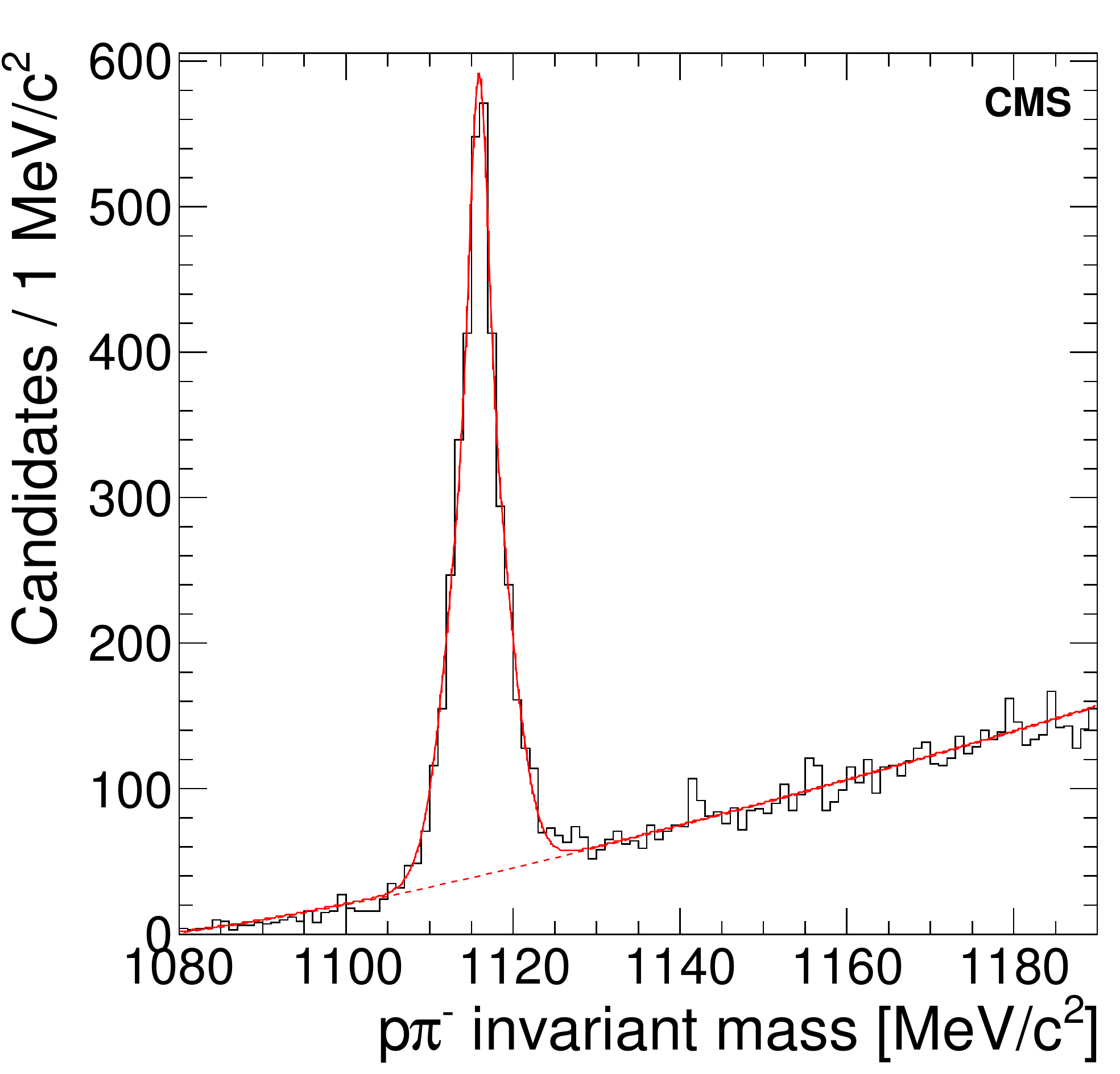}
   \label{fig:lambda_mass} 
      }}
    }
\caption{The invariant mass distributions of (a) $\pi^+\pi^-$ with a fit to the K$_\mathrm{S}^0$ and (b)
$p\pi^-$ with a fit to the $\Lambda^0$.}
\end{center}
\end{figure}

The mass resolution of the V$^0$ depends on $\eta$ as well as on the decay vertex position and 
a single Gaussian is not a sufficiently accurate functional form for the signal.  Therefore,
a double Gaussian with the same mean was used to fit the signal.  For the background shapes,
a linear background was used for $\pi^+\pi^-$ and the function $a(m-m_p-m_\pi)^b$
was used for the $p\pi^-$ spectrum where $m$ is the $p\pi^-$ invariant mass and $a$ and $b$ are 
free parameters. 
The $\pi^+\pi^-$ and $p\pi^-$ mass distributions, along with the overlaid
fits, are shown in Figs.~\ref{fig:kshort_mass} and \ref{fig:lambda_mass}, respectively.
Tables~\ref{tab:v0masses} and \ref{tab:v0sigmas} show the reconstructed V$^0$ masses and resolutions 
obtained from the data and simulation.
While the various results are close to expectations, significant discrepancies are present.  
These features can be examined as a function of track kinematic variables to better understand
the CMS tracker and magnetic field.  This work is ongoing.

\begin{table}[htbH]
\begin{center}
\caption{\label{tab:v0masses}Masses obtained from data, world average~\cite{PDG08}, and simulation (reconstructed and generated).
The uncertainties for data and simulation results are statistical only.}
\vspace{3pt}
\begin{tabular}{l|cc|cc}\hline
 & \multicolumn{4}{c}{Mass ($\!\MeVcc$)} \\
V$^0$        & Data & PDG & Simulation & Generated \\ \hline
K$_\mathrm{S}^0$ & $497.68 \pm 0.06$ & $497.61 \pm 0.02$ & $498.11 \pm 0.01$ & $497.670$ \\
$\Lambda^0$ & $1115.97 \pm 0.06$ & $1115.683 \pm 0.006$ & $1115.93 \pm 0.02$ & $1115.680$ \\ \hline
\end{tabular}
\end{center}
\end{table}

\begin{table}[htbH]
\begin{center}
\caption{\label{tab:v0sigmas}V$^0$ mass resolutions obtained from data and simulation.  The narrow and
wide Gaussian resolutions are $\sigma_1$ and $\sigma_2$, respectively.  The $\sigma_1$ fraction
is the fraction of the yield from the narrow Gaussian.  The final row gives the average resolution,
obtained from the square root of the weighted average of the two resolutions squared.  Uncertainties are statistical only.}
\vspace{3pt}
\begin{tabular}{lcccc}\hline 
Parameter & K$_\mathrm{S}^0$ Data & K$_\mathrm{S}^0$ Simulation & $\Lambda^0$ Data & $\Lambda^0$ Simulation \\ \hline
$\sigma_1 (\!\MeVcc)$ & $4.53 \pm 0.12$ & $4.47 \pm 0.04$ & $1.00 \pm 0.26$ & $1.71 \pm 0.05$ \\
$\sigma_2 (\!\MeVcc)$ & $11.09 \pm 0.41$ & $10.49 \pm 0.11$ & $3.25 \pm 0.14$ & $3.71 \pm 0.09$ \\
$\sigma_1$ fraction & $0.58 \pm 0.03$ & $0.58 \pm 0.01$ & $0.15 \pm 0.05$ & $0.44 \pm 0.03$ \\
$\overline{\sigma} (\!\MeVcc)$ & $7.99 \pm 0.14$ & $7.63 \pm 0.03$ & $3.01 \pm 0.08$ & $2.99 \pm 0.03$ \\ \hline
\end{tabular}
\end{center}
\end{table}

\subsubsection{V$^0$ Lifetime}
\begin{figure}[hbtp]
\begin{center}
    \mbox{
      \subfigure[]
{\scalebox{0.42}{
	  \includegraphics[angle=90,width=\linewidth]{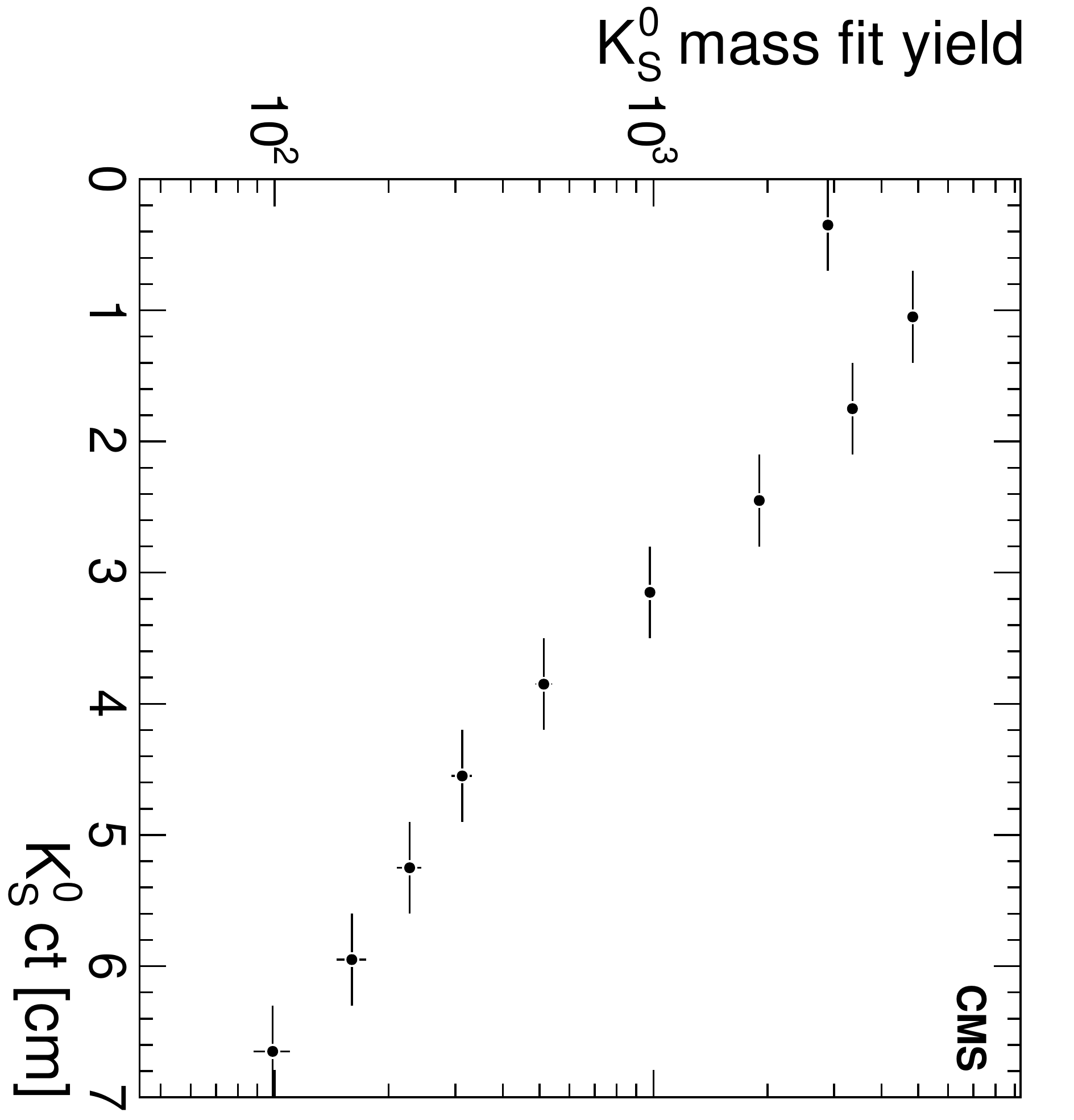}
   \label{fig:kshort_lifetime_a} 
      }}
    }
    \mbox{
      \subfigure[]
{\scalebox{0.42}{
	  \includegraphics[angle=90,width=\linewidth]{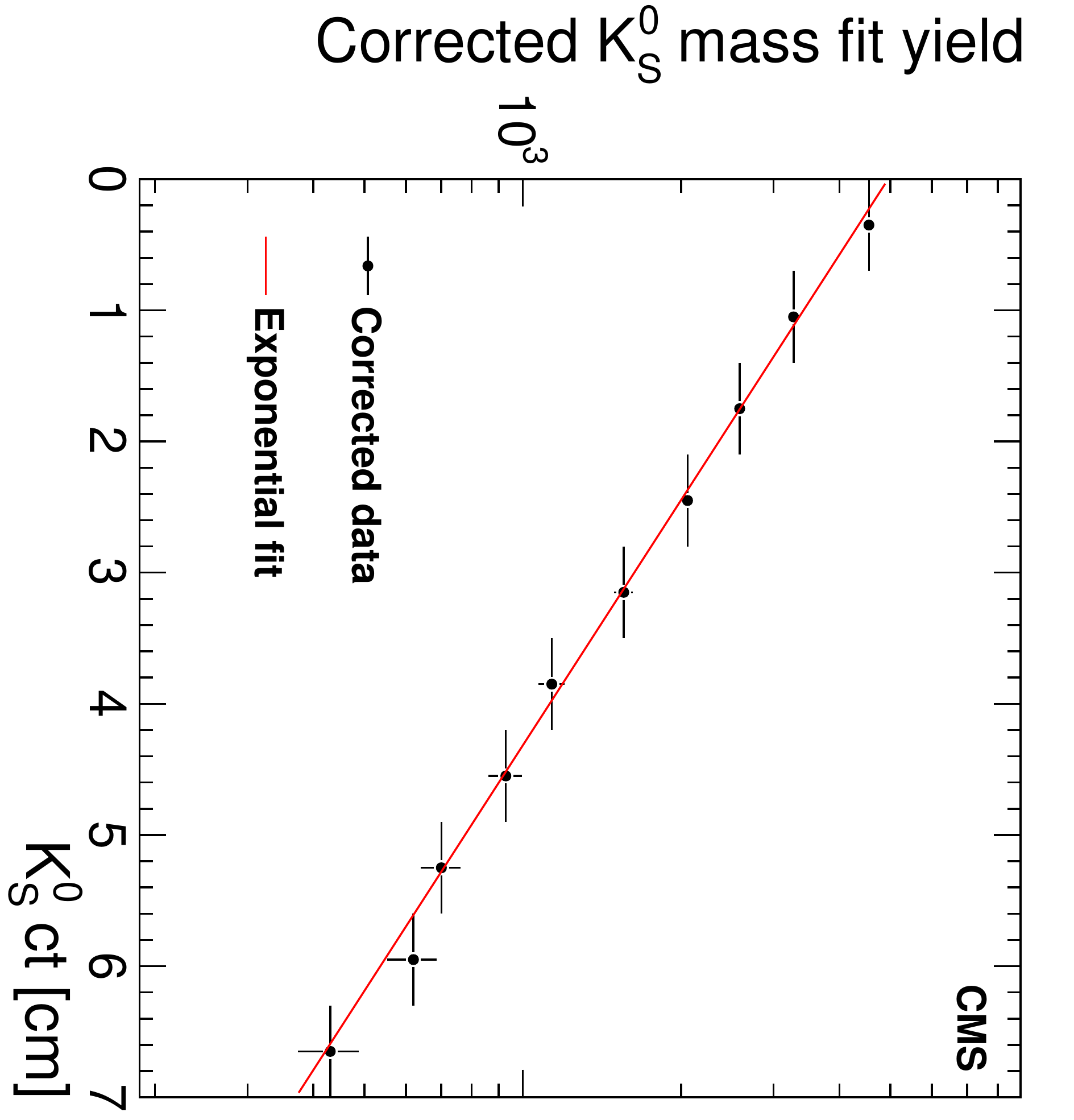}
   \label{fig:kshort_lifetime_b} 
      }}
    }
\caption{K$_\mathrm{S}^0$ $ct$ distributions for (a) uncorrected data and (b)
corrected data with an exponential fit.}
\end{center}
\end{figure}

For the $0.9\TeV$ centre-of-mass energy data and simulation, invariant mass distributions 
are made for different bins of proper decay length, $ct = mcL/p$, where $L$ is the measured
decay length.
These distributions are fitted to obtain the yield, leading to the uncorrected
$ct$ distribution as seen in Fig.~\ref{fig:kshort_lifetime_a} for the K$_\mathrm{S}^0$ data.  
The uncorrected $ct$ distribution from the simulation is divided by the generated
exponential shape given by $e^{-ct/c\tau_{Sim}}$ to obtain the correction factor versus $ct$.
The uncorrected data $ct$ distribution is divided by the correction factor to obtain the 
corrected $ct$ distribution as seen in Fig.~\ref{fig:kshort_lifetime_b} for the K$_\mathrm{S}^0$.  
This distribution is fitted with an exponential, the slope of which gives the measured lifetime.
The good fit to an exponential function ($\chi^2/\text{NDOF} = 8.1/8$) indicates that the simulation 
accurately reproduces the efficiency variation versus lifetime.
The fitted results, $\tau_{\mathrm{K}_\mathrm{S}^0} = 90.0 \pm 2.1$\,ps and $\tau_{\Lambda^0} = 271 \pm 20$\,ps 
(with $\chi^2/\text{NDOF} = 11.3/6$), are both within 1\,$\sigma$ of the world average~\cite{PDG08}.


\subsubsection{Reconstruction of K$^*(892)^-$ and $\Xi^-$}

The reconstructed sample of V$^0$ particles was exploited to reconstruct decays of other particles.

\begin{figure}[hbtp]
\begin{center}
    \mbox{
      \subfigure[]
{\scalebox{0.42}{
	  \includegraphics[width=\linewidth]{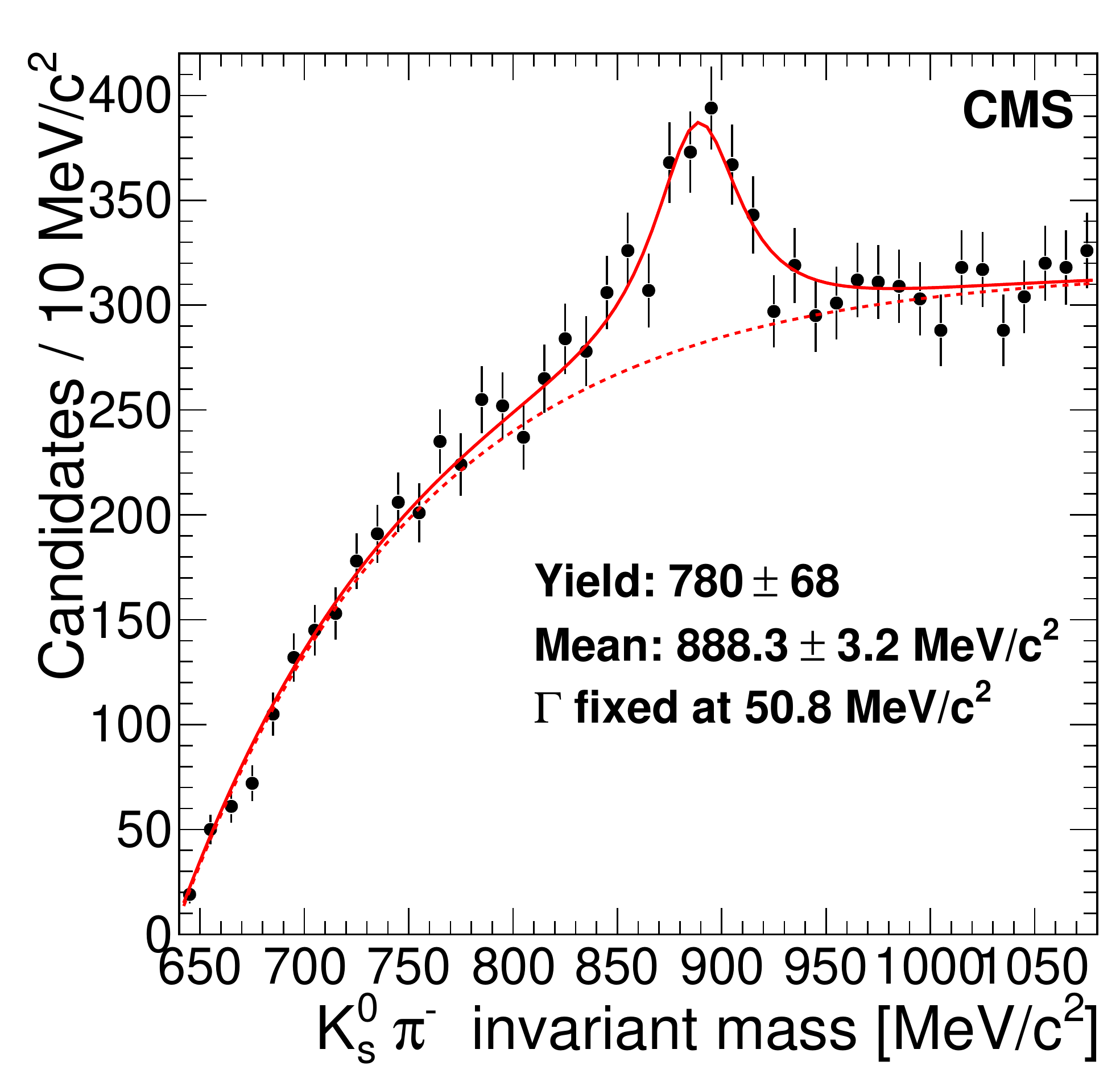}
   \label{fig:kstar_mass} 
      }}
    }
    \mbox{
      \subfigure[]
{\scalebox{0.425}{
	  \includegraphics[width=\linewidth]{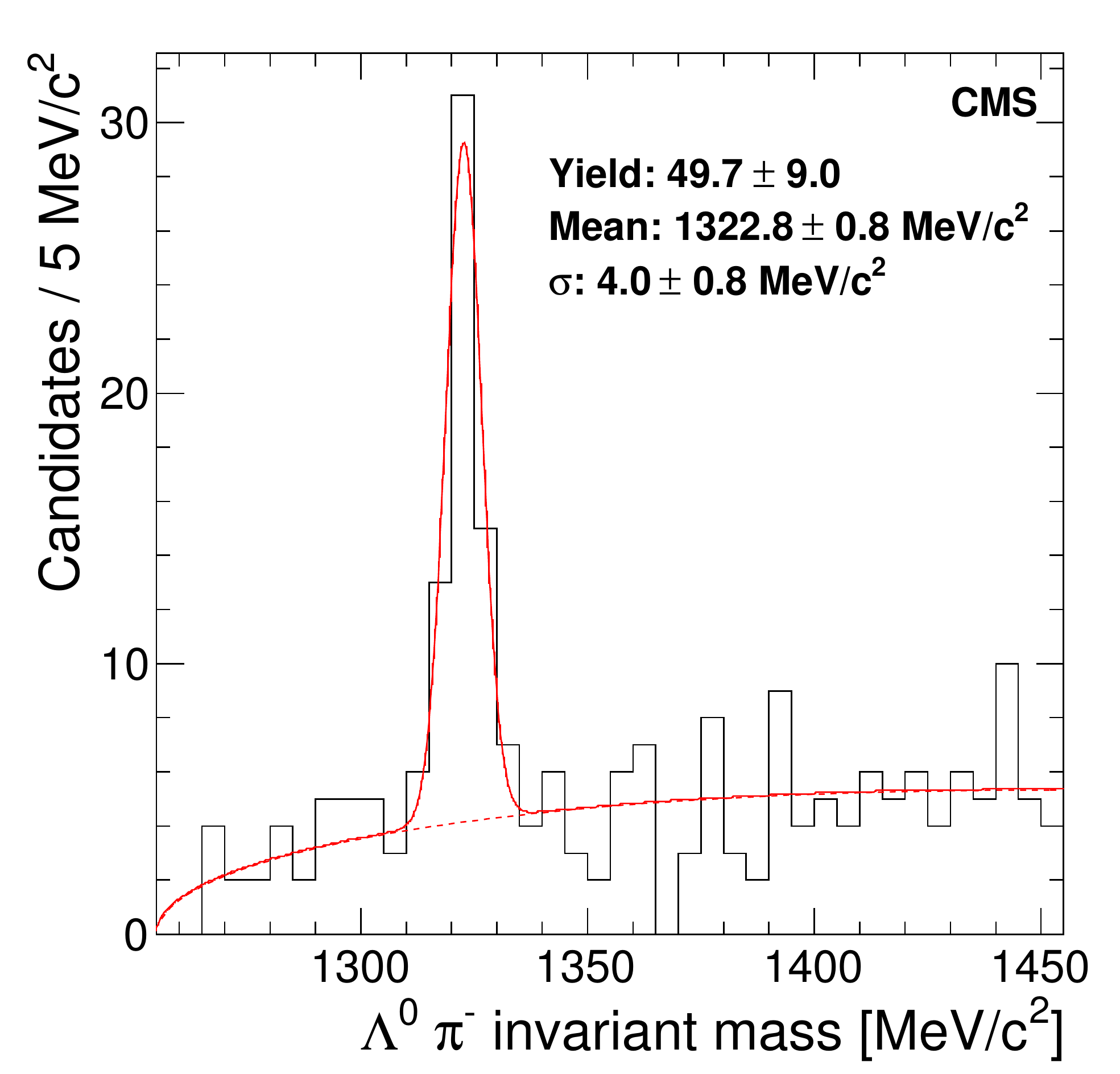}
   \label{fig:xi_mass} 
      }}
    }
\caption{Invariant mass plots of (a) K$_\mathrm{S}^0 \pi^-$ with a fit to the K$^*(892)^-$ and 
(b) $\Lambda^0\pi^-$ with a fit to the $\Xi^-$.}
\end{center}
\end{figure}

The K$_\mathrm{S}^0$ candidates are combined with charged tracks from the primary vertex to
search for the strong decay K$^*(892)^- \to K_S^0\pi^-$.  For this analysis, events
were required to contain a reconstructed primary vertex consisting of more than two tracks
and a fit probability greater than 0.5\%.  The K$_\mathrm{S}^0$ candidate must pass the same
criteria as described in Sec.~\ref{sec:V0reco}.  In addition, the requirement on the
impact parameter significance of the pions from the K$_\mathrm{S}^0$ is increased from 0.5 to 2.
The K$_\mathrm{S}^0$ candidates must also have a mass within 20\MeVcc of the nominal mass
and the K$_\mathrm{S}^0$ flight path must pass within 2\mm of the primary vertex.  The charged
track in the K$^*(892)^-$ decay must have a normalized $\chi^2$ less than 2, at least two
hits in the pixel detector, at least seven total hits, $\pt > 0.5\GeVc$, $|\eta|<2$,
and pass within 2 (3)~mm of the primary vertex in the direction transverse to (along) the
beam line.  The K$_\mathrm{S}^0\pi^-$ invariant mass is calculated using the world-average value of the
K$_\mathrm{S}^0$ mass~\cite{PDG08} and is shown in Fig.~\ref{fig:kstar_mass}.  The figure also shows an overlay
of a fit to the K$_\mathrm{S}^0\pi^-$ mass distribution.  The fit uses a Breit-Wigner
for the signal plus a threshold function for the background
\begin{equation*}
\frac{S}{\left(m^2-M_{K^*}^2\right)^2+\Gamma_{K^*}^2 M_{K^*}^2} + 
B\left[1-\exp{\left(\frac{M_K+M_\pi-m}{p}\right)}\right],
\end{equation*}
where $m$ is the K$_\mathrm{S}^0\pi^-$ invariant mass, $M_{K^*}$ and $\Gamma_{K^*}$ are the mass
and width of the K$^*(892)^-$, $M_K$ and $M_\pi$ are the world-average masses of $K^0$ and $\pi^-$,
and $S$, $B$, and $p$ are free parameters.  The K$^*$ width $(\Gamma_{K^*})$ is fixed at
the world average value of 50.8\MeVcc~\cite{PDG08}, while the K$^*$ mass $(M_{K^*})$ is a free parameter.  The
mass returned by the fit, $888.3 \pm 3.2\MeVcc$, is consistent with the world average value of $891.66 \pm
0.26\MeVcc$~\cite{PDG08}.

The $\Xi^-$ was reconstructed through its decay to
$\Lambda^0\pi^-$.  The $\Xi^-$ is a long-lived baryon, with a decay topology different from that 
of the K$^*(892)^-$: the $\pi^-$ from the $\Xi^-$ decay should be detached
from the primary vertex rather than originating from it. The $\Lambda^0$
candidates were reconstructed as described in Sec.~\ref{sec:V0reco} except that a looser
transverse significance cut of 10 (rather than 15) was applied.  $\Lambda^0$ candidates
with a mass within 8\MeVcc of the world-average value were combined with charged tracks with the
same sign as the pion in the $\Lambda^0$ decay.  The $\Lambda^0\pi^-$ fit used a
$\Lambda^0$ mass constraint and the vertex was required to have a fit probability better
than 1\%.  All three tracks involved in the decay were required to have at least 6 valid
hits and a 3D impact parameter with respect to the primary vertex greater than 3$\sigma$.  The
resulting mass plot, shown in Fig.~\ref{fig:xi_mass}, is fit with a single Gaussian
for the signal and a background shape of $Aq^{(1/2)}+Bq^{(3/2)}$ where $q = m-M_\Lambda -
M_\pi$, $m$ is the $\Lambda^0 \pi^-$ invariant mass, and $A$ and $B$ are free parameters.  
The measured mass of $1322.8 \pm 0.8\MeVcc$ is close to the world
average value of $1321.71 \pm 0.07\MeVcc$~\cite{PDG08}.  The resolution of $4.0 \pm 0.8\MeVcc$ 
is consistent with the simulation result of $3.6 \pm 0.4\MeVcc$.

\subsection{Particle Identification Using Measured Energy Losses}
\label{sec:dedx}

Estimating the energy loss $(dE/dx)$ of a particle by means of charge collected by the CMS silicon strip tracker
is described in Sec.~\ref{sec:strips_dedx}.  In this section, applications of $dE/dx$ measurements are used to identify protons
and kaons produced in $\Lambda^0$ and $\phi$ decays.

\subsubsection{$dE/dx$ Verification with $\Lambda \to p\pi^-$ Decays}
\begin{figure}[hbtp]
\begin{center}
    \mbox{
      \subfigure[]
{\scalebox{0.42}{
	  \includegraphics[width=\linewidth]{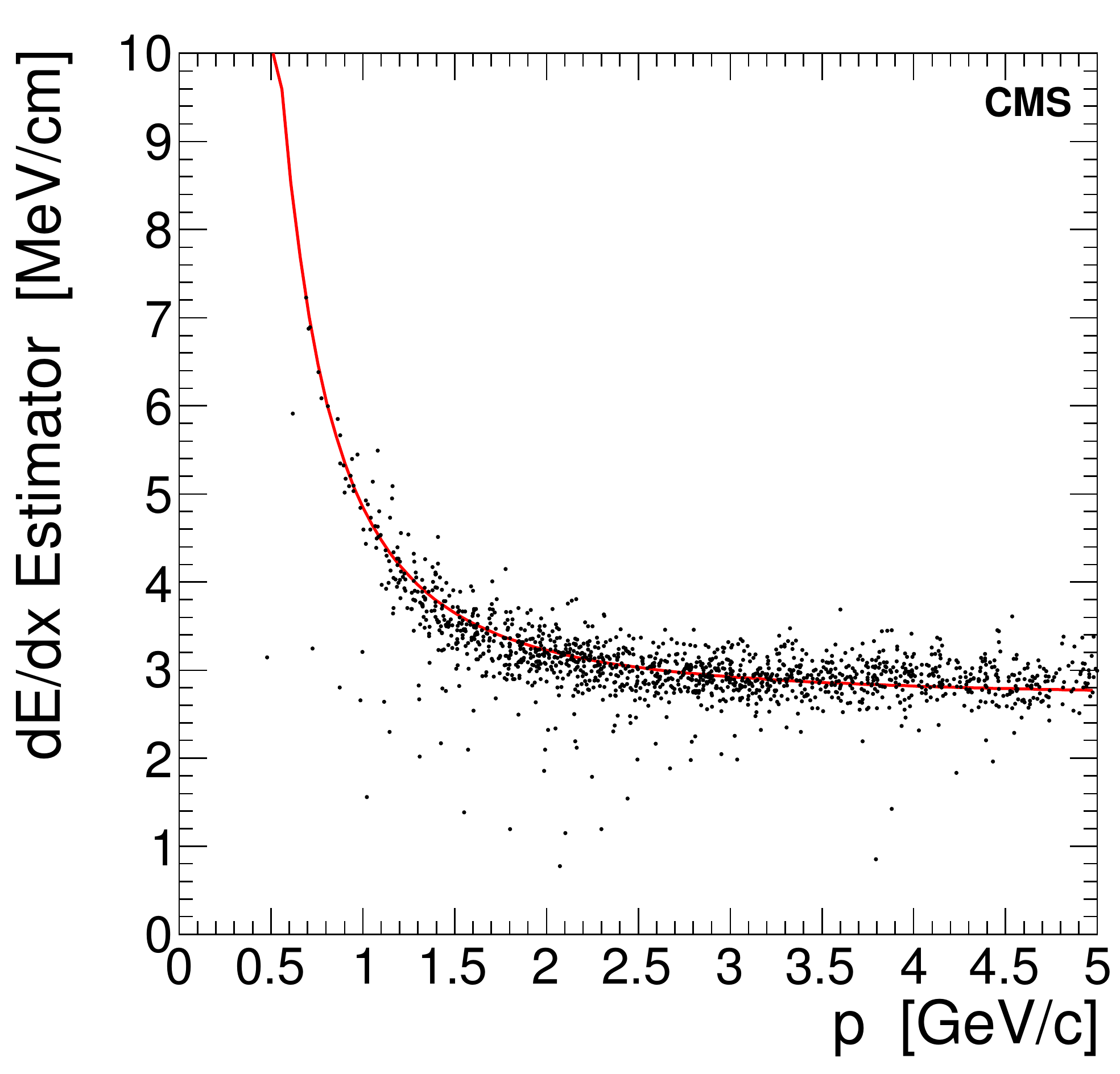}
   \label{fig:lambda_estimator_hard} 
      }}
    }
    \mbox{
      \subfigure[]
{\scalebox{0.42}{
	  \includegraphics[width=\linewidth]{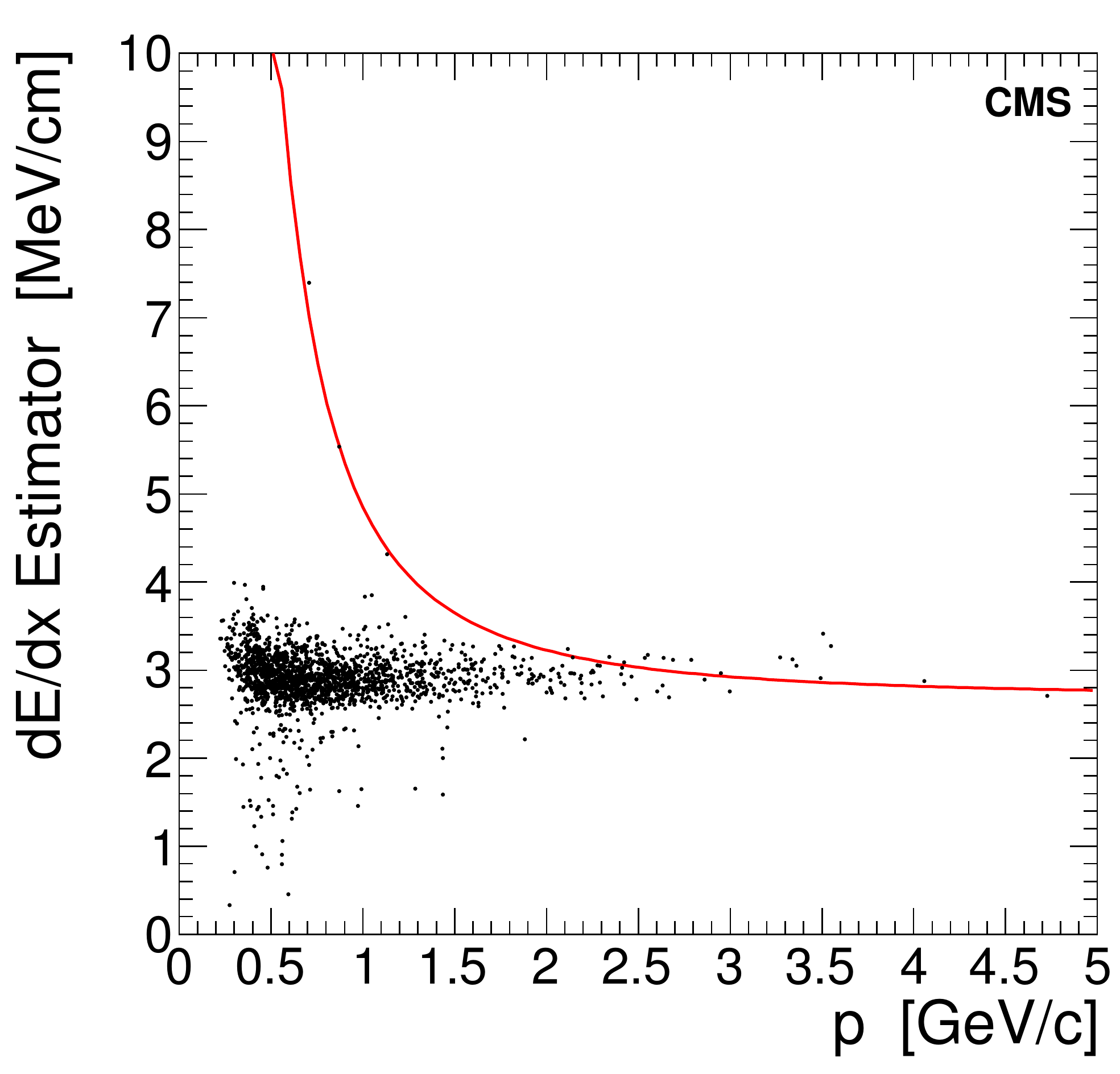}
   \label{fig:lambda_estimator_soft} 
      }}
    }
\caption{Estimated energy loss as a function of the momentum for (a) the highest momentum track
and (b) the lowest momentum track for the $\Lambda^0$ candidate decay products.
The superimposed curves comes from the proton fit in the inclusive track sample shown in Fig.~\ref{fig:dedx_p}.}
\end{center}
\end{figure}

The kinematics of the $\Lambda^0 \to p\pi^-$ decay requires $p_p > p_\pi$ for all
$\Lambda^0$ particles reconstructed at CMS\@.  This provides a clean source of protons and
pions which can be used to check the $dE/dx$ results.  We apply the same selection as in
Section~\ref{sec:V0reco}, and plot the $dE/dx$ distribution as a function of the momentum for
tracks associated to V$^0$ candidates in the mass range 1.08--1.16\GeVcc, separately for
the highest momentum tracks (Fig.~\ref{fig:lambda_estimator_hard}) and the lowest momentum tracks (Fig.~\ref{fig:lambda_estimator_soft}).
As expected, the highest momentum tracks are generally found near the proton curve while the lowest momentum
tracks are generally inconsistent with the proton curve.  The few exceptions are consistent with background under 
the $\Lambda^0$ peak.

\subsubsection{Reconstruction of $\phi(1020) \to \mathrm{K}^+\mathrm{K}^-$}
The $\phi(1020) \to \mathrm{K}^+\mathrm{K}^-$ decay was reconstructed using data taken at 0.9\TeV centre-of-mass energy.  
The candidate kaon tracks come from the collection of \textit{highPurity} tracks and are
required to have $\pt > 0.5\GeVc$, normalized $\chi^2 < 2$, at least five hits, $|\eta|<2$, 
and a transverse impact parameter with respect to the reconstructed beamspot smaller than 3\mm.
Finally, for tracks with $p<1\GeVc$, the track must have a measured $dE/dx$ consistent with the
kaon hypothesis (see Eq.~\ref{eq:bethebloch}): $K (M^\textrm{min}/p)^2+C < dE/dx < K (M^\textrm{max}/p)^2+C$.  
The parameters of the $dE/dx$ cut for kaons are those extracted from a fit to the $dE/dx$ vs.\ $p$
distribution, as described in Sec.~\ref{sec:strips_dedx}.  We use a compatibility window of $\pm200\MeVcc$
around the K mass, with $M^\textrm{min}$ and $M^\textrm{max}$ being lower and upper boundaries of this window.

\begin{figure}[hbtp]
\begin{center}
    \mbox{
      \subfigure[]
{\scalebox{0.47}{
	  \includegraphics[width=\linewidth]{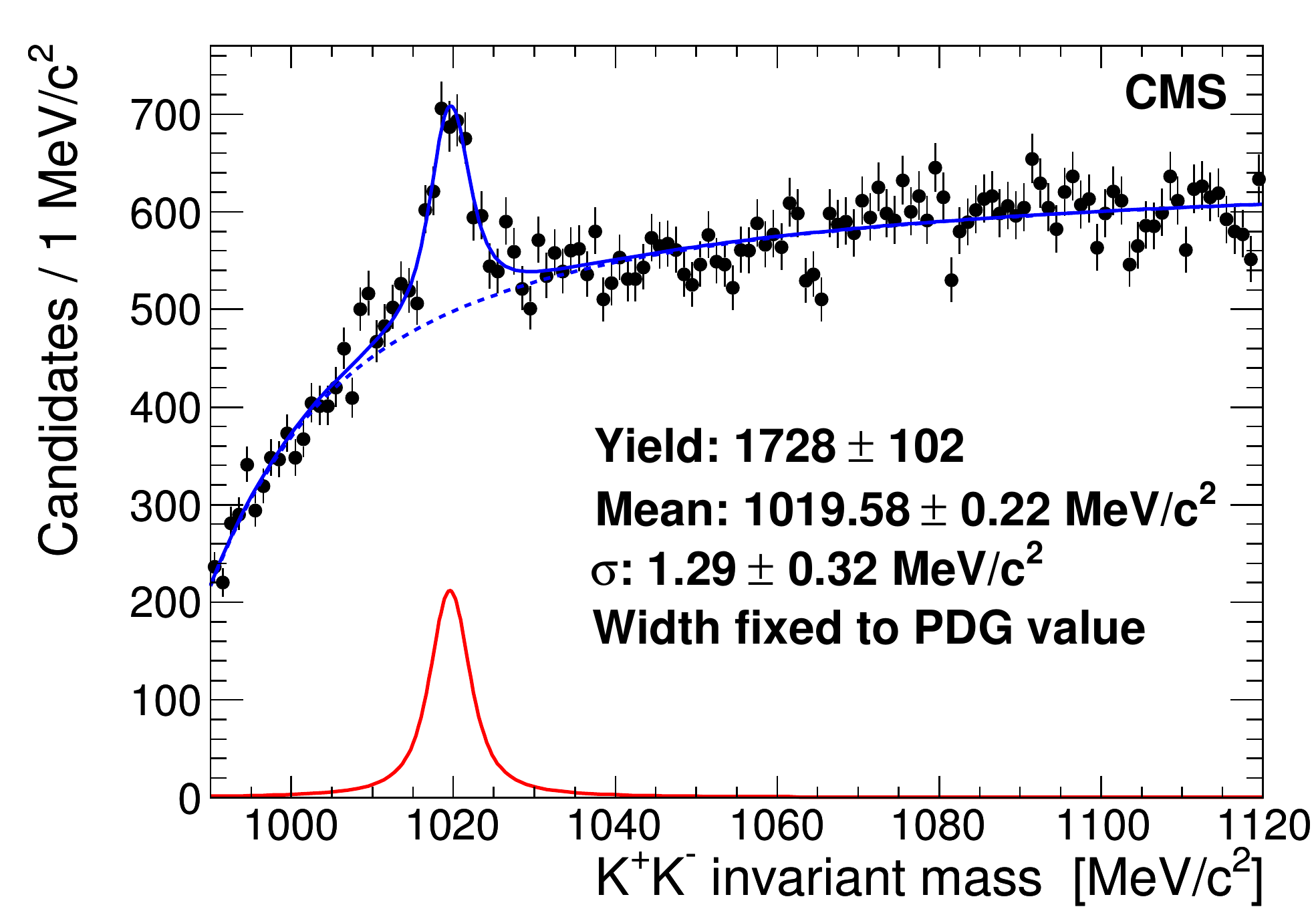}
   \label{fig:phi_mass_signal} 
      }}
    }
    \mbox{
      \subfigure[]
{\scalebox{0.47}{
	  \includegraphics[width=\linewidth]{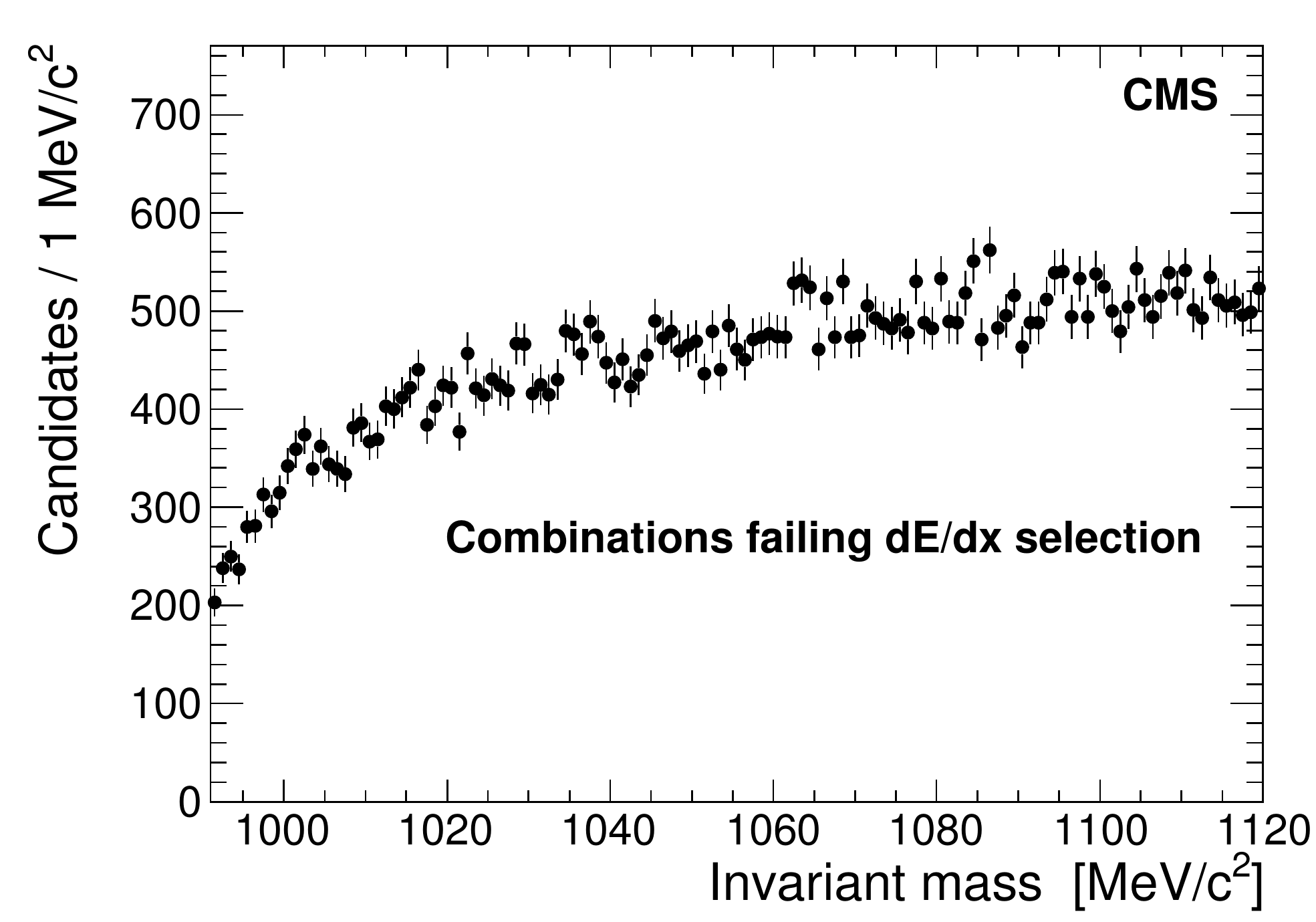}
   \label{fig:phi_mass_background} 
      }}
    }
\caption{$\mathrm{K}^+\mathrm{K}^-$ invariant mass distribution, with (a) both kaons satisfying the $dE/dx$ requirement and with (b) at least 
one particle failing that requirement.  In (a) a fit to the $\phi(1020)$ hypothesis is shown.}
\end{center}
\end{figure}

The fit of the mass spectra of pairs of tracks accepted by the $dE/dx$ selection used the sum of two
normalized functions: a convolution of a relativistic Breit-Wigner shape with a Gaussian for the 
$\phi$ signal and an arctangent function for the background.  The mass plot and overlaid fit are 
shown in Fig.~\ref{fig:phi_mass_signal}.  The fitted $\phi$ mass of $1019.58 \pm 0.22\MeVcc$ is in agreement
with the world-average value of $1019.455 \pm 0.020\MeVcc$.  The resolution found in data is $1.29 \pm 0.32\MeVcc$,
in agreement with the value found in simulation, $1.41\MeVcc$.  Candidates in which at least
one track fails the $dE/dx$ requirement are shown in Fig.~\ref{fig:phi_mass_background} where only background
is observed, indicating that the $dE/dx$ requirement has a high efficiency to select $\phi(1020)$ candidates.

\subsection{Reconstruction of Photon Conversions and Nuclear Interactions}
\label{sec:interactions}

While the tracker is essential for finding charged particles and measuring their momenta,
the tracker material is also a source for interactions.  For photons, interactions
with the tracker material can produce $e^+e^-$ conversion pairs, while for hadrons,
nuclear interactions can produce multiple hadrons.  Photon conversions in the Tracker
reduce the efficiency for low-energy-photon finding by the electromagnetic calorimeter,
while nuclear interactions reduce track finding efficiency and can affect the resolution
of many hadronic observables such as jets or missing transverse energy.  Thus,
identification of conversions and nuclear interactions can be used to improve many aspects of the
event reconstruction.  Furthermore, studies of conversions and interactions can be used to
improve our understanding of the material in the Tracker.

\begin{figure}[hbtp]
\begin{center}
    \mbox{
      \subfigure[]
{\scalebox{0.313}{
	  \includegraphics[angle=90,width=\linewidth]{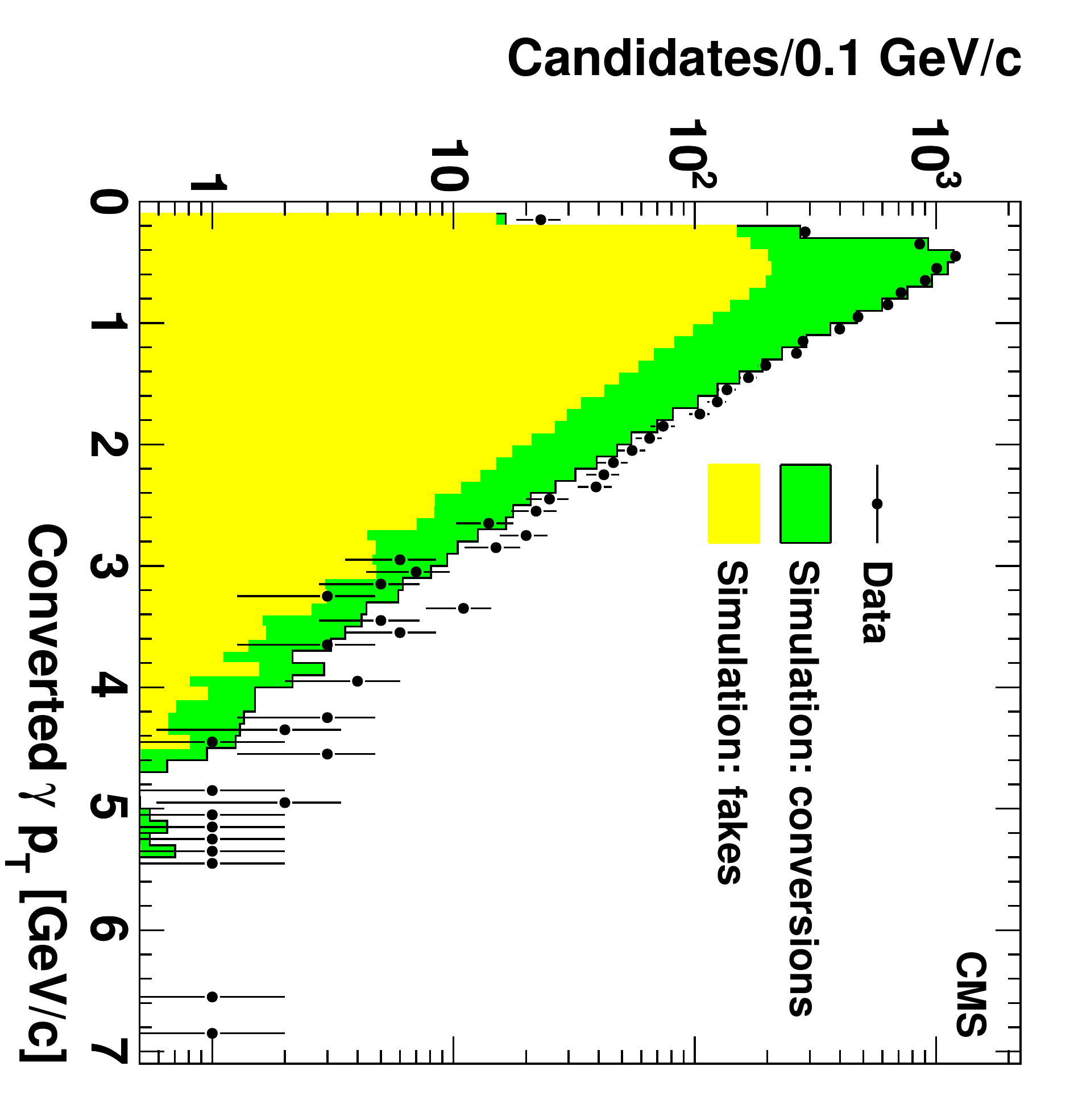}
   \label{fig:conversion_pt} 
      }}
    }
    \mbox{
      \subfigure[]
{\scalebox{0.313}{
	  \includegraphics[angle=90,width=\linewidth]{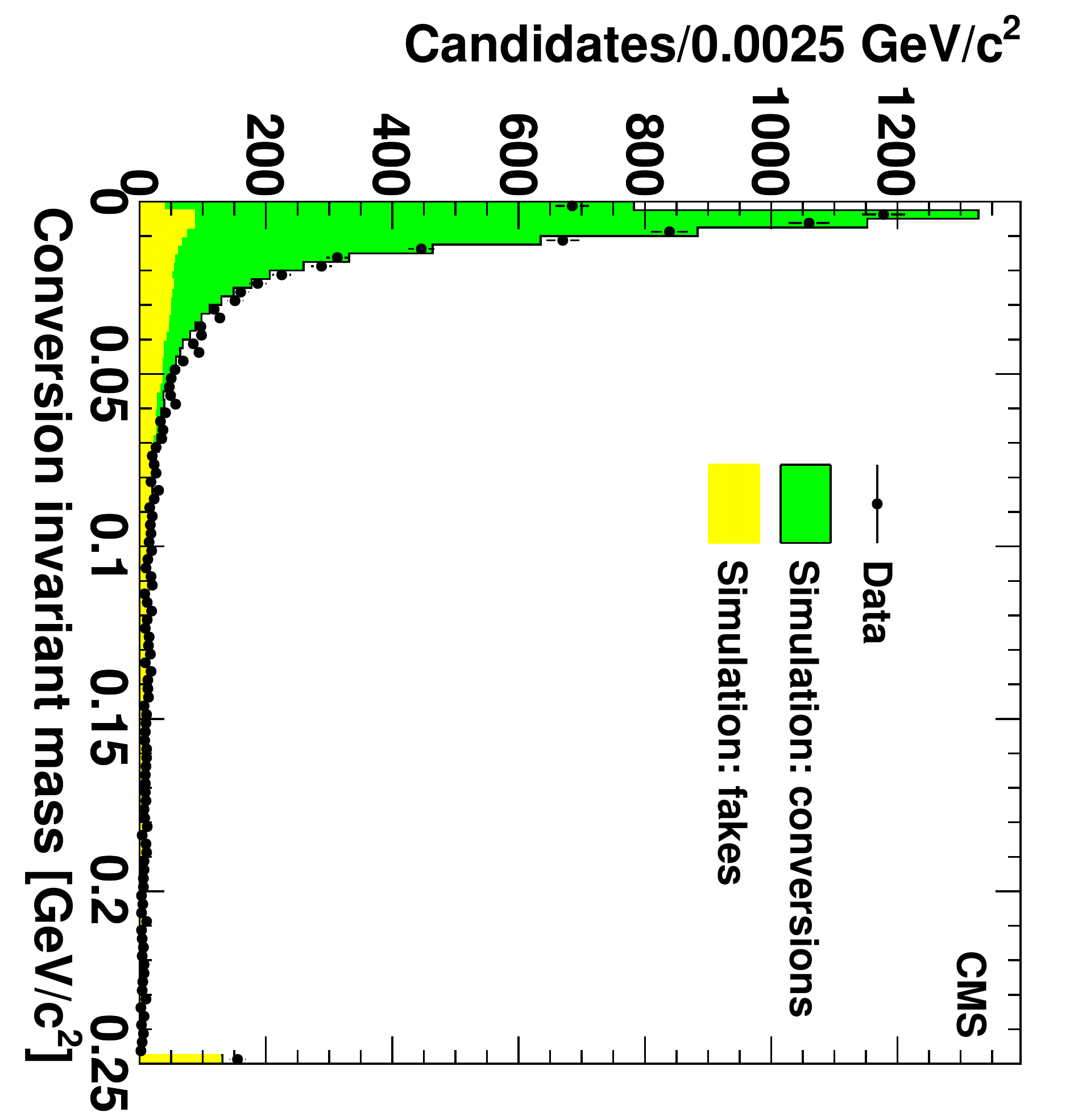}
   \label{fig:conversion_mass} 
      }}
    }
    \mbox{
      \subfigure[]
{\scalebox{0.313}{
	  \includegraphics[width=\linewidth]{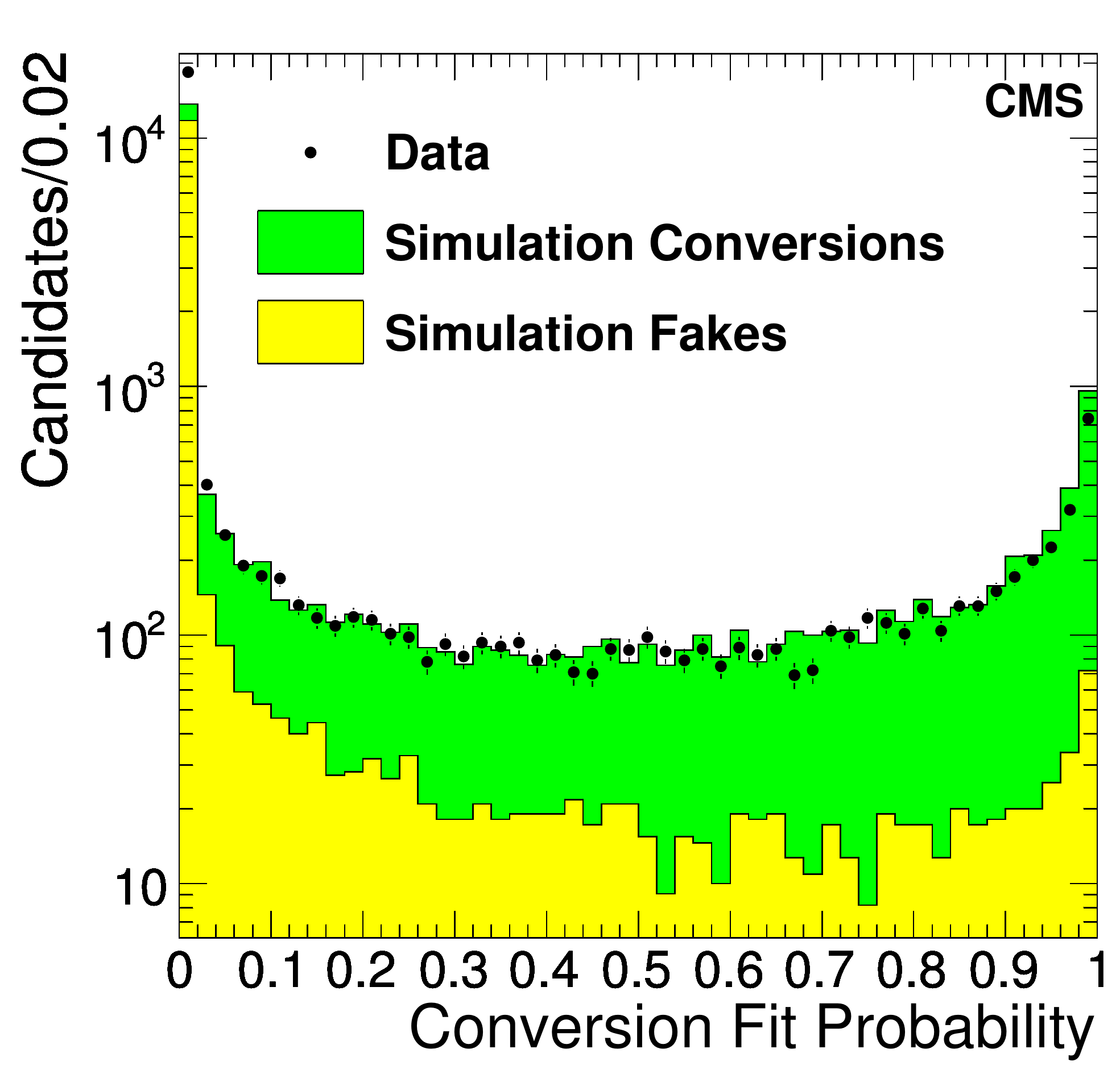}
   \label{fig:conversion_prob} 
      }}
    }
\caption{Comparisons of data photon conversions (points) and real and fake photon conversion from simulation (filled histograms) showing:
(a) distributions of the reconstructed \pt of the converted photons from the first method,
(b) the invariant mass of the $e^+e^-$ pairs from the first method, and (c)
the distribution of the vertex $\chi^2$ probability from the second method.
The last bin of (b) is the overflow bin.}
\end{center}
\end{figure}

The electrons and positrons from converted photons can be identified by the
electromagnetic calorimeter and used as seeds for track
reconstruction~\cite{CMS_NOTE_2006_005}.  In the minimum bias events collected in
December 2009, however, the photons have a soft spectrum as seen in
Fig.~\ref{fig:conversion_pt} and therefore the conversion pairs are unlikely to reach the
electromagnetic calorimeter.  These conversion pairs can still be reconstructed by using
tracker-seeded conversion reconstruction techniques, made possible by the iterative tracking 
algorithm described in Section~\ref{sec:reconstruction} which extends the capability of 
reconstructing low-\pt and detached tracks.  The essential signature of a
massless conversion photon is the two parallel tracks at the production vertex, in both 
the transverse and longitudinal planes.
The reconstructed invariant mass, shown in Fig.~\ref{fig:conversion_mass},
shows the effect of the mass resolution, which is well modelled by the simulation.
Two different conversion reconstruction approaches have been used.  Both methods fit two
oppositely charged tracks to a common 3D vertex with the constraint that the two tracks
are parallel at the vertex.  The methods differ mainly in the preselection of the track
pairs.  The first method, from which Figs.~\ref{fig:conversion_pt} and \ref{fig:conversion_mass} are derived,
requires both tracks have at least 3 hits and normalized $\chi^2$ less than 10 and at least one
track with 5 or more hits.  The tracks are required to have positive charge-signed transverse
impact parameter, positive distance of minimum approach in 2D (i.e., the two full track
circles have one or no intersection in the transverse plane), small $z$ separation at
their innermost point ($|\Delta z|<5$\cm) if they are in the barrel, and a small opening
angle in both the transverse ($\Delta \phi <0.2$) and longitudinal plane
($\Delta\cot\theta<0.1$ where $\theta$ is the polar angle relative to the $z$ axis).  The vertex fit must have a $\chi^2$
probability better than $5\times 10^{-3}$ and be located inside the innermost hits
on the tracks.  To increase efficiency, the second method takes \textit{all} tracks with a 
$\chi^2$ probability above $10^{-6}$ and requires a vertex with fit probability greater
than $10^{-6}$, radius greater than 2\cm, and at most one hit per track inside of the
vertex position.  The $\chi^2$ probability from the second
method is shown in Fig.~\ref{fig:conversion_prob} with good agreement between data and 
simulation.  

The nuclear interaction finder starts from the full list of tracks described in
Section~\ref{sec:reconstruction}.  For each pair of tracks, the distance of closest approach is
computed and if the two tracks are close enough they are considered linked together.  A
recursive finder produces blocks of tracks linked together from which a rough estimate of
the displaced vertex position is computed.  Finally, the tracks from a block are refitted
together with a displaced vertex as a common constraint.  $V^{0}$ decays and photon conversions
are removed from the resulting sample of displaced vertices.  A tight selection is applied
to the remaining vertices to remove fake tracks and pairs from the
primary vertex. The resulting sample of significantly displaced vertices in the radial
direction ($r>2.5$\cm) is called the nuclear interactions sample.
In the data, $80\%$ of nuclear interactions are reconstructed with 2 tracks and $20\%$ with 3
tracks. In the first case, a $30\%$ combinatorial fake rate is expected from the simulation, 
while in the second case the fake rate is negligible.

\begin{figure}[hbtp]
\begin{center}
    \mbox{
      \subfigure[]
{\scalebox{0.42}{
	  \includegraphics[width=\linewidth]{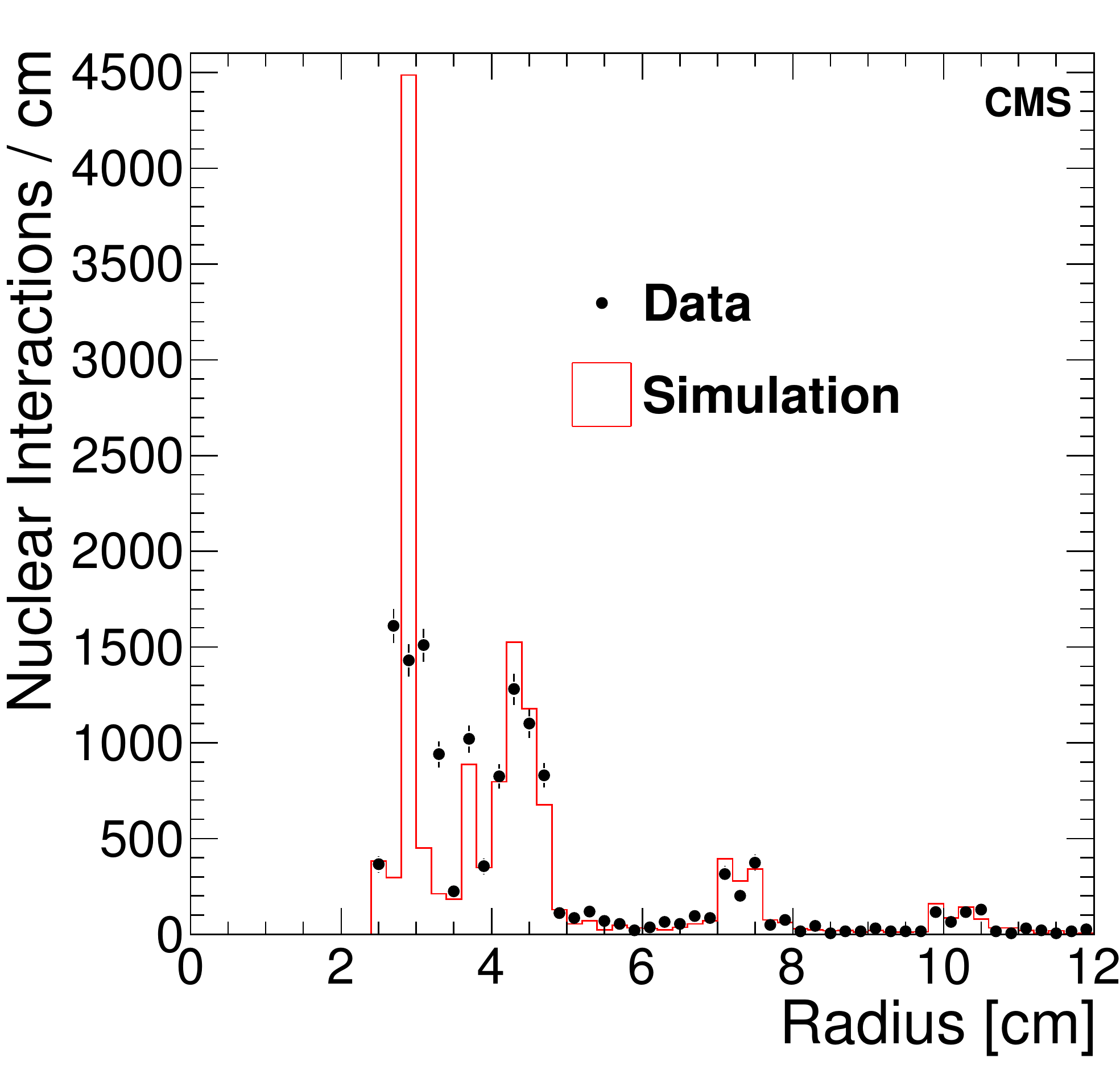}
   \label{fig:interactions_r} 
      }}
    }
    \mbox{
      \subfigure[]
{\scalebox{0.42}{
	  \includegraphics[width=\linewidth]{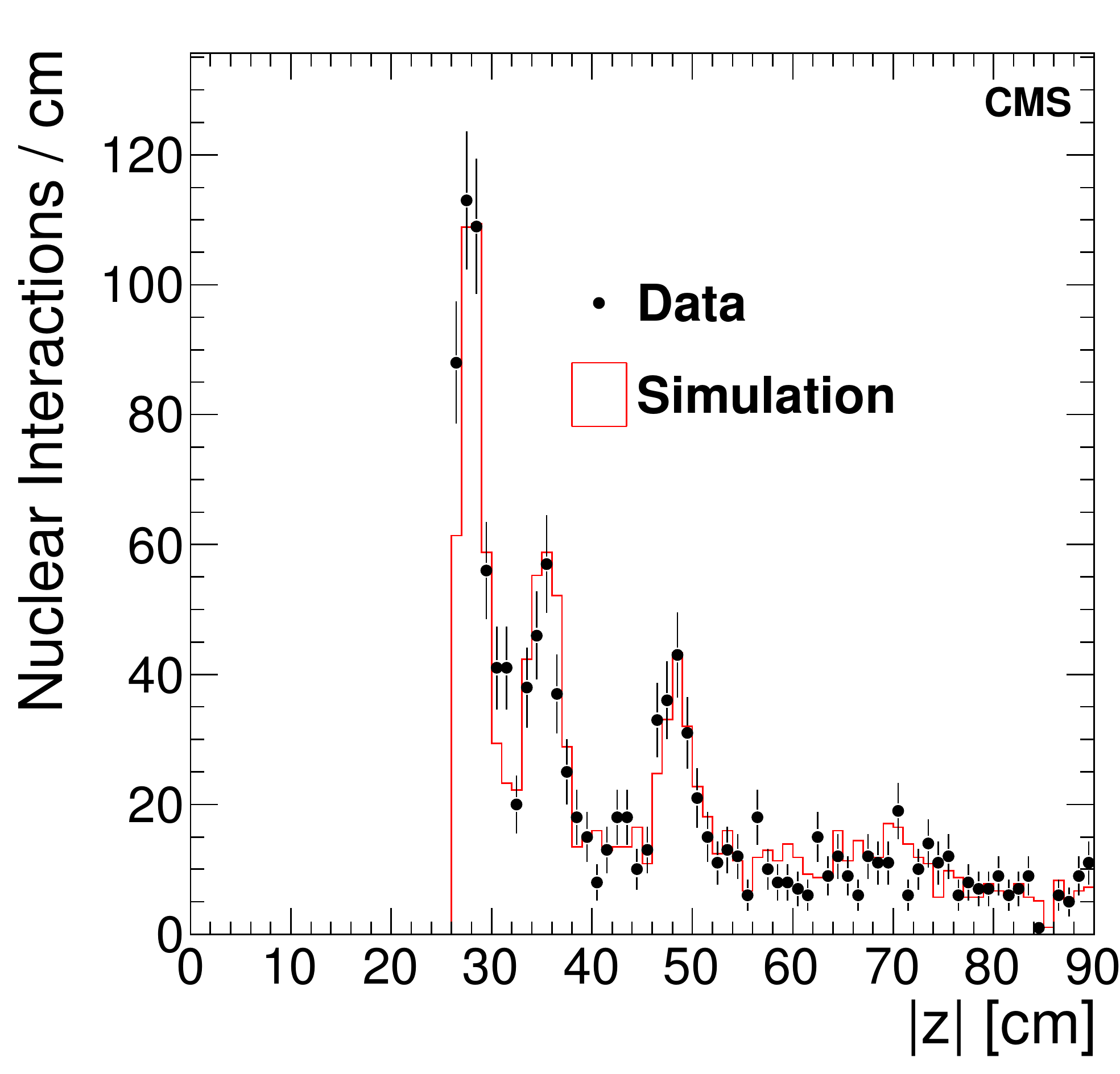}
   \label{fig:interactions_z} 
      }}
    }
\caption{Distributions of nuclear interaction vertices versus (a) radial position $r$
for $|z|<26$\cm and (b) versus the magnitude of the longitudinal coordinate $|z|$ for
$|z|>26$\cm and $r<19$\cm. The simulation histogram is normalized to the
total number of nuclear interactions found in data in the full $z$ range.}
\end{center}
\end{figure}

The distribution of nuclear interaction positions provides a means of observing the material
in the detector and validating the simulation of the material.  The distribution of
radial position $r$ of the nuclear vertices, compared to the simulation, is shown in
Fig.~\ref{fig:interactions_r}.  The beam pipe at a radius of 3\cm, as well as the three barrel pixel layers
at average radii of 4.3, 7.3, and 10.2~cm, are clearly seen.
The radius is measured relative to the centre of the pixel detector.  In the version of the
simulation used here, this is also the centre of the beam pipe.  In reality, the 
beam pipe centre is offset from the pixel detector centre resulting in a smeared distribution
versus radius.
Nevertheless, there is good agreement between the data
and the simulation for the relative rate of nuclear interactions in the different barrel pixel
structures and the beam pipe.  This indicates a consistent description of the material
distribution in this region.  
The material distribution in the endcap pixel detector is studied by selecting nuclear
interactions with $|z|>26$\cm and $r < 19$\cm. The longitudinal position $|z|$ of the
nuclear vertices, compared to the simulation, is shown in
Fig.~\ref{fig:interactions_z}.  The pixel barrel flange
($|z|<30$\cm) and the two pixel disks can be clearly distinguished. The tail up to 1\,m is from pixel services.

%

\subsection{Study of $b$-tag Related Observables}
\label{sec:btag}

The measurement of impact parameters and the reconstruction of secondary vertices have been tested
with the limited event sample of December 2009.  At higher collision energy these objects will
provide the main observables used in b-tagging algorithms.

The 2009 data contain only a few well-defined jets and mainly tracks at momenta below
those typically used in b-tagging. To test the reconstruction on a sufficiently large
sample, a few changes to the reconstruction chain have been applied with respect to what is
described in Ref.~\cite{CMS_PAS_BTV_09_001}.  As described in
Ref.~\cite{CMS_PAS_JME_10_001}, jet reconstruction is performed using the anti-\kt jet
clustering algorithm~\cite{Cacciari:2008gp} on objects obtained from the CMS particle flow
reconstruction~\cite{CMS_PAS_PFT_09_001,CMS_PAS_PFT_10_001}.  To recover low-momentum
jets, the cone size is increased to 0.7 and the minimum $\pt$ is reduced to $3\GeVc$.
The b-tagging algorithms are run on tracks associated with these jets.  
The track selection is also changed relative to
Ref.~\cite{CMS_PAS_BTV_09_001}; the minimum $\pt$ requirement is removed and 7 rather than
8 hits are required.

\begin{figure}[hbtp]
\begin{center}
    \mbox{
      \subfigure[]
{\scalebox{0.47}{
	  \includegraphics[width=\linewidth]{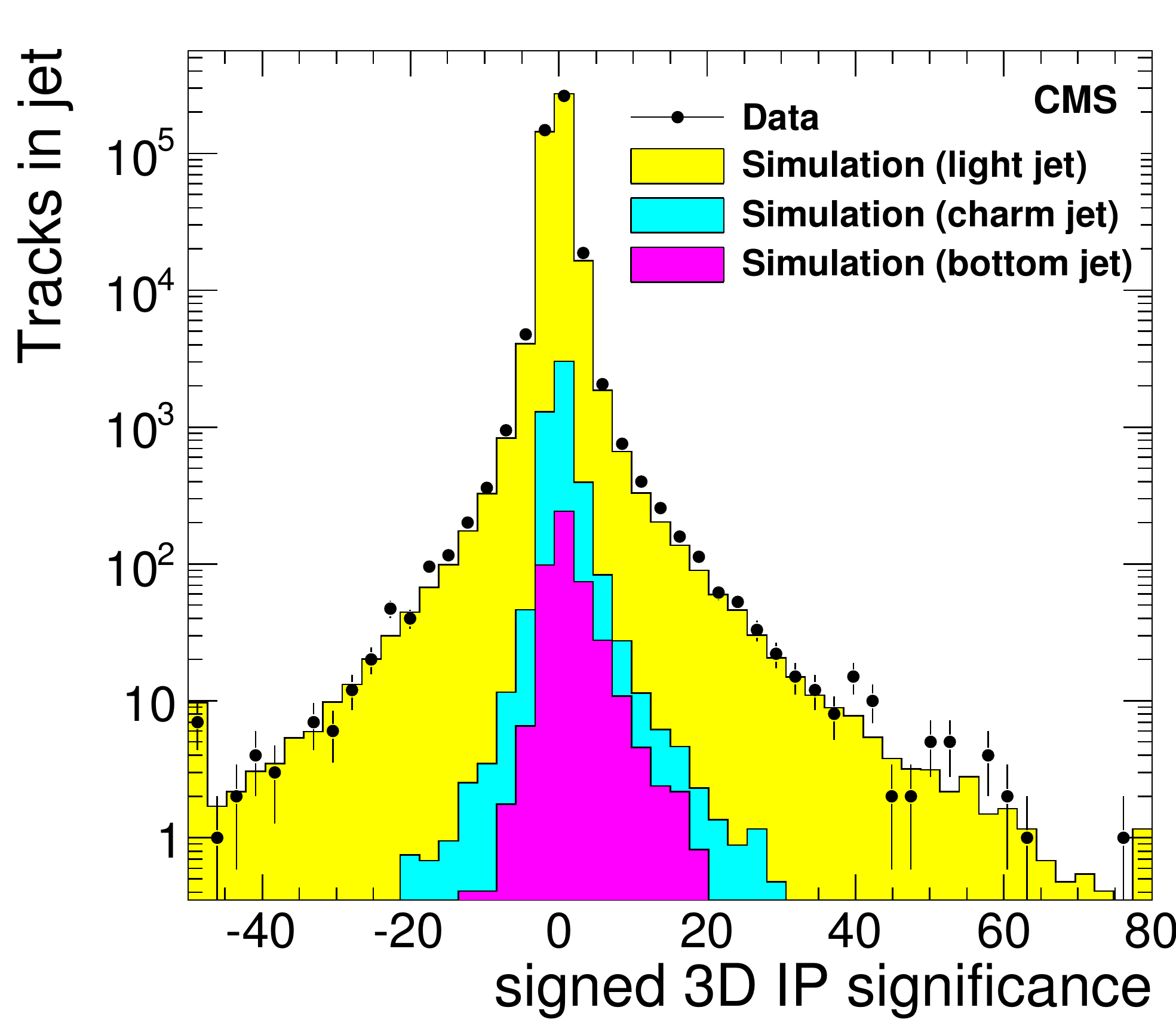}
   \label{fig:btag_ip} 
      }}
    }
    \mbox{
      \subfigure[]
{\scalebox{0.47}{
	  \includegraphics[width=\linewidth]{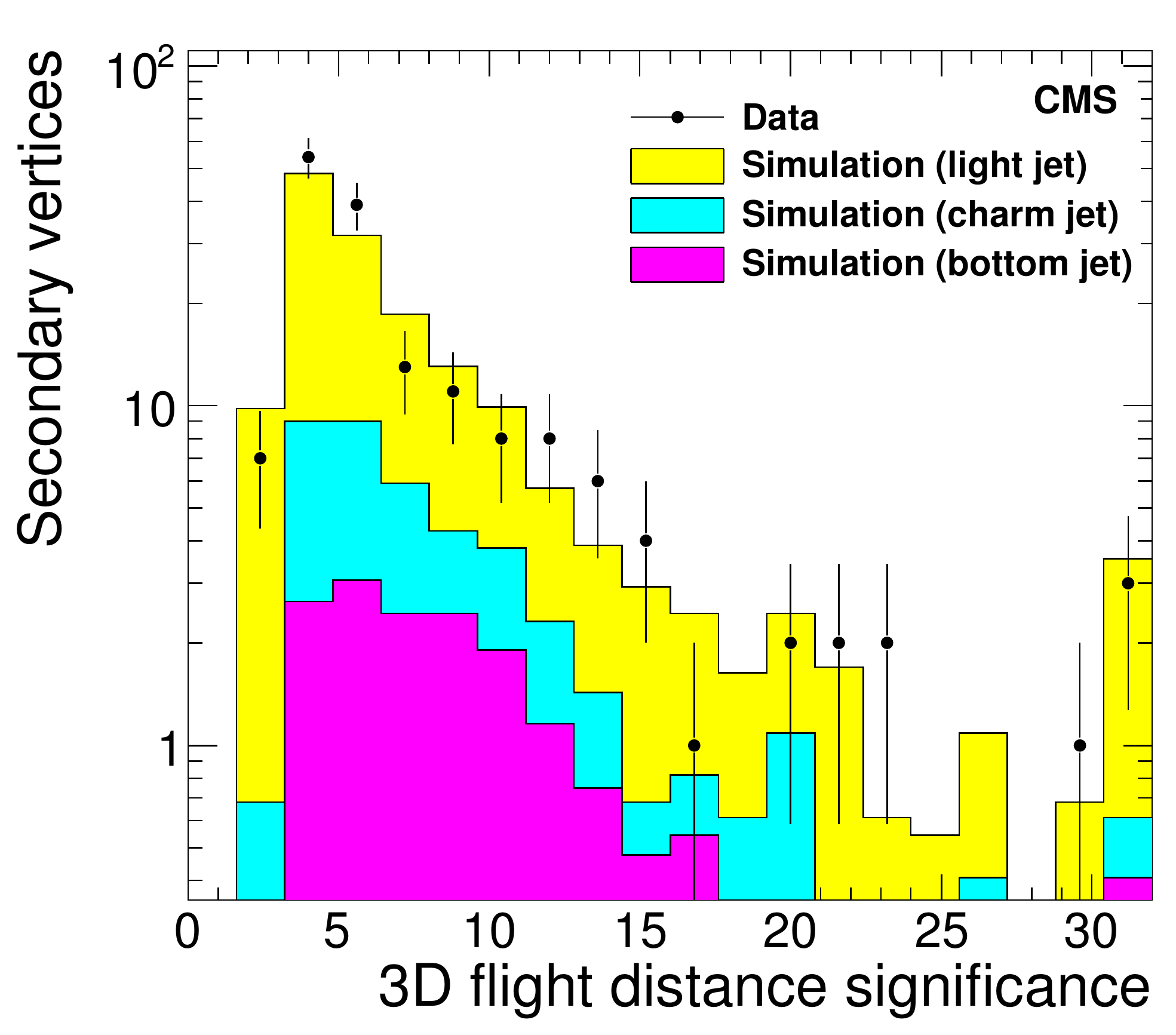}
   \label{fig:btag_sv} 
      }}
    }
    \caption{Distribution of (a) the significance of the three-dimensional impact parameter for all tracks 
in a jet and (b) the significance of the three-dimensional displacement of the secondary vertex.  
The data are shown as full circles while the simulation contributions from light flavour, 
charm, and bottom are shown as different-shaded histograms. 
The outermost bins contain the respective histogram underflow/overflow.}
\end{center}
\end{figure}

The impact parameter is computed with respect to the reconstructed primary vertex and the
distributions are compared between data and a minimum bias simulation reconstructed with the
same algorithm settings.  Figure~\ref{fig:btag_ip} shows the three-dimensional
impact parameter significance distribution for all tracks in a jet.

\begin{figure}[hbtp]
  \begin{center}
    \includegraphics[width=0.5\textwidth]{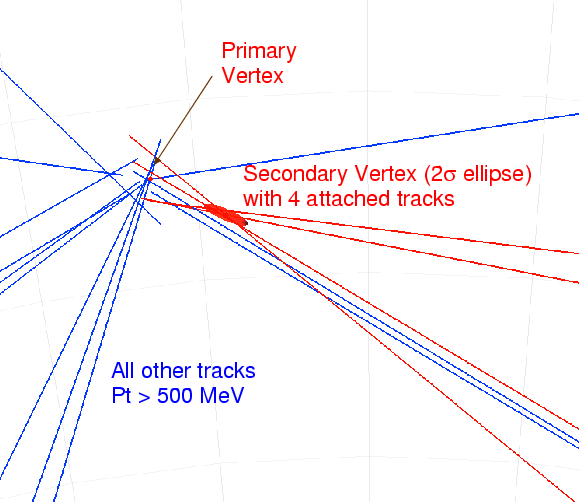}
    \caption{Display of an event with a four-track secondary vertex.  The vertex is separated from the primary vertex
by 7$\sigma$ and the invariant mass of the four particles is 1.64\GeVcc, assuming they are all pions.}
    \label{fig:fourtrackvertex}
  \end{center}
\end{figure}

The secondary-vertex reconstruction using the tracks associated to jets has also been 
slightly modified compared to the algorithm described in Ref.~\cite{CMS_PAS_BTV_09_001}. The differences are a
looser track selection, a relaxed vertex-to-jet direction compatibility, the use of track
refitting in the secondary-vertex fit, and the use of the primary-vertex constraint rather than
the beamspot.  In addition, to suppress K$_\mathrm{S}^0$ candidates, the transverse secondary-vertex separation must be
less than 2.5\cm and the secondary-vertex invariant mass more than 15\MeVcc from the nominal K$_\mathrm{S}^0$ mass.
The significance of the distance between primary and secondary vertices is compared to what is expected from
a simulation of minimum bias events in Fig.~\ref{fig:btag_sv}.  While many two- and three-track vertices are
reconstructed, only one four-track vertex is found in the data.  This event is shown in Fig.~\ref{fig:fourtrackvertex}.

\section{Conclusion}
\label{sec:conclusion}
The all-silicon CMS Tracker was designed for LHC physics. During its
conception, emphasis was placed on providing fine granularity, large
pseudorapidity coverage, and a number of redundant measurements to
facilitate the pattern recognition. Meeting these design criteria has
resulted in excellent tracking performance for the large-volume
detector operating in a 3.8\,T magnetic field in the early collision
running. The proximity of the pixel detector to the LHC beam line
permits precise reconstruction of primary and secondary vertices.  The
extended commissioning with cosmic rays in 2008 and 2009 allowed most
calibrations to be completed and provided a good initial alignment for
most of the detector.  This allowed fast and reliable operation of the
Tracker in the first LHC collisions in December 2009.

The performance of the Tracker has been studied using the collision
data at centre-of-mass energies of 0.9 and 2.36\,TeV\@.  The tracking
and vertexing resolutions are in agreement with the expected design
performance in minimum bias events, for the level of the alignment
achieved.  Studies of the decays of K$_\mathrm{S}^0, \Lambda^0$, $\Xi^-$, and
K$^*(892)^-$ test the capability to reconstruct displaced vertices and
agree well with predictions from simulation.  In particular,
measurements of V$^0$ mass, resolution, and lifetime provide strong
tests of our understanding of the magnetic field, Tracker material,
and detector performance.  Our knowledge of the Tracker material is
also evident in the agreement between data and simulation for photon
conversions and nuclear interactions.  Energy loss measurements in the
Tracker, made possible by the calibration of the silicon strip
analogue readout chain, provide good particle identification at low
momentum as seen in the reconstructed $\phi$ and $\Lambda$ decays.
Finally, the alignment parameters determined in the cosmic ray studies
are already adequate for the precise determination of impact
parameters and the reconstruction of secondary vertices. This will
ultimately be exploited for b-hadron physics and top-quark studies.

In conclusion, the CMS Tracker has been commissioned to take data at
the LHC\@. New collision data will allow more precise alignment and
calibration, which will enable the study of the new vista of particle
physics that is now opening up at the LHC.

\section*{Acknowledgements}
\label{sec:acknowledgments}
We wish to congratulate our colleagues in the CERN accelerator departments for the
excellent performance of the LHC machine. We thank the technical and administrative staff
at CERN and other CMS institutes for their devoted efforts during the design, construction
and operation of CMS. The cost of the detectors, computing infrastructure, data
acquisition and all other systems without which CMS would not be able to operate was
supported by the financing agencies involved in the experiment. We are particularly
indebted to: the Austrian Federal Ministry of Science and Research; the Belgium Fonds de
la Recherche Scientifique, and Fonds voor Wetenschappelijk Onderzoek; the Brazilian
Funding Agencies (CNPq, CAPES, FAPERJ, and FAPESP); the Bulgarian Ministry of Education
and Science; CERN; the Chinese Academy of Sciences, Ministry of Science and Technology,
and National Natural Science Foundation of China; the Colombian Funding Agency
(COLCIENCIAS); the Croatian Ministry of Science, Education and Sport; the Research
Promotion Foundation, Cyprus; the Estonian Academy of Sciences and NICPB; the Academy of
Finland, Finnish Ministry of Education, and Helsinki Institute of Physics; the Institut
National de Physique Nucl\'eaire et de Physique des Particules~/~CNRS, and Commissariat
\`a l'\'Energie Atomique, France; the Bundesministerium f\"ur Bildung und Forschung,
Deutsche Forschungsgemeinschaft, and Helmholtz-Gemeinschaft Deutscher For\-schungs\-zent\-ren,
Germany; the General Secretariat for Research and Technology, Greece; the National
Scientific Research Foundation, and National Office for Research and Technology, Hungary;
the Department of Atomic Energy, and Department of Science and Technology, India; the
Institute for Studies in Theoretical Physics and Mathematics, Iran; the Science
Foundation, Ireland; the Istituto Nazionale di Fisica Nucleare, Italy; the Korean Ministry
of Education, Science and Technology and the World Class University program of NRF, Korea;
the Lithuanian Academy of Sciences; the Mexican Funding Agencies (CINVESTAV, CONACYT, SEP,
and UASLP-FAI); the Pakistan Atomic Energy Commission; the State Commission for Scientific
Research, Poland; the Funda\c{c}\~ao para a Ci\^encia e a Tecnologia, Portugal; JINR
(Armenia, Belarus, Georgia, Ukraine, Uzbekistan); the Ministry of Science and Technologies
of the Russian Federation, and Russian Ministry of Atomic Energy; the Ministry of Science
and Technological Development of Serbia; the Ministerio de Ciencia e Innovaci\'on, and
Programa Consolider-Ingenio 2010, Spain; the Swiss Funding Agencies (ETH Board, ETH
Zurich, PSI, SNF, UniZH, Canton Zurich, and SER); the National Science Council, Taipei;
the Scientific and Technical Research Council of Turkey, and Turkish Atomic Energy
Authority; the Science and Technology Facilities Council, UK; the US Department of Energy,
and the US National Science Foundation.

Individuals have received support from the Marie-Curie IEF program (European Union); the
Leventis Foundation; the A. P. Sloan Foundation; the Alexander von Humboldt Foundation;
the Associazione per lo Sviluppo Scientifico e Tecnologico del Piemonte (Italy); the
Belgian Federal Science Policy Office; the Fonds pour la Formation \`a la Recherche dans
l'\'industrie et dans l'\'Agriculture (FRIA-Belgium); and the Agentschap voor Innovatie
door Wetenschap en Technologie (IWT-Belgium).

\bibliography{auto_generated}   
\clearpage
\appendix

%
%

\cleardoublepage\appendix\section{The CMS Collaboration \label{app:collab}}\begin{sloppypar}\hyphenpenalty=5000\widowpenalty=500\clubpenalty=5000\textbf{Yerevan Physics Institute,  Yerevan,  Armenia}\\*[0pt]
V.~Khachatryan, A.M.~Sirunyan, A.~Tumasyan
\vskip\cmsinstskip
\textbf{Institut f\"{u}r Hochenergiephysik der OeAW,  Wien,  Austria}\\*[0pt]
W.~Adam, T.~Bergauer, M.~Dragicevic, J.~Er\"{o}, C.~Fabjan, M.~Friedl, R.~Fr\"{u}hwirth, V.M.~Ghete, J.~Hammer\cmsAuthorMark{1}, S.~H\"{a}nsel, M.~Hoch, N.~H\"{o}rmann, J.~Hrubec, M.~Jeitler, G.~Kasieczka, W.~Kiesenhofer, M.~Krammer, D.~Liko, I.~Mikulec, M.~Pernicka, H.~Rohringer, R.~Sch\"{o}fbeck, J.~Strauss, A.~Taurok, F.~Teischinger, W.~Waltenberger, G.~Walzel, E.~Widl, C.-E.~Wulz
% ...
\vskip\cmsinstskip
% ...
1:~~Also at CERN, European Organization for Nuclear Research, Geneva, Switzerland\\
% ...

\end{sloppypar}
\end{document}